\documentclass[a4paper,11pt]{article}
\usepackage{jcappub}
\usepackage{amssymb}
\usepackage{graphicx}
\usepackage{amsmath}
\usepackage{hyperref}

\title{\boldmath Bubble wall velocity from number density current in (non)equilibrium}

\author[1,2]{Zi-Yan Yuwen,}
\emailAdd{yuwenziyan@itp.ac.cn}
\author[3]{Jun-Chen Wang,}
\emailAdd{junchenwang@stu.pku.edu.cn}
\author[1,4]{Shao-Jiang Wang}
\emailAdd{schwang@itp.ac.cn (Corresponding author)}

\affiliation[1]{CAS Key Laboratory of Theoretical Physics, Institute of Theoretical Physics, Chinese Academy of Sciences, Beijing 100190, China}
\affiliation[2]{School of Physical Sciences, University of Chinese Academy of Sciences (UCAS), Beijing 100049, China}
\affiliation[3]{School of Physics, Peking University, Beijing 100871, China}
\affiliation[4]{Asia Pacific Center for Theoretical Physics (APCTP), Pohang 37673, Korea}

\abstract{Cosmological first-order phase transitions (FOPTs) serve as comprehensive probes into our early Universe with associated generations of stochastic gravitational waves and superhorizon curvature perturbations or even primordial black holes. In characterizing the FOPT, phenomenological parameters like transition temperatures, strength factors, bubble separations, and energy budgets can be easily extracted from the macroscopic equilibrium features of the underlying particle physics models except for the terminal wall velocity of the bubble expansion, making it the last key parameter to be determined most difficultly due to the non-equilibrium nature of the microscopic transition model. In this paper, we propose a new model-independent approach to calculate the bubble wall velocity by virtue of an extra junction condition from the conservation and violation of the total number density current across the shock front (if any) and bubble wall, respectively.}

\begin{document}
\maketitle
\flushbottom

\section{Introduction}\label{sec:introduction}

The cosmological first-order phase transition (FOPT)~\cite{Mazumdar:2018dfl,Hindmarsh:2020hop,Caldwell:2022qsj,Athron:2023xlk} is one of the most inspiring phenomenon in the early Universe to probe new physics~\cite{Cai:2017cbj,Bian:2021ini} with associated productions of stochastic gravitational-wave backgrounds (SGWBs)~\cite{Caprini:2015zlo,Caprini:2019egz}, density-induced curvature perturbations at super-horizon scales~\cite{Liu:2022lvz,Elor:2023xbz,Lewicki:2024ghw,Cai:2024nln}, and even the formations of primordial black holes via colliding~\cite{Hawking:1982ga,Moss:1994pi,Moss:1994iq,Jung:2021mku}, squeezing~\cite{Baker:2021nyl,Baker:2021sno,Cline:2022xhx,Kawana:2021tde,Lu:2022paj,Kawana:2022lba,Huang:2022him}, or accumulating~\cite{Kodama:1982sf,Liu:2021svg,Hashino:2021qoq,He:2022amv,Kawana:2022olo,Gouttenoire:2023naa,Lewicki:2023ioy,Kanemura:2024pae,Flores:2024lng,Cai:2024nln} mechanisms, as well as the generations of baryogenesis~\cite{Cohen:1990it,Cohen:1990py,Cohen:1993nk,Cohen:1994ss,Cohen:2012zza} and primordial magnetic fields~\cite{Hogan:1983zz,Quashnock:1988vs,Vachaspati:1991nm,Cheng:1994yr,Di:2020kbw,Yang:2021uid}. In particular, the recent GW detections have put strong constraints on the cosmological FOPTs in practice with the real data. For example, a strong FOPT at the LIGO-Virgo band can be marginally ruled out~\cite{Romero:2021kby,Huang:2021rrk,Jiang:2022mzt,Badger:2022nwo} even after taking into account the GW contributions from both wall collisions and fluid motions (dominantly sound waves)~\cite{Yu:2022xdw}. Note here that a strong FOPT usually occurs for a 
simultaneous nucleation of bubbles~\cite{Leitao:2015fmj,Megevand:2016lpr,Megevand:2017vtb,Jinno:2017ixd,Cutting:2018tjt} when the percolation temperature is significantly lower than the nucleation temperature of the first bubble~\cite{Kobakhidze:2017mru,Cai:2017tmh}. 

The bubble wall velocity plays a key role in determining the properties of FOPTs. For example, the GW energy density spectra from wall collisions~\cite{Witten:1984rs,Hogan:1986qda,Kosowsky:1991ua,Kosowsky:1992rz,Kosowsky:1992vn,Kamionkowski:1993fg,Huber:2008hg,Caprini:2007xq,Caprini:2009fx}, fluid motions including sound wave~\cite{Hogan:1986qda,Hindmarsh:2013xza,Hindmarsh:2015qta,Hindmarsh:2017gnf} and turbulences~\cite{Witten:1984rs,Kamionkowski:1993fg,Kosowsky:2001xp,Dolgov:2002ra,Nicolis:2003tg,Caprini:2006jb,Gogoberidze:2007an,Caprini:2009yp,Niksa:2018ofa,Pol:2019yex,Brandenburg:2021tmp,Brandenburg:2021bvg} crucially depend on the terminal wall velocity of bubble expansion. Specifically, the peak amplitudes of GW spectra from wall collisions~\cite{Jinno:2016vai,Jinno:2017fby,Konstandin:2017sat,Cutting:2020nla,Cutting:2018tjt} and fluid motions~\cite{Hindmarsh:2013xza,Hindmarsh:2015qta,Hindmarsh:2017gnf,Caprini:2009yp,Niksa:2018ofa,Pol:2019yex,Brandenburg:2021tmp,Brandenburg:2021bvg} depend cubically and linearly on the terminal wall velocity, respectively, while the peak frequencies from wall collisions~\cite{Jinno:2016vai,Jinno:2017fby,Konstandin:2017sat,Cutting:2020nla,Cutting:2018tjt}  and fluid motions~\cite{Hindmarsh:2013xza,Hindmarsh:2015qta,Hindmarsh:2017gnf,Caprini:2009yp,Niksa:2018ofa,Pol:2019yex,Brandenburg:2021tmp,Brandenburg:2021bvg} depend almost constantly and inversely on the terminal wall velocity, respectively. The efficiency factors for converting the released vacuum energy into kinetic energies of wall collisions~\cite{Ellis:2019oqb,Ellis:2020nnr,Cai:2020djd,Lewicki:2022pdb} and fluid motions~\cite{Espinosa:2010hh,Cai:2018teh,Giese:2020rtr,Giese:2020znk,Wang:2020nzm,Tenkanen:2022tly,Wang:2022lyd,Wang:2023jto} also depend explicitly on the terminal wall velocity. In particular, as a complementary description to the original sound shell model~\cite{Hindmarsh:2016lnk,Hindmarsh:2019phv,Guo:2020grp} with freely propagating sound shells long after bubble percolations, the recently proposed hydrodynamic sound shell model~\cite{Cai:2023guc} captures a long-missing contribution from forced propagating sound shells before bubble percolations, whose double-shell contribution gradually dominates over the single-shell contribution with decreasing wall velocity, leading to a wide dome around the peak or even likely the emergence of double peaks when further lowering the wall velocity. Furthermore, the recognition of the infrared $k^3$-scaling in the GW energy density spectrum~\cite{Cai:2023guc} expected by causality~\cite{Cai:2019cdl} was later confirmed by both numerical simulation~\cite{Sharma:2023mao} and analytic calculation~\cite{RoperPol:2023dzg}, where the original $k^9$-scaling turns out to show up near the peak only when the lifetime of sound shells is much longer than that of bubble walls. In all of the examples presented above, the bubble wall velocity is of essential importance. However, the determination of the bubble wall velocity from either microscopic particle physics or macroscopic hydrodynamics is rather difficult, making it the last key parameter to be drawn from the general features of the cosmological FOPT.  

The microscopic approach aims at precisely calculating the pressure recoil~\cite{Turok:1992jp} directly on the bubble wall imposed by the interacting particles from the ambient plasma around the wall. This wall pressure recoil is usually referred to as the friction force in the literature, which is most conveniently computed in the ultra-relativistic limit to extract its asymptotic dependence on the Lorentz factor $\gamma_w\equiv(1-\xi_w^2)^{-1/2}$ of the terminal wall velocity $\xi_w$. When eventually balancing the driving force in the ultra-relativistic regime, this $\gamma_w$-dependent friction force would finally lead to an estimation for the terminal wall velocity~\cite{Moore:2000wx}. This microscopic approach gives rise to the $\gamma_w^0$-dependence at the leading order (LO) from particle transmissions and reflections~\cite{Bodeker:2009qy}, $\gamma_w^1$-dependence at the next-to-leading order (NLO) from transition splitting of a fermion emitting a soft vector boson~\cite{Bodeker:2017cim}, $\gamma_w^2$-dependence from resumming multiple soft gauge bosons scattering to all orders~\cite{Hoche:2020ysm}, and $\gamma_w^1$-dependence from resumming real and virtual gauge emissions at all leading-log orders~\cite{Gouttenoire:2021kjv}. The effect of a finite wall width has been considered in Ref.~\cite{Ai:2023suz} for the light-to-heavy transition splitting process that leads to a logarithmic dependence on $\gamma_w$. A recent successive study~\cite{Long:2024sqg} has also considered the finite-wall-width effect and further refined their previous work~\cite{Hoche:2020ysm} by reproducing the $\gamma_w^1$-dependence with their semiclassical current radiation formalism for soft emissions. The effects of longitudinal modes of emissions have not been fully appreciated until Refs.~\cite{GarciaGarcia:2022yqb,Azatov:2023xem} for the estimations of friction at LO and NLO, respectively, leading to a non-monotonic dependence on $\gamma_w$~\cite{GarciaGarcia:2022yqb}. 

Nevertheless, the above microscopic approach is preferably implemented in the ultra-relativistic regime when the interacting time scale for particles passing through or reflected away from the wall is much shorter than the relaxation/thermalization time scale in the plasma, in which case it is better to use the ballistic approximation~\cite{BarrosoMancha:2020fay,Azatov:2020ufh,Balaji:2020yrx} or even the equilibrium distributions to directly compute the momentum transfer~\cite{Bodeker:2009qy}. However, for a relatively smaller terminal wall velocity so that the interaction time can be larger than the relaxation time to thermalize the distributions, a more straightforward approach can be adopted by solving the Boltzmann equations~\cite{Joyce:1994zt,Joyce:1994zn,Moore:1995ua,Moore:1995si,Huber:2013kj,Konstandin:2014zta,Cline:2020jre,Laurent:2020gpg,Dorsch:2021ubz,Dorsch:2021nje,DeCurtis:2022hlx,Laurent:2022jrs,DeCurtis:2023hil,Dorsch:2023tss,DeCurtis:2024hvh} with appropriate assumptions on the deviation from the equilibrium distribution, but it is only feasible for a handful of specific models (see, e.g.~\cite{John:2000zq,Cline:2000nw,Carena:2000id,Huber:2001xf,Carena:2002ss,Konstandin:2005cd,Cirigliano:2006dg,Kozaczuk:2015owa,Dorsch:2018pat,Friedlander:2020tnq,Wang:2020zlf,Lewicki:2021pgr,Cline:2021iff,Jiang:2022btc}). Moreover, for a much smaller terminal wall velocity, the above Boltzmann approach becomes even more difficult to play with, as this small wall velocity is usually preferred in the strongly-coupled FOPTs~\cite{Li:2023xto,Wang:2023lam,Kang:2024xqk} as indicated from holographic numerical simulations~\cite{Bea:2021zsu,Bigazzi:2021ucw,Janik:2022wsx,Bea:2022mfb} of holographic models~\cite{Bigazzi:2020avc,Ares:2020lbt,Bigazzi:2020phm,Zhu:2021vkj,Ares:2021ntv,Ares:2021nap,Morgante:2022zvc,Cai:2022omk,He:2022amv,Zhao:2022uxc,Li:2023mpv,He:2023ado}. Remarkably, the first-order hydrodynamics works perfectly fine with the holographic numerical simulations except at the wall position~\cite{Bea:2021zsu,Janik:2022wsx}, across which the delicate contributions to the pressure differences near or away from the wall can still be exactly captured by the first-order hydrodynamics~\cite{Wang:2022txy,Wang:2023kux}.

The macroscopic approaches, on the other hand, devote to determine the terminal wall velocity in terms of other equilibrium characteristics from hydrodynamics of bubble expansion~\cite{Enqvist:1991xw,Dine:1992wr,Liu:1992tn,Laine:1993ey,Ignatius:1993qn,Carrington:1993ng,Heckler:1994uu,Kurki-Suonio:1995rrv,KurkiSuonio:1996rk}. One of the macroscopic approaches is to directly simulate the bubble wall expansion~\cite{Lewicki:2022nba,Krajewski:2023clt,Krajewski:2024gma} with some phenomenological parameterizations~\cite{Megevand:2009ut,Megevand:2009gh,Espinosa:2010hh,Leitao:2010yw,Huber:2011aa,Megevand:2012rt,Megevand:2013hwa,Huber:2013kj,Megevand:2013hwa,Megevand:2013yua,Megevand:2014yua,Leitao:2014pda,Megevand:2014dua,Leitao:2015ola} for the friction term that interconnects the conservation equations of the energy-momentum tensors between the scalar wall and plasma fluid. However, such phenomenological parameterizations for the friction term in the equation of motions (EoMs) are only borrowed from previous estimations in the ultra-relativistic regime for their asymptotic $\gamma_w$ dependence~\cite{Bodeker:2009qy,Bodeker:2017cim,Hoche:2020ysm,Gouttenoire:2021kjv,Long:2024sqg,GarciaGarcia:2022yqb}, and hence they have never been rigorously verified for the entire process of time evolution during which the bubble wall starts to evolve from scratch in the numerical simulation. 
The other macroscopic approach is to impose an extra junction condition across the bubble wall, when combined with the usual junction condition from the conservation of the total energy-momentum tensor across the bubble wall, leading to an extra constraint from which the terminal wall velocity can be solved model-independently in terms of other equilibrium characteristics of hydrodynamics. One such a recent attempt is to impose an extra junction condition, $s\bar{\gamma}\bar{v}=\mathrm{const}.$~\cite{Ai:2021kak,Ai:2023see,Ai:2024shx}, across the bubble wall from a conserved entropy density current equation, $\nabla_\mu(su^\mu)=0$, assuming local equilibrium near the wall, where $\bar{\gamma}\equiv(1-\bar{v}^2)^{-1/2}$ is the Lorentz factor the fluid velocity $\bar{v}$ in the rest frame of bubble wall and $su^\mu$ is the entropy density current in the rest frame of plasma fluid. 

However, as shown in Appendix~\ref{app:EntropyFlow}, this extra junction condition is explicitly broken across the shock front regardless of wall geometries, whereas the entropy density current should be conserved across the shock front followed directly from the conservation of total energy-momentum tensor if the thin-wall approximation is assumed. Therefore, as long as we work with hydrodynamics under the thin-wall approximation, we cannot use this extra junction condition as an appropriate matching condition across the shock front, nor should we apply it to the bubble wall in local equilibrium. In fact, the junction condition should be derived rigorously from an integral form (weak form) of the conservation equation. For example, the usual junction conditions from the conserved energy-momentum tensor $T^{\mu\nu}$ can be rigorously derived from the weak-form conservation equation $\int_\mathcal{V}\nabla_\mu(T^{\mu\nu}\lambda_\nu)\mathrm{d}^4x=0$, whose integrand is a total 4-derivative that can be fully integrated out via Stokes's theorem into some junction conditions. However, the weak-form conservation equation of the entropy density current, $\int_\mathcal{V}T\nabla_\mu(su^\mu)\mathrm{d}^4x=0$, is not an integration over any total 4-derivative to arrive at any forms of junction conditions. Note that the temperature $T$ as a prefactor in the integrand cannot be simply removed from the weak-form conservation equation since the temperature profile is discontinuous across the bubble-wall/shock-front interfaces as long as the thin-wall approximation is used for hydrodynamics.

To propose another weak-form conservation equation with its integration over a total 4-derivative analog to the entropy density current conservation in local equilibrium, we recall that in hydrodynamics, the complete set of conservation equations contains not only the conservation of the total energy-momentum tensor, $\nabla_\mu T^{\mu\nu}=0$, but also the conservation of the number density current, $\nabla_\mu N^\mu=0$ with $N^\mu\equiv nu^\mu$. Only in our case with an FOPT, the scalar wall also contributes to the number density current in addition to plasma particles. This is because, with the thin-wall approximation, both the symmetric and broken phases are in exactly false and true vacuum states, respectively, on top of which the scalar excitations can be regarded as some scalar-wall particles interacting with plasma particles. Hence, the phase-transition scalar field plays two-fold roles: the background part acts as an external force while the excitation part also contributes to collision terms on the left- and right-hand sides of the Boltzmann equation, respectively. Therefore, the thin-wall scalar field should also contribute to the number density current in addition to the plasma particles. The resulting total number density current can be proved to be conserved provided that the local dynamical equilibrium can be maintained with the cross-section of any given process counterbalanced by its inverse process with the same cross-section, which is exactly the case when crossing the shock front well within the symmetric phase with thermal equilibrium in the early Universe. However, different from the conservation across the shock front in the symmetric phase, the total number density current can be violated across the bubble wall due to asymmetric cross-sections between the symmetric and broken phases.

In this paper, we first propose the junction condition in Sec.~\ref{sec:setup} from the conservation and violation of the total number density current across the shock front (if any) and bubble wall, respectively, and then calculate the bubble wall velocity in Sec.~\ref{sec:BubbleWallVelocity} for the weak/Jouguet deflagration and weak detonation modes, correspondingly. The last section is devoted for conclusions and discussions.

\section{Hydrodynamics}\label{sec:setup}

In this section, we will first review in Sec.~\ref{subsec:EoMs} the first-order hydrodynamics~\cite{Espinosa:2010hh} to describe the late-time stage of bubble expansion from a fast FOPT, and then introduce in Sec.~\ref{subsec:JunctionConditions} our junction conditions from the conservation equations of the total energy-momentum tensor and number density current.

\subsection{Equation of motions}\label{subsec:EoMs}

The conventional working assumption for using the first-order hydrodynamics to depict the vacuum bubble expansion within the thermal plasma fluid is to focus on the late-time stage of a fast FOPT as we have summarized in Ref.~\cite{Wang:2023kux}. The FOPT is fast if it can be completed within one Hubble time so that the background expansion can be neglected and treated as a flat spacetime to set up global coordinates at the bubble nucleation site with $x^\mu=(t,z,x=0,y=0),(t,\rho,\varphi=0,z=0), (t,r,\theta=0,\varphi=0)$ for planar, cylindrical, and spherical walls, respectively, where all the last two components can be set to zero by virtue of geometric symmetries of the wall moving along the $x^1$ direction. The late-time stage of bubble expansion is considered so that (i) the initial bubble radius is negligible and hence the bubble expansion is self-similar in $\xi\equiv x^1/x^0$; (ii) the bubble wall width becomes thinner and thinner and hence the thin-wall approximation can be adopted; (iii) the bubble wall velocity has approached asymptotically a constant terminal value and hence the bubble expansion has reached a steady state in the self-similar frame. With above assumptions, the scalar-plasma system of bubble wall expansion in thermal plasma can be simplified into a wall-fluid system.

\paragraph{The scalar-plasma system} 
For any scalar field $\phi$ with a vacuum potential $V_0(\phi)$ interacting with the thermal-plasma particles of species $i$ with degeneracy factor $g_i$ (accounting for spin, internal degrees of freedom, etc.) and distribution function $f_i(x^\mu,k^\mu)$, the total energy-momentum tensor is presumed to be separable, $T^{\mu\nu}=T_\phi^{\mu\nu}+T_f^{\mu\nu}$, with
\begin{align}
T_\phi^{\mu\nu}&=\nabla^\mu\phi\nabla^\nu\phi+g^{\mu\nu}\left[-\frac12(\nabla\phi)^2-V_0(\phi)\right],\label{eq:Tphi}\\
T_f^{\mu\nu}&=\sum\limits_{i=\mathrm{B,F}}g_iT_{f_i}^{\mu\nu}=\sum\limits_{i=\mathrm{B,F}}g_i\int\mathcal{D}k\,k^\mu k^\nu f_i(x,k),\label{eq:Tpla}
\end{align}
where the integral measure $\mathcal{D}k$ is defined to yield a Lorentz-invariant 3-momentum integral with a positive energy $E_i(\mathbf{k})=\sqrt{\mathbf{k}^2+m_i^2}$ for the particle with a mass $m_i(\phi)$,
\begin{align}
\int\mathcal{D}k\equiv\int\frac{\mathrm{d}^3\mathbf{k}}{(2\pi)^3}\frac{1}{E_i(\mathbf{k})}=\int\frac{\mathrm{d}^4k}{(2\pi)^4}\frac{2}{\sqrt{-g}}(2\pi)\delta(k^2+m_i^2)\Theta(k^0).
\end{align}
For thermal plasma without bulk viscosity and shear as well as diffusion and dissipation, its stress tensor can be rearranged into a perfect-fluid form in the first-order hydrodynamics as
\begin{align}\label{eq:plasmafluid}
T_f^{\mu\nu}
=\sum\limits_{i=\mathrm{B,F}}g_i\int\frac{\mathrm{d}^3\mathbf{k}}{(2\pi)^3}\left[\left(E_i+\frac{\mathbf{k}^2}{3E_i}\right)u^\mu u^\nu+g^{\mu\nu}\frac{\mathbf{k}^2}{3E_i}\right]f_i
\equiv(e_f+p_f)u^\mu u^\nu+p_fg^{\mu\nu},
\end{align}
where the energy density and pressure of this thermal plasma are defined in the local rest frame of fluid element as
\begin{align}
e_f=\sum\limits_{i=\mathrm{B,F}}g_i\int\frac{\mathrm{d}^3\mathbf{k}}{(2\pi)^3}E_i(\mathbf{k})f_i,\quad
p_f=\sum\limits_{i=\mathrm{B,F}}g_i\int\frac{\mathrm{d}^3\mathbf{k}}{(2\pi)^3}\frac{\mathbf{k}^2}{3E_i(\mathbf{k})}f_i.
\end{align}

The particle number density current $N_f^\mu\equiv(n_f,\mathbf{N}_f)$ is made of the particle number density $n_f$ and particle number flux $\mathbf{N}_f$, which can be uniformly defined as
\begin{align}
N_f^\mu=\sum\limits_{i=\mathrm{B,F}}g_i N_{f_i}^\mu=\sum\limits_{i=\mathrm{B,F}}g_i\int\mathcal{D}k\,k^\mu f_i(x,k).
\end{align}
As $N_f^0=n_f$ and $\mathbf{N}_f=0$ in the rest frame of plasma particles, the only covariant way to express the particle number density current is $N_f^\mu=n_fu^\mu$. Similarly, there is also a particle entropy density current defined by
\begin{align}
S_f^\mu=\sum\limits_{i=\mathrm{B,F}}g_iS_{f_i}^\mu=\sum\limits_{i=\mathrm{B,F}}g_i\int\mathcal{D}k\,k^\mu[\pm(1\pm f_i)\ln(1\pm f_i)-f_i\ln f_i],
\end{align}
whose temporal component is simply the particle entropy density $S_f^0=s$ in the rest frame of plasma particles. Based on the same argument above, the only covariant way to express the particle entropy density current is $S_f^\mu=s_fu^\mu$.

For each particle of species $i$ moving along its worldline $x^\mu(\lambda)$ with the affine parameter $\lambda$ defined by $k^\mu=\mathrm{d}x^\mu/\mathrm{d}\lambda$ for a massless particle and $\lambda=\tau/m_i$ for a massive particle with $\mathrm{d}\tau^2=-\mathrm{d}s^2=-g_{\mu\nu}\mathrm{d}x^\mu\mathrm{d}x^\nu$, the distribution function of particle species $i$ is governed by the corresponding relativistic Boltzmann equation,
\begin{align}
\frac{\mathrm{D}}{\mathrm{d}\lambda}f_i(x^\mu,k^\mu)
=\left(\frac{\mathrm{D}x^\mu}{\mathrm{d}\lambda}\nabla_\mu+\frac{\mathrm{D}k^\mu}{\mathrm{d}\lambda}\frac{\partial}{\partial k^\mu}\right)f_i
=\left(k^\mu\nabla_\mu-\frac12\nabla^\mu\phi\frac{\mathrm{d}m_i^2}{\mathrm{d}\phi}\frac{\partial}{\partial k^\mu}\right)f_i=C[f_i],
\end{align}
where the directional derivative of 4-momentum is given by the geodesic equation $\mathrm{D}k^\mu/\mathrm{d}\lambda=\mathrm{d}k^\mu/\mathrm{d}\lambda+\Gamma^\mu_{\nu\sigma}k^\nu k^\sigma=m_iF_i^\mu$ with an external force $F_i^\mu=-m'_i(\phi)\nabla^\mu\phi$ depending on the background scalar field $\phi(x^\mu)$. Here the collision term $C[f_i]$ is complicated by nature but it at least renders a monotonically increasing particle entropy density current, $\nabla_\mu S_f^\mu\geq0$, as required by the second law of thermodynamics of any isolated system, which is equivalent to
\begin{align}
\sum\limits_{i=\mathrm{B,F}}g_i\int\mathcal{D}k\,\left(\ln \frac{f_i}{1\pm f_i}\right)C[f_i]\leq0
\end{align}
by the Boltzmann H-theorem with the Boltzmann equation in the absence of external force. Here the equality is reached when approaching local equilibrium with a vanishing collision term, in which case the distribution function can be solved as $f_{i=\mathrm{B,F}}^\mathrm{eq}=[\exp(E_i(\mathbf{k})/T)\mp1]^{-1}$.

\paragraph{Particle number density current} For the $n\to n$ scattering process, the vanishing 4-integral over the collision term directly leads to the conservation of particle number density current (see, for example, Eq.(4.27) of Ref.~\cite{Hindmarsh:2020hop}),
\begin{align}
0&=\sum\limits_{i=\mathrm{B,F}}g_i\int\mathcal{D}k\,C[f_i]
=\sum\limits_{i=\mathrm{B,F}}g_i\int\mathcal{D}k\,k^\mu\nabla_\mu f_i
=\nabla_\mu\left(\sum\limits_{i=\mathrm{B,F}}g_i\int\mathcal{D}k\,k^\mu f_i\right)
=\nabla_\mu N_f^\mu, \label{eq:dNfnoF}
\end{align}
for any solution of the Boltzmann equation $k^\mu\nabla_\mu f_i=C[f_i]$ in the absence of external force $F_i^\mu=-m'_i(\phi)\nabla^\mu\phi=0$ from the scalar field. The same conclusion also holds in the presence of scalar-field external force for the same $n\to n$ scattering process  (we omit the degeneracy factors and summation symbols over particle species for clarity and simplicity that can be easily filled in),
\begin{align}
0&=\int\mathcal{D}k\,C[f]=\int\mathcal{D}k\left(k^\mu\nabla_\mu+mF^\mu\frac{\partial}{\partial k^\mu}\right)f\label{eq:intCf}\\
&=\nabla_\mu\left(\int\mathcal{D}k\,k^\mu f\right)+\int\mathcal{D}k\,mF^\mu\frac{\partial f}{\partial k^\mu}\\
&=\nabla_\mu N_f^\mu+\int\mathcal{D}k\,\frac{\mathrm{D}k^\mu}{\mathrm{d}\lambda}\frac{\partial f}{\partial k^\mu}\\
&=\nabla_\mu N_f^\mu+\int\mathcal{D}k\,\frac{\mathrm{d}x^\nu}{\mathrm{d}\lambda}\nabla_\nu k^\mu\frac{\partial f}{\partial k^\mu}\\
&=\nabla_\mu N_f^\mu+\int\mathcal{D}k\left(\frac{\mathrm{d}f}{\mathrm{d}\lambda}-\frac{\mathrm{d}x^\nu}{\mathrm{d}\lambda}\nabla_\nu f\right)\\
&=\nabla_\mu N_f^\mu+\frac{\mathrm{d}}{\mathrm{d}\lambda}\int\mathcal{D}k\,f-\frac{\mathrm{d}x^\nu}{\mathrm{d}\lambda}\nabla_\nu\int\mathcal{D}k\,f\\
&=\nabla_\mu N_f^\mu+\frac{\mathrm{d}n_f}{\mathrm{d}\lambda}-\frac{\mathrm{d}x^\nu}{\mathrm{d}\lambda}\nabla_\nu n_f\\
&=\nabla_\mu N_f^\mu,
\end{align}
where we have used the Boltzmann equation $(k^\mu\nabla_\mu+mF^\mu\partial/\partial k^\mu)f=C[f]$ in the first line, the definition of number density current $N_f^\mu=\int\mathcal{D}k\,k^\mu f$ in the second line, the geodesic equation $\mathrm{D}k^\mu/\mathrm{d}\lambda=(\mathrm{d}x^\nu/\mathrm{d}\lambda)\nabla_\nu k^\mu=mF^\mu$ in the third and fourth lines, the chain rule $\mathrm{d}f/\mathrm{d}\lambda=(\mathrm{d}x^\nu/\mathrm{d}\lambda)\nabla_\nu f+(\mathrm{d}x^\nu/\mathrm{d}\lambda)\nabla_\nu k^\mu(\partial f/\partial k^\mu)$ in the fifth line, the definition of number density $n_f=\int\mathcal{D}k\,f$ in the sixth line, and the chain rule $\mathrm{d}n_f/\mathrm{d}\lambda=(\mathrm{d}x^\nu/\mathrm{d}\lambda)\nabla_\nu n_f$ in the seventh line. However, for a general $n\to m$ process that does not necessarily conserve the particle number ($n\neq m$) like the particle splitting or annihilation process, the integration over the collision term can be nonzero, $\int\mathcal{D}k\,C[f]\neq0$, and hence the particle number density current is generally not conserved. Nevertheless, if the local dynamical equilibrium can be maintained so that each process can be counterbalanced by its inverse process, such an integration over the collision term will still vanish as we will prove shortly below this paragraph, and the number density current can be conserved in local dynamical equilibrium.

\paragraph{Local dynamical equilibrium} Consider a collision process with $n$ reactant particles transforming into $m$ product particles with a cross-section $\sigma^2_{n,m}(\mathbf{q},\mathbf{p})$, where $\mathbf{q} = ({q}_1, ... , {q}_n)$ and $\mathbf{p} = ({p}_1, ... , {p}_m)$ are the momenta of reactant and product particles, respectively. Next, we evaluate the collision term consisting of both the loss term and the gain term,
\begin{align}
    I_{\mathrm{loss}} &= \sigma^2_{n,m}
    (\mathbf{q},\mathbf{p}) \delta
    \left(\sum_{i=1}^n q_i - \sum_{j=1}^m p_j\right) \prod_{i=1}^n(1+f(q_i)) \prod_{j=1}^m f(p_j), \\
    I_{\mathrm{gain}} &= \sigma^2_{n,m}
    (\mathbf{q},\mathbf{p}) \delta
    \left(\sum_{i=1}^n q_i - \sum_{j=1}^m p_j\right) \prod_{j=1}^m(1+f(p_j)) \prod_{i=1}^n f(q_i).
\end{align}
The collision term for the first product particle $p_1$ is given by an integral of the difference $I_{\mathrm{gain}}-I_{\mathrm{loss}}$ over the momenta of other particles after taking into consideration of all possible $n\to m$ processes,
\begin{align}
    I_\mathrm{col}(p_1) = \sum_{n,m} \int\prod_{i=1}^n \mathcal{D} q_i \prod_{j=2}^m \mathcal{D} p_j \times (I_{\mathrm{gain}}-I_{\mathrm{loss}}).
\end{align}
Then, integrating the collision term over $p_1$ results in
\begin{align}
    \int \mathcal{D} p_1 I_\mathrm{col}(p_1) = 2\sum_{n>m}&\int\prod_{i=1}^n \mathcal{D} q_i \prod_{j=1}^m \mathcal{D} p_j \left( \sigma^2_{n,m}(\mathbf{q},\mathbf{p}) - 
    \sigma^2_{m,n}(\mathbf{p},\mathbf{q}) \right) \nonumber\\
    &\times\left(
    \prod_{j=1}^m(1+f(p_j)) \prod_{i=1}^n f(q_i) - \prod_{i=1}^n(1+f(q_i)) \prod_{j=1}^m f(p_j)
    \right).
\end{align}
For local dynamical equilibrium with detailed balance, any $n\to m$ process should always balance its inverse process $m\to n$ with the same cross-section, $\sigma^2_{n,m}(\mathbf{q},\mathbf{p}) = \sigma^2_{m,n}(\mathbf{p},\mathbf{q})$, and thus the collision integral should simply vanish,
$\int \mathcal{D} p_1 I_\mathrm{col}(p_1) = 0$, and hence conserving the particle number density current. The same argument also holds for multi-component perfect fluids as long as the cross-sections of a process and its inverse process keep the same value. Furthermore, with the thin-wall approximation, both the symmetric and broken phases are in exactly false and true vacuum states, respectively, on top of which the scalar-field excitations interacting with plasma particles should also contribute to the collision terms. Then, the vanishing integral of all collision terms from local dynamical equilibrium among not only plasma particles but also with scalar-wall particles would necessarily conserve the total number density current. Here, the local dynamical equilibrium condition can be easily fulfilled in thermal equilibrium in the early Universe, which is exactly the case for the region across the shock front well within the symmetric phase. Nevertheless, the cross-sections are usually different when crossing the bubble wall between the symmetric and broken phases, and hence the number density current is generally NOT conserved across the bubble wall.

\paragraph{The equation of motion} The EoM of this scalar-plasma system simply follows from the conservation equation of the total energy-momentum tensor, $\nabla_\mu T^{\mu\nu}=\nabla_\mu(T_\phi^{\mu\nu}+T_f^{\mu\nu})=0$, but each of which will be broken by a transfer flow $\mathcal{F}^\mu$ with opposite signs,
\begin{align}
\nabla_\mu T_\phi^{\mu\nu}&\equiv[\nabla_\mu\nabla^\mu\phi-V'_0(\phi)]\nabla^\nu\phi=+\mathcal{F}^\nu,\\
\nabla_\mu T_f^{\mu\nu}&\equiv\sum\limits_{i=\mathrm{B,F}}g_i\int\mathcal{D}k\,k^\mu k^\nu\nabla_\mu f_i=-\mathcal{F}^\nu.
\end{align}
In the absence of external force $F_i^\mu=-m'_i(\phi)\nabla^\mu\phi=0$ from the scalar field, the energy-momentum tensor of the plasma fluid should be conserved on its own,
\begin{align}
0&=\sum\limits_{i=\mathrm{B,F}}g_i\int\mathcal{D}k\,k^\nu C[f_i]
=\sum\limits_{i=\mathrm{B,F}}g_i\int\mathcal{D}k\,k^\nu k^\mu\nabla_\mu f_i=\nabla_\mu\left(\sum\limits_{i=\mathrm{B,F}}g_i\int\mathcal{D}k\,k^\mu k^\nu f_i\right)=\nabla_\mu T_f^{\mu\nu}.\label{eq:dTfnoF}
\end{align}
In the presence of the external force $F_i^\mu=-m'_i(\phi)\nabla^\mu\phi$, we can derive the transfer flow from
\begin{align}\label{eq:dTfwithF}
0&=\sum\limits_{i=\mathrm{B,F}}g_i\int\mathcal{D}k\,k^\nu C[f_i]
=\sum\limits_{i=\mathrm{B,F}}g_i\int\mathcal{D}k\left(k^\mu k^\nu\nabla_\mu f_i+m_iF_i^\mu k^\nu\frac{\partial f_i}{\partial k^\mu}\right)\nonumber\\
&=\nabla_\mu\left(\sum\limits_{i=\mathrm{B,F}}g_i\int\mathcal{D}k\,k^\mu k^\nu f_i\right)-\left(\sum\limits_{i=\mathrm{B,F}}g_i\int\mathcal{D}k\,m_iF_i^\mu\frac{\partial k^\nu}{\partial k^\mu}f_i\right)\nonumber\\
&=\nabla_\mu T_f^{\mu\nu}-\sum\limits_{i=\mathrm{B,F}}g_i\int\mathcal{D}k\,m_iF_i^\nu f_i
\end{align}
as
\begin{align}
\nabla_\mu T_f^{\mu\nu}
=-\sum\limits_{i=\mathrm{B,F}}g_im_im'_i(\phi)\nabla^\nu\phi\int\mathcal{D}k\,f_i
=-\nabla^\nu\phi\sum\limits_{i=\mathrm{B,F}}g_i\frac{\mathrm{d}m_i^2}{\mathrm{d}\phi}\int\frac{\mathrm{d}^3\mathbf{k}}{(2\pi)^2}\frac{f_i}{2E_i(\mathbf{k})}
\equiv-\mathcal{F}^\nu.
\end{align}
It is easy to check that the equilibrium part of transfer flow without the $\nabla^\nu\phi$ prefactor from the equilibrium part of the total distribution function $f_i(x,k)=f_i^\mathrm{eq}(E_i)+\delta f_i(x,k)$ exactly reproduces the field derivative of the finite-temperature correction $\Delta V_T(\phi,T)$ to the total effective potential $V_\mathrm{eff}(\phi,T)=V_0(\phi)+\Delta V_T(\phi,T)$ as shown explicitly in Ref.~\cite{Konstandin:2014zta}~\footnote{We demonstrate a simplified derivation below without specifying and summing over the particle species.
\begin{align}
\frac{\partial\Delta V_T}{\partial\phi}
&=\frac{\mathrm{d}m^2}{\mathrm{d}\phi}\frac{\mathrm{d}}{\mathrm{d}m^2}\left[-\int\frac{\mathrm{d}^3\mathbf{k}}{(2\pi)^3}\frac{\mathbf{k}^2}{3E}f(E)\right]
=-\frac13\frac{\mathrm{d}m^2}{\mathrm{d}\phi}\int\frac{\mathrm{d}^3\mathbf{k}}{(2\pi)^3}\left[f(E)\frac{\mathrm{d}}{\mathrm{d}m^2}\frac{\mathbf{k}^2}{E}+\frac{\mathbf{k}^2}{E}\frac{\mathrm{d}f(E)}{\mathrm{d}m^2}\right]\\
&=-\frac13\frac{\mathrm{d}m^2}{\mathrm{d}\phi}\int\frac{\mathrm{d}^3\mathbf{k}}{(2\pi)^3}\left[-\frac{\mathbf{k}^2}{E^2}\frac{f(E)}{2E}+\frac{\mathbf{k}^2}{E}\frac{1}{2E}\frac{\mathrm{d}f(E)}{\mathrm{d}E}\right]
=-\frac13\frac{\mathrm{d}m^2}{\mathrm{d}\phi}\int\frac{\mathrm{d}^3\mathbf{k}}{(2\pi)^3}\frac{\mathbf{k}^2}{E}\frac{\mathrm{d}}{\mathrm{d}E}\frac{f(E)}{2E}\\
&=-\frac13\frac{\mathrm{d}m^2}{\mathrm{d}\phi}\int\frac{\mathrm{d}^3\mathbf{k}}{(2\pi)^3}\frac{k}{2}\frac{\mathrm{d}}{\mathrm{d}k}\frac{f(E)}{E}
=-\frac13\frac{\mathrm{d}m^2}{\mathrm{d}\phi}\int_0^\infty\frac{4\pi\mathbf{k}^2\mathrm{d}k}{(2\pi)^3}\frac{k}{2}\frac{\mathrm{d}}{\mathrm{d}k}\frac{f(E)}{E}\\
&=-\frac{2\pi}{(2\pi)^3}\frac{\mathrm{d}m^2}{\mathrm{d}\phi}\int_0^\infty\frac{k^3}{3}\mathrm{d}\frac{f(E)}{E}
=-\frac{1}{(2\pi)^2}\frac{\mathrm{d}m^2}{\mathrm{d}\phi}\left[\left(\frac{k^3}{3}\frac{f(E)}{E}\right)\bigg|_0^\infty-\int_0^\infty\mathbf{k}^2\mathrm{d}k\,\frac{f(E)}{E}\right]\\
&=-\frac{1}{(2\pi)^2}\frac{\mathrm{d}m^2}{\mathrm{d}\phi}\frac{k^3}{3\sqrt{\mathbf{k}^2+m^2}}\frac{1}{\exp(\sqrt{\mathbf{k}^2+m^2}/T)\mp1}\bigg|_{k\to0}^{k\to\infty}+\frac{\mathrm{d}m^2}{\mathrm{d}\phi}\int_0^\infty\frac{4\pi\mathbf{k}^2\mathrm{d}k}{(2\pi)^3}\frac{f(E)}{2E}\\
&=\frac{\mathrm{d}m^2}{\mathrm{d}\phi}\int\frac{\mathrm{d}^3\mathbf{k}}{(2\pi)^3}\frac{f(E)}{2E},
\end{align}
where we have abbreviated $f^\mathrm{eq}\equiv f(E)$, $|\mathbf{k}|^2=\mathbf{k}^2$, and $|\mathbf{k}|=k$, and used $\mathrm{d}E/\mathrm{d}m^2=1/2E$ and $E\mathrm{d}E=k\mathrm{d}k$,}
\begin{align}\label{eq:VTfeq}
\frac{\partial\Delta V_T}{\partial\phi}
=\sum\limits_{i=\mathrm{B,F}}g_i\left[-\frac{\mathrm{d}m_i^2}{\mathrm{d}\phi}\int\frac{\mathrm{d}^3\mathbf{k}}{(2\pi)^3}\frac{\mathrm{d}}{\mathrm{d}m_i^2}\frac{\mathbf{k}^2f_i^\mathrm{eq}(E_i)}{3E_i(\mathbf{k})}\right]
=\sum\limits_{i=\mathrm{B,F}}g_i\frac{\mathrm{d}m_i^2}{\mathrm{d}\phi}\int\frac{\mathrm{d}^3\mathbf{k}}{(2\pi)^3}\frac{f_i^\mathrm{eq}(E_i)}{2E_i(\mathbf{k})},
\end{align}
so that the transfer flow should be of a form like
\begin{align}
\mathcal{F}^\mu=\nabla^\mu\phi\sum\limits_{i=\mathrm{B,F}}g_i\frac{\mathrm{d}m_i^2}{\mathrm{d}\phi}\int\frac{\mathrm{d}^3\mathbf{k}}{(2\pi)^2}\frac{f_i^\mathrm{eq}+\delta f_i}{2E_i(\mathbf{k})}
=\nabla^\mu\phi\left(\frac{\partial\Delta V_T}{\partial\phi}-\frac{\partial p_{\delta f}}{\partial\phi}\right),
\end{align}
where a pressure-like term $p_{\delta f}$ defined by the non-equilibrium part of the transfer flow via
\begin{align}
-\frac{\partial p_{\delta f}}{\partial\phi}=\sum\limits_{i=\mathrm{B,F}}g_i\frac{\mathrm{d}m_i^2}{\mathrm{d}\phi}\int\frac{\mathrm{d}^3\mathbf{k}}{(2\pi)^2}\frac{\delta f_i}{2E_i(\mathbf{k})}
\end{align}
does not necessarily coincide with the non-equilibrium pressure term $\delta p_f$ defined by $p_f=-\Delta V_T+\delta p_f$ via
\begin{align}
-\frac{\partial\delta p_f}{\partial\phi}
=\sum\limits_{i=\mathrm{B,F}}g_i\left[-\frac{\mathrm{d}m_i^2}{\mathrm{d}\phi}\int\frac{\mathrm{d}^3\mathbf{k}}{(2\pi)^3}\frac{\mathrm{d}}{\mathrm{d}m_i^2}\frac{\mathbf{k}^2\delta f_i}{3E_i(\mathbf{k})}\right]
\end{align}
since $\delta f_i$ does not depend on $E_i$ alone as $f_i^\mathrm{eq}$ does in deriving Eq.~\eqref{eq:VTfeq}.
The final EoMs of this scalar-plasma system then read
\begin{align}
\nabla_\mu\nabla^\mu\phi-\frac{\partial V_\mathrm{eff}}{\partial\phi}&=-\frac{\partial p_{\delta f}}{\partial\phi},\label{eq:EOMscalar}\\
\nabla_\mu T_f^{\mu\nu}+\nabla^\nu\phi\frac{\partial \Delta V_T}{\partial\phi}&=\nabla^\nu\phi\frac{\partial p_{\delta f}}{\partial\phi}.\label{eq:EOMplasma}
\end{align}

\paragraph{The wall-fluid system} So far, we have not used any of the assumptions (flat space time, self-similar expansion, thin-wall approximation, and steady-state expansion) introduced at the beginning of this subsection since all the discussions above are introduced in general for any scalar field coupled to the thermal plasma in non-equilibrium. For the specific scalar-plasma system describing a steady-state self-similar thin-wall expansion in flat spacetime, we can take the gradient of $\phi$ to be parallel to the fluid 4-velocity $u^\mu$ with a proportionality factor $\psi$, that is, $\nabla^\mu\phi=\psi u^\mu$. From the time-like normalization condition $u_\mu u^\mu=-1$, we can determine $\psi=\sqrt{-(\nabla\phi)^2}$. Then, the energy-momentum tensor of this scalar wall becomes
\begin{align}
T_\phi^{\mu\nu}=\nabla^\mu\phi\nabla^\nu\phi+g^{\mu\nu}\left[-\frac12(\nabla\phi)^2-V_0\right]=\psi^2u^\mu u^\nu+g^{\mu\nu}\left(\frac12\psi^2-V_0\right)\equiv w_\phi u^\mu u^\nu+p_\phi g^{\mu\nu},
\end{align}
where the scalar energy density and pressure are $e_\phi=\frac12\psi^2+V_0$ and $p_\phi=\frac12\psi^2-V_0$, respectively, and the enthalpy reads $w_\phi=e_\phi+p_\phi=\psi^2$. Under the thin-wall approximation, the scalar pressure is everywhere $-V_0$ except at the wall position traced by $r=\xi_w t$, only around which there exists non-equilibrium effect and hence the $\delta p_\phi=\frac12\psi^2$ term can be regarded as a non-equilibrium contribution to the scalar pressure. Therefore, the total pressure reads~\footnote{We mention a byproduct in passing that, if the effective action for the scalar wall can take a form as $S_\mathrm{eff}=\int\mathrm{d}^4x[-\frac12(\nabla\phi)^2+p]$, where the effective potential is corrected by the non-equilibrium effect~\eqref{eq:totalpressure} across the wall, then the EoM reads $\nabla^2\phi=-\frac{\partial p}{\partial\phi}=\frac{\partial V_\mathrm{eff}}{\partial\phi}-\frac{\partial\delta p_\phi}{\partial\phi}-\frac{\partial\delta p_f}{\partial\phi}=\frac{\partial V_\mathrm{eff}}{\partial\phi}-\frac{\partial\delta p_f}{\partial\phi}$ since $\frac{\partial\delta p_\phi}{\partial\phi}=0$, which, after compared to the scalar EoM~\eqref{eq:EOMscalar}, $\nabla^2\phi=\frac{\partial V_\mathrm{eff}}{\partial\phi}-\frac{\partial p_{\delta f}}{\partial\phi}$, would lead to an unexpected identification,
\begin{align}
\frac{\partial\delta p_f}{\partial\phi}=\frac{\partial p_{\delta f}}{\partial\phi}\Rightarrow\nabla_\mu T_f^{\mu\nu}=\nabla^\nu\phi\frac{\partial p_f}{\partial\phi}.
\end{align}}
\begin{align}\label{eq:totalpressure}
p=p_\phi+p_f=(-V_0+\delta p_\phi)+(-\Delta V_T+\delta p_f)=-(V_0+\Delta V_T)+(\delta p_\phi+\delta p_f)=-V_\mathrm{eff}+\delta p.
\end{align}

It is always difficult to specify the non-equilibrium contributions if we discuss the energy-momentum tensors from the scalar-wall and plasma fluid separately. Since both the scalar wall the plasma fluid admit a perfect-fluid form for their energy-momentum tensors, the total energy-momentum tensor must also be of a perfect-fluid form,
\begin{align}
T^{\mu\nu}=T_f^{\mu\nu}+T_\phi^{\mu\nu}=(e+p)u^\mu u^\nu+pg^{\mu\nu}=wu^\mu u^\nu+pg^{\mu\nu},
\end{align}
where the total energy density, pressure, and enthalpy read $e=e_f+e_\phi=e_f+V_0+\frac12\psi^2$, $p=p_f+p_\phi=p_f-V_0+\frac12\psi^2$, and  $w=w_f+w_\phi=w_f+\psi^2$, respectively. Hence, it would be more convenient to discuss the wall-fluid system as a whole with the first-order hydrodynamics. The fluid four-velocity takes the form $u^\mu=\gamma(1,\mathbf{v})=\gamma(1,v,0,0)$ with $\gamma(v)=(1-v^2)^{-1/2}$ in the coordinate $x^\mu=(t,z,x=0,y=0),(t,\rho,\varphi=0,z=0), (t,r,\theta=0,\varphi=0)$ for planar, cylindrical, and spherical walls, respectively, expanding along the $x^1$ direction with a constant terminal velocity $\xi_w$. As the bubble expansion is self-similar, the fluid velocity $v(\xi)$ depends only on the self-similar coordinate $\xi=x^1/x^0$ alone so are the other thermodynamical quantities such as $e(\xi)$, $p(\xi)$, and $w(\xi)$. With an abbreviation $\mu(\zeta, v(\xi))\equiv(\zeta-v)/(1-\zeta v)$ for the Lorentz transformation of the fluid velocity $v(\xi)$ in the bubble-center frame ($x^\mu$-coordinate system originated from the bubble center) into a rest frame comoving with a velocity $\zeta$, we can define a wall-frame fluid velocity $\bar{v}(\xi)=\mu(\xi_w,v(\xi))$ so that fluid velocity reads $u^\mu=\bar{\gamma}(1,-\bar{v},0,0)$ in the rest frame of an expanding wall with a Lorentz factor $\bar{\gamma}=(1-\bar{v})^{-1/2}$, where the minus sign is introduced to ensure a positive $\bar{v}$. Therefore, the total energy-momentum tensor in the rest frame of the wall admits only the following non-vanishing components,
\begin{align}
T^{\bar{x}^0\bar{x}^0}&=w\bar{\gamma}^2-p,\\
T^{\bar{x}^1\bar{x}^1}&=w\bar{\gamma}^2\bar{v}^2+p,\\
T^{\bar{x}^0\bar{x}^1}&=T^{\bar{x}^1\bar{x}^0}=-w\bar{\gamma}^2\bar{v}.
\end{align}
A similar form can also be obtained for the total energy-momentum tensor in a rest frame of the shock front by replacing all quantities with an overbar symbol with an overtilde symbol, where the shock-frame fluid velocity is defined by $\tilde{v}(\xi)=\mu(\xi_{sh},v(\xi))$ with $\xi_{sh}$ the plasma-frame velocity for the shock front, if any.

\paragraph{Hydrodynamic equations} Having established the perfect-fluid ansatz for the wall-fluid system, the hydrodynamic equation~\cite{Espinosa:2010hh} emerges as the EoM from the conservation of total energy-momentum tensor, $ \nabla_\mu T^{\mu\nu} = 0 $. Projecting this equation parallel and perpendicular to the fluid four-velocity direction with $u^\mu = \gamma(v) (1,\Vec{v})$ and $\tilde{u}^\mu = \gamma(v) (v, \Vec{v}/v)$ results in
\begin{align}
    u_\nu \nabla_\mu T^{\mu\nu} &= w\nabla_\mu u^\mu + u^\mu \nabla_\mu p = 0 , \label{eq:EoMu}\\
    \tilde{u}_\nu \nabla_\mu T^{\mu\nu} &= w \tilde{u}_\nu u^\mu \nabla_\mu u^\nu + \tilde{u}^\mu \nabla_\mu p = 0, \label{eq:EoMutilde}
\end{align}
where the time-like normalization $u_\mu u^\mu = - \tilde{u}_\mu \tilde{u}^\mu = -1$ and perpendicular condition $\tilde{u}_\mu u^\mu = 0$ are used. To actually solve above EoMs, one has to expand them in a specific coordinate frame. We will first restrict ourself to the spherical-wall case hereafter to be more specific and realistic, and hence all the quantities are independent of $\theta$ and $\varphi$ because of the spherical symmetry. We then define the bubble-center frame $(t,r)$ where $t$ is the time elapse since the bubble nucleation and $r$ is the distance to the bubble center. Since we have assumed the bubble expansion is self-similar without any characteristic length scale, we next define the self-similar frame $(\tau,\xi)$ where $\tau$ is identical to $t$ of bubble-center frame and $\xi = r/t$ is the self-similar coordinate. Recall that we have also assumed the bubble expansion is in steady state in the self-similar frame, then the time derivatives in $\tau$ in the self-similar frame will simply vanish. Therefore, the Jacobi matrices from the bubble-center frame to the self-similar frame with $\tau = t$ and $\xi = r/t$ for the tangent vectors and 1-forms are 
\begin{align}
    \left[ \begin{array}{c} \partial_\tau \\ \partial_\xi \end{array} \right] =
    \left[ \begin{array}{cc} 1 & \xi \\ 0 & t \end{array} \right] 
    \left[ \begin{array}{c} \partial_t \\ \partial_r \end{array} \right] , \quad 
    \left[ \begin{array}{c} \mathrm{d}\tau \\ \mathrm{d}\xi \end{array} \right] =
    \left[ \begin{array}{cc} 1 & 0 \\ -\xi /t & 1/t \end{array} \right] 
    \left[ \begin{array}{c} \mathrm{d} t \\ \mathrm{d} r \end{array} \right] , \quad
\end{align}
where we only show the radial parts since the angular parts are identity matrices. Finally, in the self-similar frame, the EoMs from Eqs.~\eqref{eq:EoMu} and~\eqref{eq:EoMutilde} become
\begin{align} \label{eq:EoMuCoor}
    \frac{\gamma}{t} \left( (\xi - v)\frac{\mathrm{d} e}{\mathrm{d} \xi} - \gamma^2 w (1- \xi v)\frac{\mathrm{d} v}{\mathrm{d} \xi} - 2 \frac{w v}{\xi} \right) &= 0 \\
    \frac{\gamma}{t} \left( (1 - \xi v)\frac{\mathrm{d} p}{\mathrm{d} \xi} - \gamma^2 w (\xi - v)\frac{\mathrm{d} v}{\mathrm{d} \xi} \right) &= 0 \label{eq:EoMubarCoor}
\end{align}
which can be further rearranged by division and summation to reach the following forms,
\begin{align}  
    2 \frac{v}{\xi} &= \gamma^2 (1-\xi v) \left( \frac{\mu(\xi,v)^2}{c_s^2} - 1 \right) \frac{\mathrm{d} v}{\mathrm{d} \xi}, \label{eq:EoMutxi}\\
    \frac{\mathrm{d} w}{\mathrm{d} \xi} &= w\gamma^2 \mu(\xi,v) \left( \frac{1}{c_s^2} + 1 \right) \frac{\mathrm{d} v}{\mathrm{d} \xi}.\label{eq:EoMubartxi}
\end{align}
Here $c_s^2 = \mathrm{d} p / \mathrm{d} e = \partial_\xi p / \partial_\xi e$ is the square of sound velocity and $\mu(\xi,v) = (\xi - v)/(1 - \xi \cdot v)$ is the fluid velocity seen by an observer in a frame with velocity $\xi$. One thing that should be reminded is that, when carrying from Eqs.~\eqref{eq:EoMuCoor} and~\eqref{eq:EoMubarCoor} to Eqs.~\eqref{eq:EoMutxi} and~\eqref{eq:EoMubartxi}, the factors like $\gamma/t$ and $w$ have been divided, which are discontinuous functions of $\xi$ at the bubble wall and the shock front. This may cause some ambiguities, thus, it is necessary to revisit these equations more carefully when we deal with the junction conditions in the vicinity of the bubble wall and the shock front. Apart from the junction conditions that connect the EoMs across the bubble wall and shock front, one would need to further specify the equation of state to solve the hydrodynamic equations. For the sake of better illustration of our approach to calculate the bubble wall velocity, we simply choose the bag equation of state where the wall-fluid system is further reduced to a system made of vacuum energy and relativistic radiations,
\begin{align}
e = e_f + e_\phi = a T^4 + \epsilon, \quad p = p_f + p_\phi = \frac{1}{3} a T^4 - \epsilon,
\end{align}
where $e_f = a T^4$ and $p_f=\frac{1}{3}e_f$ are the leading-order thermal corrections (massless radiations) derived from the finite-temperature quantum field theory, and $\epsilon \equiv V_0$ is the vacuum energy density of the scalar field $\phi$. The coefficient $a$ of this vacuum-radiation system is defined to capture the relativistic degrees of freedom in the energy-momentum tensor by~\cite{Jackiw:1974cv,Dolan:1973qd},
\begin{align} \label{eq:def_a}
    a = \frac{\pi^2}{30} \left( \sum_{\mathrm{light~Bosons}} g_B + \frac{7}{8} \sum_{\mathrm{light~Fermions}} g_F \right)
\end{align}
where $g_B$ and $g_F$ denote the degrees of freedom of relativistic Bosons and Fermions, respectively. For more general cases with the equation of state beyond the simple bag model, we reserve the corresponding studies in computing the bubble wall velocity for future works.

\begin{figure}
    \centering
    \includegraphics[width=0.4\textwidth]{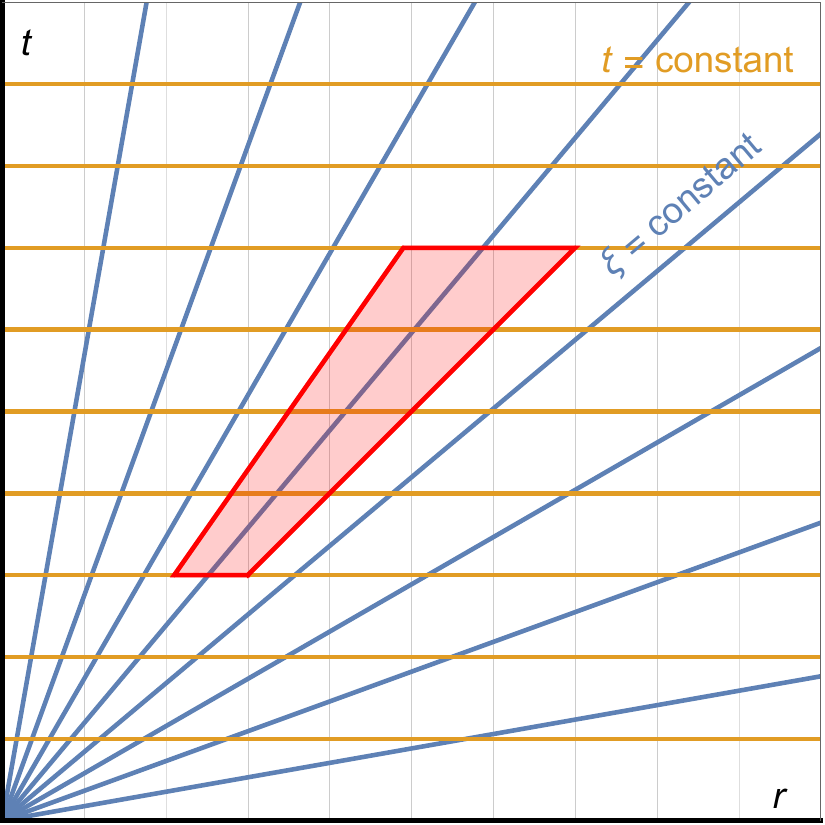}
    \quad
    \includegraphics[width=0.4\textwidth]{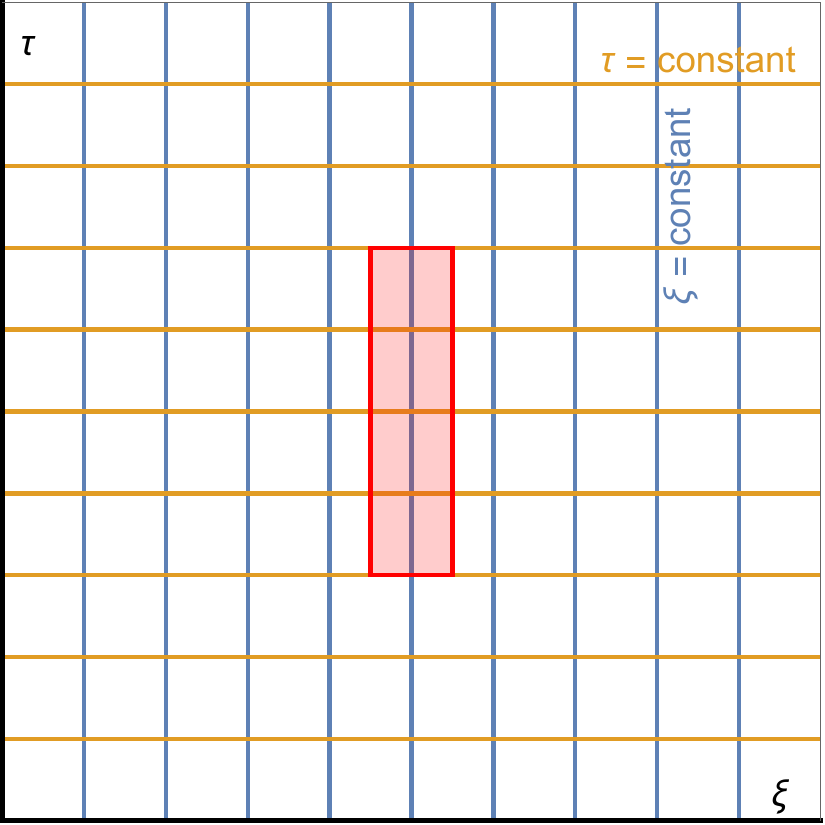}
    \caption{Left: the constant $\xi$ lines and constant $t$ lines in the bubble center frame; Right: the constant $\xi$ lines and constant $\tau$ (namely $t$) lines in the self-similar frame. A trapezoid in the bubble center frame is squeezed into a rectangle in the self-similar frame.}
    \label{fig:Coordinate_transform}
\end{figure}

\subsection{Junction conditions}\label{subsec:JunctionConditions}

To solve hydrodynamic equations, one has to specify the equation of state, which only assigns values to the energy density and pressure everywhere except at the bubble wall and shock front, if any, since the non-equilibrium contributions from $\delta p_\phi$ and $\delta p_f$ are unknown at these discontinuous interfaces. However, by regarding the scalar wall-plasma fluid as a whole with an uniform perfect-fluid ansatz, one can eventually solve the hydrodynamic equations across these discontinuous interfaces by maintaining the conservation of total energy-momentum tensor across these interfaces, which simply leads to the usual junction conditions in connecting the components of total energy-momentum tensor across these junction interfaces. 

\paragraph{Differential conservation equation} The traditional way to derive these junction conditions seems to follows directly from the differential equation $\nabla_\mu T^{\mu\nu}=0$ after taking the time derivative in the rest frame of the junction interface to vanish. For example, the junction conditions across the bubble wall are derived as follows,
\begin{align}
\partial_\mu T^{\mu0}=\partial_{\bar{t}}T^{\bar{t}\bar{t}}+\partial_{\bar{r}}T^{\bar{r}\bar{t}}=\partial_{\bar{r}}T^{\bar{r}\bar{t}}=0
&\Rightarrow w_-\bar{\gamma}_-^2\bar{v}_-=w_+\bar{\gamma}_+^2\bar{v}_+,\label{eq:Wall1stJunction}\\
\partial_\mu T^{\mu1}=\partial_{\bar{t}}T^{\bar{t}\bar{r}}+\partial_{\bar{r}}T^{\bar{r}\bar{r}}=\partial_{\bar{r}}T^{\bar{r}\bar{r}}=0
&\Rightarrow w_-\bar{\gamma}_-^2\bar{v}_-^2+p_-=w_+\bar{\gamma}_+^2\bar{v}_+^2+p_+,\label{eq:Wall2ndJunction}
\end{align}
while the junction conditions across the shock front are derived as follows,
\begin{align}
\partial_\mu T^{\mu0}=\partial_{\tilde{t}}T^{\tilde{t}\tilde{t}}+\partial_{\tilde{r}}T^{\tilde{r}\tilde{t}}=\partial_{\tilde{r}}T^{\tilde{r}\tilde{t}}=0
&\Rightarrow w_L\tilde{\gamma}_L^2\tilde{v}_L=w_R\tilde{\gamma}_R^2\tilde{v}_R,\label{eq:Shock1stJunction}\\
\partial_\mu T^{\mu1}=\partial_{\tilde{t}}T^{\tilde{t}\tilde{r}}+\partial_{\tilde{r}}T^{\tilde{r}\tilde{r}}=\partial_{\tilde{r}}T^{\tilde{r}\tilde{r}}=0
&\Rightarrow w_L\tilde{\gamma}_L^2\tilde{v}_L^2+p_L=w_R\tilde{\gamma}_R^2\tilde{v}_R^2+p_R,\label{eq:Shock2ndJunction}
\end{align}
where $w_\pm$, $\bar{v}_\pm$, and $\bar{\gamma}_\pm\equiv\gamma(\bar{v}_\pm)$ are the enthalpy, wall-frame fluid velocity, and corresponding Lorentz factors just in front and back of the bubble wall, respectively, while $w_{L/R}$, $\tilde{v}_{L/R}$, and $\tilde{\gamma}_{L/R}\equiv\gamma(\tilde{v}_{L/R})$ are the enthalpy, shock-frame fluid velocity, and corresponding Lorentz factors just in back and in front of the shock-wave front, respectively. Similarly, any conserved current $\nabla_\mu j^\mu=0$ can also result in a junction condition across the junction surface directly from the differential form of the conservation equation provided that the time derivative in the rest frame of the junction interface can be set zero. For example, assuming the conservation of entropy density current $S^\mu = su^\mu$ or number density current $N^\mu = nu^\mu$ would lead to
\begin{align}
    \partial_\mu S^\mu = \partial_{\bar{t}}S^{\bar{t}} + \partial_{\bar{r}}S^{\bar{r}} = \partial_{\bar{r}}S^{\bar{r}} = 0 &\Rightarrow s_- \bar{\gamma}_- \bar{v}_- = s_+ \bar{\gamma}_+ \bar{v}_+,  \\
    \partial_\mu N^\mu = \partial_{\bar{t}}N^{\bar{t}} + \partial_{\bar{r}}N^{\bar{r}} = \partial_{\bar{r}}N^{\bar{r}} = 0 &\Rightarrow n_- \bar{\gamma}_- \bar{v}_- = n_+ \bar{\gamma}_+ \bar{v}_+,\\
    \partial_\mu S^\mu = \partial_{\tilde{t}}S^{\tilde{t}} + \partial_{\tilde{r}}S^{\tilde{r}} = \partial_{\tilde{r}}S^{\tilde{r}} = 0 &~\Rightarrow~ s_L \tilde{\gamma}_L \tilde{v}_L = s_R \tilde{\gamma}_R \tilde{v}_R,  \\
    \partial_\mu j^\mu = \partial_{\tilde{t}}N^{\tilde{t}} + \partial_{\tilde{r}}N^{\tilde{r}} = \partial_{\tilde{r}}N^{\tilde{r}} = 0 &~\Rightarrow~ n_L \tilde{\gamma}_L \tilde{v}_L = n_R \tilde{\gamma}_R \tilde{v}_R.
\end{align}

However, it is far from obvious to set the time derivative of some quantity in the rest frame of the junction interface to be simply zero. Here we fill in the detailed argument. Take the bubble wall as an example of the junction interface, the Lorentz transformation between the bubble-center frame and local wall frame reads $\bar{t} = \gamma_w t - \gamma_w \xi_w r$ and $\bar{r} = -\gamma_w \xi_w t + \gamma_w r$ with $\xi_w$ the bubble wall velocity in the bubble-center frame and $\gamma_w\equiv(1-\xi_w^2)^{-1/2}$ the corresponding Lorentz factor. The key argument here is that the bubble expansion is of a steady state in the self-similar frame $(\tau=t, \xi=r/t)$ so that any quantity $Q$ constructed from fluid variables ($e, p, v, T, \cdots$) is time-independent $\partial_\tau Q=0$ in the self-similar frame. However, a time-independent $Q$ in the self-similar frame is not always time-independent in the wall frame. There are only two exceptional cases as seen shortly below from
\begin{align}
    0 = \partial_\tau Q = (\partial_t + \xi \partial_r) Q = \gamma_w \left[ (1 - \xi \xi_w) \partial_{\bar{t}} + (\xi - \xi_w) \partial_{\bar{r}} \right]Q.
\end{align}
The first exceptional case is to look at the homogeneous region with $\partial_{\bar{r}}Q=0$ to diminish the second term, in which case the steady-state expansion condition $\partial_\tau Q=0$ in the self-similar frame ensures a time-independent $Q$ in the wall frame by $\partial_{\bar{t}}Q = \gamma_w^{-1}(1- \xi \xi_w)^{-1}\partial_\tau Q =0$ within that homogeneous region. However, this is not the case for imposing the junction condition since the junction condition is not imposed in the homogeneous region but at the junction interface, in vicinity of which $Q$ can be highly inhomogeneous. Hence comes the second exceptional case when evaluating $Q$ at the bubble wall $\xi= r_w(t)/t = \xi_w$ so as to also vanish the second term $(\xi-\xi_w)\partial_{\bar{r}}Q|_\mathrm{wall}$ provided that $\partial_{\bar{r}}Q|_\mathrm{wall}$ is not singular around the wall, then the steady-state expansion condition $\partial_\tau Q_{\mathrm{wall}}=0$ in the self-similar frame also ensures a time-independent $Q_{\mathrm{wall}}$ in the wall frame by $\partial_{\bar{t}}Q|_{\mathrm{wall}} = \gamma_w \partial_\tau Q |_{\mathrm{wall}} = 0$ at the wall. However, all fluid variables can receive singular contributions to their spatial derivatives when crossing the junction interface unless they can change continuously through that interface, which is the case when the thin-wall approximation is not assumed. But if the thin-wall approximation can be indeed abandoned, one should also not use it in hydrodynamics for self-consistency, which is currently unavailable for the case with a FOPT.

As long as working with hydrodynamics under thin-wall approximation~\cite{Espinosa:2010hh}, we cannot ignore this ambiguity but propose shortly below an integrated conservation equation method. With this method, we need not assume a vanishing time derivative in the rest frame of the junction interface since only the spatial-derivative components are extracted out by the normal vector perpendicular to the junction interface to construct the junction condition. Note that the junction conditions obtained from the integrated conservation equation method can still be equivalent to those obtained from the differential conservation equation. The key difference only arises when the differential conservation equation to be integrated is not exactly a total 4-derivative. The usual conservation equation $T\nabla_\mu(su^\mu)=0$ of entropy density current is one such example, where the inhomogeneous $T$ profile seems to be canceled out from the differential conservation equation to arrive at a junction condition that is actually broken but not supposed to be violated across the shock front as shown in the Appendix~\ref{app:EntropyFlow}.

\paragraph{Integrated conservation equation: interface frame}
We first introduce the weak-form conservation equation method in the interface frame for the conservation of total energy-momentum tensor. Take the bubble wall as an example of the junction interface, we first consider a sufficiently small piece of the bubble wall $\Sigma$ so that it can be treated as a planar wall moving in the $z$ direction by $(t, z=\xi_w t, x=0,y=0)$ from $t=t_i$ to $t=t_f$ with a wall velocity $\xi_w$. Then, we choose a 4-volume $\mathcal{V}$ enclosed by the following four surfaces: 
\begin{enumerate}
\item the surface $\Sigma_+$ in front of the wall traced by $z=(\xi_w+\varepsilon)t$;
\item the surface $\Sigma_-$ behind the wall traced by $z=(\xi_w-\varepsilon)t$;
\item the initial surface $\Sigma_i$ at $t=t_i$ connecting $\Sigma_\pm$ by $(\xi_w-\varepsilon)t<z<(\xi_w+\varepsilon)t$;
\item the final surface $\Sigma_f$ at $t=t_f$ connecting $\Sigma_\pm$ also by $(\xi_w-\varepsilon)t<z<(\xi_w+\varepsilon)t$.
\end{enumerate}
Next, we can calculate the following integral by the Stokes's theorem,
\begin{align}\label{eq:TIntegralStokes}
\int_\mathcal{V}\nabla_\mu(T^{\mu\nu}\lambda_\nu)\mathrm{d}^4x=\oint_{\partial\mathcal{V}=\Sigma_+\bigcup\Sigma_-\bigcup\Sigma_i\bigcup\Sigma_f} n_\mu T^{\mu\nu}\lambda_\nu\mathrm{d}^3x,
\end{align}
for an arbitrary vector $\lambda_\nu$ with a finite derivative $\nabla_\mu\lambda_\nu$ so that $\nabla_\mu(T^{\mu\nu}\lambda_\nu)=\nabla_\mu T^{\mu\nu}\lambda_\nu+T^{\mu\nu}\nabla_\mu\lambda_\nu=T^{\mu\nu}\nabla_\mu\lambda_\nu$ is also finite, which is crucial when shrinking the 4-volume $\mathcal{V}$ down to zero as seen shortly below. Also note that this is the only place where we have used the conservation equation of the total energy-momentum tensor $\nabla_\mu T^{\mu\nu}=0$. Here $n_\mu$ is the normal vector to the boundary $\partial\mathcal{V}$. Finally, we can shrink the volume $\mathcal{V}\to0$ by setting $\varepsilon\to0$ so that $\Sigma_{i,f}\to0$ and $\Sigma_\pm\to\Sigma$. Now, the left-hand-side of~\eqref{eq:TIntegralStokes} with a finite integrand $T^{\mu\nu}\nabla_\mu\lambda_\nu$ will also vanish as the volume $\mathcal{V}$ approaching zero,
\begin{align}
\int_{\mathcal{V}\to0}\nabla_\mu(T^{\mu\nu}\lambda_\nu)\mathrm{d}^4x\to0,
\end{align}
while the right-hand-side of~\eqref{eq:TIntegralStokes} receives contributions only from the surfaces $\Sigma_\pm\to\Sigma$,
\begin{align}
\oint_{\partial\mathcal{V}\to\Sigma_\pm\to\Sigma} n_\mu T^{\mu\nu}\lambda_\nu\mathrm{d}^3x
&\to\int_{\Sigma_-
\to\Sigma}n_\mu T^{\mu\nu}\lambda_\nu\mathrm{d}^3x
+\int_{\Sigma_+\to\Sigma}n_\mu T^{\mu\nu}\lambda_\nu\mathrm{d}^3x\nonumber\\
&\equiv\int_\Sigma n_\mu^-T_-^{\mu\nu}\lambda_\nu\mathrm{d}^3x
+\int_\Sigma n_\mu^+T_+^{\mu\nu}\lambda_\nu\mathrm{d}^3x,
\end{align}
so that this integral in the vicinity of the wall,
\begin{align}
\int_\Sigma(n_\mu^-T_-^{\mu\nu}+n_\mu^+T_+^{\mu\nu})\lambda_\nu\mathrm{d}^3x=0,
\end{align}
is valid for an arbitrary $\lambda_\nu$, that is to say, $n_\mu^- T_-^{\mu\nu}+n_\mu^+ T_+^{\mu\nu}=0$ for the normal vectors $n_\mu^\pm$ of the surfaces $\Sigma_\pm\to\Sigma$. Therefore, in the rest frame of the wall with $n_\mu^\pm=(0,\pm1,0,0)$, the total energy-momentum tensor $T^{\mu\nu}=wu^\mu u^\nu+p\eta^{\mu\nu}$ with wall-frame fluid velocity $u^\mu=\bar{\gamma}(1,-\bar{v},0,0)$ can be conserved across the wall provided with following junction conditions,
\begin{align}
n_z^-T_-^{zt}+n_z^+T_+^{zt}=0\Rightarrow T_-^{zt}=T_+^{zt}&~\Rightarrow~  w_-\bar{\gamma}_-^2\bar{v}_-=w_+\bar{\gamma}_+^2\bar{v}_+,\label{eq:junctionwall1}\\
n_z^-T_-^{zz}+n_z^+T_+^{zz}=0\Rightarrow T_-^{zz}=T_+^{zz}& ~\Rightarrow~  w_-\bar{\gamma}_-^2\bar{v}_-^2+p_-=w_+\bar{\gamma}_+^2\bar{v}_+^2+p_+.\label{eq:junctionwall2}
\end{align}
Following the same procedure, we can also derive the usual junction conditions across the shock front,
\begin{align}
w_L\tilde{\gamma}_L^2\tilde{v}_L&=w_R\tilde{\gamma}_R^2\tilde{v}_R,\label{eq:junctionshock1}\\
w_L\tilde{\gamma}_L^2\tilde{v}_L^2+p_L&=w_R\tilde{\gamma}_R^2\tilde{v}_R^2+p_R.\label{eq:junctionshock2}
\end{align}

\begin{figure}
    \centering
    \includegraphics[width=0.5\textwidth]{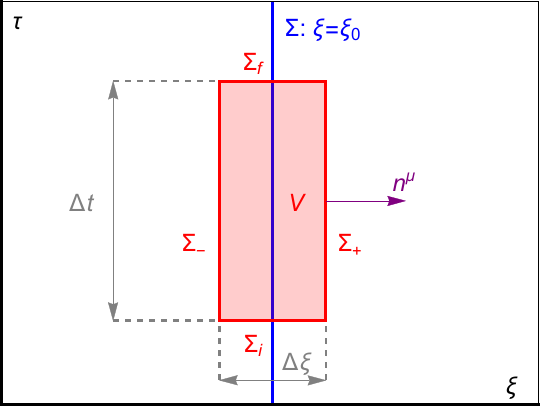}
    \caption{The interface $\Sigma$ (blue solid) is covered by its neighborhood $V$ (red box) in the self-similar frame. All the variables are assumed to be constant along the $\xi = \mathrm{const}$ surface $\Sigma_+$ (or $\Sigma_-$).}
    \label{fig:SigmaAndV}
\end{figure}

\paragraph{Integrated conservation equation: self-similar frame}
As we have assumed the bubble expansion is in a steady state in the self-similar frame, it could be more formally straightforward to work in the self-similar frame. In the self-similar frame, the junction interface $\Sigma$ corresponds to a constant $\xi=\xi_0$ surface with $\xi_0=\xi_w$ for the bubble wall and $\xi_0=\xi_{sh}$ for the shock front, and the box $\mathcal{V}$ is chosen as small as possible to cover a small fraction of $\Sigma$ as shown in Fig.~\ref{fig:SigmaAndV}. The box $\mathcal{V}$ is of size $\Delta t \Delta \xi$ (apart from the angular part), and since all the variables are only functions of $\xi$, the integration over $t$ could be easily performed and pulled out. Setting $\lambda_\mu=(1,0)$ in the bubble center frame, and using the fact that the normal vector $n_\mu = \gamma_0^{-1}(-\xi_0,1)$ in the bubble center frame, the right-hand side of Eq.~\eqref{eq:TIntegralStokes} can be worked out explicitly as
\begin{align}
    \begin{aligned}
        0 &=\int_{\partial\mathcal{V}} \mathrm{d}^3x~ T^{\mu\nu} \lambda_\nu n_\mu 
        = 4\pi \left( \int_{\Sigma_+} \mathrm{d} t - \int_{\Sigma_-} \mathrm{d} t + \int_{\Sigma_f} t\mathrm{d} \xi - \int_{\Sigma_i} t\mathrm{d} \xi  \right) T^{\mu\nu} \lambda_\nu n_\mu \\
        &= 4\pi \left( \int_{\xi = \xi_0 + \frac{\Delta \xi}{2}} - \int_{\xi = \xi_0-\frac{\Delta \xi}{2}} \right) \mathrm{d} t ~\left\{\gamma_0^{-1} [ w\gamma^2(v-\xi_0) + p \xi_0] \right\} +  4\pi \Delta t \int_{\xi_0 - \frac{\Delta \xi}{2}}^{\xi_0 + \frac{\Delta \xi}{2}} \mathrm{d} \xi\,(w\gamma^2 - p) \\
        &= 4\pi \Delta t \gamma_0^{-1} [w\gamma^2(v-\xi_0) + p \xi_0]\bigg|_{\xi=\xi_0 - \frac{\Delta \xi}{2}}^{\xi=\xi_0 + \frac{\Delta \xi}{2}} +  4\pi \Delta t \int_{\xi_0 - \frac{\Delta \xi}{2}}^{\xi_0 + \frac{\Delta \xi}{2}} \mathrm{d} \,\xi (w\gamma^2 - p). 
    \end{aligned}
\end{align}
The factor $4\pi$ comes from the integral over the angular part. After taking the $\Delta\xi\to 0$ limit, the second term in the last line is simply vanished. Since $4\pi \Delta t \gamma_0^{-1}$ is a constant, one finally reaches the junction condition,
\begin{align} \label{eq:JClambda1}
    \Delta (w\gamma^2(v-\xi_0) + p \xi_0) = 0.
\end{align}
Following the same procedure and setting $\lambda_\mu = (0,1)$, one can get another junction condition,
\begin{align} \label{eq:JClambda2}
    \Delta (w\gamma^2 v (v-\xi_0) + p) = 0.
\end{align}
After some linear combinations, they are actually equivalent to the two junction conditions we obtained before at the bubble wall and shock front,
\begin{align}
\Delta \left(w \gamma^2 \frac{(\xi_0-v)(1- \xi_0 v)}{1-\xi_0^2} \right) &= \begin{cases}
\Delta (w\bar{\gamma}^2 \bar{v}) =0, & \xi_0=\xi_w;\\
\Delta (w\tilde{\gamma}^2 \tilde{v}) =0, & \xi_0=\xi_{sh},\\
\end{cases}\label{eq:JC1}\\
\Delta \left(w \gamma^2 \frac{(\xi_0-v)^2}{(1-\xi_0)^2} + p \right) & = \begin{cases}
\Delta (w\bar{\gamma}^2 \bar{v}^2 + p) =0, &\xi_0=\xi_w;\\
\Delta (w\tilde{\gamma}^2 \tilde{v}^2 + p) =0, &\xi_0=\xi_{sh}.\\
\end{cases}\label{eq:JC2}
\end{align}

\paragraph{Junction conditions from a conserved current}

With all the above junction conditions, we can ensure a global conservation $\nabla_\mu T^{\mu\nu}=0$ not only inside and outside of the bubble and sound shell but also across the bubble wall and shock-wave front. Similarly, for any vector current $j^\mu$ with its conservation equation of a total-derivative form $\nabla_\mu j^\mu=0$, we can integrate it totally by the Stokes's theorem as
\begin{align}
\int_\mathcal{V}\nabla_\mu j^\mu\mathrm{d}^4x = \oint_{\partial\mathcal{V}=\Sigma_+\bigcup\Sigma_-\bigcup\Sigma_i\bigcup\Sigma_f} n_\mu j^\mu \mathrm{d}^3x,
\end{align}
which, after taking the limit $\mathcal{V}\to 0$, renders
\begin{align}
\int_\Sigma(n_\mu^-j_-^{\mu}+n_\mu^+j_+^{\mu})\mathrm{d}^3x=0 ~\Rightarrow~ 
n_\mu^-j_-^{\mu}+n_\mu^+j_+^{\mu} = 0.
\end{align}
Substituting $n_\mu^\pm=(0,\pm1,0,0)$ in the rest frame of the junction interface simply extracts the junction conditions as $j_-^{\bar{z}}=j_+^{\bar{z}}$ for the bubble-wall interface and $j_-^{\tilde{z}}=j_+^{\tilde{z}}$ for the shock-front interface. Then, the remaining task is to find such a conserved current with a total-derivative form to arrive at an extra junction condition, which, after combined with the usual junction conditions from $\nabla_\mu T^{\mu\nu}=0$, can model-independently solve for the bubble wall velocity.

One naive guess is the entropy density current $S^\mu\equiv su^\mu$, whose conservation $\nabla_\mu S^\mu=0$ is only reached in local equilibrium from the Boltzmann H-theorem. However, as shown in Appendix~\ref{app:EntropyFlow} independent of wall geometries, the derived junction condition $s_L\tilde{\gamma}_L\tilde{v}_L=s_R\tilde{\gamma}_R\tilde{v}_R$ is explicitly broken across the shock front, whereas no net entropy generation process is expected to occur there in the symmetric phase. This clearly disfavors the junction condition $s_L\tilde{\gamma}_L\tilde{v}_L=s_R\tilde{\gamma}_R\tilde{v}_R$ as an appropriate one across the shock front, so does its counterpart $s_-\bar{\gamma}_-\bar{v}_-=s_+\bar{\gamma}_+\bar{v}_+$ across the bubble wall. The origin of this contradiction can be traced back to the differential conservation equation $\nabla_\mu(su^\mu)=0$, where the temperature profile has been implicitly assumed to be homogeneous in local equilibrium. However, this is exactly not the case for the bubble expansion at the junction interface, where the temperature profile can still be inhomogeneous across the junction interface even assuming local equilibrium~\footnote{This is best demonstrated in Refs.~\cite{Wang:2022txy,Wang:2023kux} that, for the hydrodynamic backreaction force (comprising the thermal force and friction force) to counterbalance the driving force from the effective-potential difference, the friction force from non-equilibrium effect can be vanished in local equilibrium, but at the mean time the thermal force from the temperature gradient cannot be vanished in order to counterbalance the driving force.}. The actual differential conservation equation for the entropy density current should be $T\nabla_\mu(su^\mu)=0$ as proved in Appendix~\ref{app:EntropyFlow}  directly from the conservation of total energy-momentum tensor under the thin-wall approximation, where the inhomogeneous temperature profile cannot be simply canceled out before integrating this differential conservation equation across the junction interface. Unfortunately, this differential conservation equation cannot be totally integrated out by the Stokes's theorem, hence no junction condition can be obtained.

\paragraph{Junction conditions from number density current} The next natural candidate is the number density current $N^\mu\equiv nu^\mu$, whose conservation $\nabla_\mu N^\mu = 0$ is guaranteed with local dynamical equilibrium as proved in Sec.~\ref{subsec:EoMs}. For the reason to be clear shortly below this paragraph, here we define a total number density $n\equiv n_f+n_\phi$ to include not only the contribution of particles from the plasma fluid but also the contribution of particles from the false/true vacua under the thin-wall approximation. With a bag EoS, the number density from plasma particles reads $n_f=b T^3$ with $b$ given by
\begin{align}\label{eq:b}
    b = \frac{\zeta(3)}{\pi^2} \left( \sum_{\mathrm{light~Bosons}} g_B + \frac{3}{4} \sum_{\mathrm{light~Fermions}} g_F \right)
\end{align}
if the plasma fluid is modeled as a relativistic perfect fluid. The junction condition can be obtained by integrating the conservation equation of the total number density current as
\begin{align} \label{eq:TaubJC}
 (n u^\mu)_+ n_\mu^+ + (n u^\mu)_- n_\mu^- =0,
\end{align}
leading to the following junction conditions across the bubble wall and shock front \footnote{We also recognize this junction condition at the shock front as the one first studied by Taub~\cite{Taub:1948zz} for relativistic shocks and well developed by Thorne~\cite{Thorne:1973} from a differential perspective. Recent work extends them to general relativistic cases~\cite{Mallick:2022ubp}. Here we do not consider the back-reaction of the fluid and thus the special relativistic junction conditions suit our study well.}, 
\begin{align}
    n_z^-N_-^{z}+n_z^+N_+^{z}=0 &~\Rightarrow~ n_- \bar{\gamma}_- \bar{v}_- = n_+ \bar{\gamma}_+ \bar{v}_+,\\
    n_z^LN_L^{z}+n_z^RN_R^{z}=0 &~\Rightarrow~ n_L \tilde{\gamma}_L \tilde{v}_L = n_R \tilde{\gamma}_R \bar{v}_R,
\end{align}
respectively. The same junction condition from the total number density current can also be derived from the integrated conservation equation in the self-similar frame as
\begin{align} \label{eq:JCnumber}
    \Delta (n\gamma\cdot (\xi_0 -v)) = 0,
\end{align}
which is exactly equivalent to what we have obtained in the interface frame,
\begin{align}
    \Delta \left( - \frac{1}{\sqrt{1-\xi_0^2}} n \gamma(\xi_0 - v) \right)=\begin{cases}
    \Delta (n \bar{\gamma} \bar{v}) =  0, & \xi_0=\xi_w,\\
    \Delta (n \tilde{\gamma}\tilde{v})=0, & \xi_0=\xi_{sh}.
    \end{cases}.
\end{align}
It is worth noting that the local dynamical equilibrium is usually violated across the bubble wall between the symmetric and broken phases as the cross-section of a given process from one phase to the other cannot counterbalance its inverse process of opposite direction. Therefore, the junction condition from the total number density current across the bubble wall usually reads $n_- \bar{\gamma}_- \bar{v}_- \neq n_+ \bar{\gamma}_+ \bar{v}_+$, hence, we introduce a phenomenological parameter to characterize such a deviation from local equilibrium in the interface frame or self-similar frame as
\begin{align}\label{eq:JCGammaw}
\frac{n_+ \bar{\gamma}_+ \bar{v}_+}{n_- \bar{\gamma}_- \bar{v}_-} =\Gamma_w, \quad\text{or}\quad \frac{n_+ \gamma_+ (\xi_w -v_+)}{n_- \gamma_- (\xi_w -v_-)} = \Gamma_w,
\end{align}
respectively. In principle, $\Gamma_w$ should be calculable in terms of some thermodynamic and hydrodynamic quantities in equilibrium near the junction interface if the scalar-wall contribution to the total number density current can be appropriately defined from a field-theoretic perspective, which will be reserved for future study. Yet its precise definition would not affect our model-independent proposal for the bubble wall velocity.

\paragraph{Necessity to introduce the total number density current} In the case of FOPT, if the local dynamical equilibrium is only assumed among plasma particles away from the bubble wall, then the number density current from these plasma particles alone can be proved to be conserved across the shock front as shown in Sec.~\ref{subsec:EoMs}, yet its junction condition across the shock front would necessarily imply the absence of the shock front and correspondingly also the absence of bubble wall in the first place. This can be seen by applying the junction condition $(n_f)_L\tilde{\gamma}_L\tilde{v}_L=(n_f)_R\tilde{\gamma}_R\tilde{v}_R$ across the shock front in the deflagration case, which, after combined with the usual junction condition from conserved energy-momentum tensor,
\begin{align}
\frac{(n_f)_R\tilde{\gamma}_R\tilde{v}_R}{(n_f)_L\tilde{\gamma}_L\tilde{v}_L}=&1=\frac{b_RT_R^3\tilde{\gamma}_R\tilde{v}_R}{b_LT_L^3\tilde{\gamma}_L\tilde{v}_L},\\
\frac{w_R\tilde{\gamma}_R^2\tilde{v}_R}{w_L\tilde{\gamma}_L^2\tilde{v}_L}=&1=\frac{a_RT_R^4\tilde{\gamma}_R^2\tilde{v}_R}{a_LT_L^4\tilde{\gamma}_L^2\tilde{v}_L},
\end{align}
would lead to a relation (obviously $b_L=b_R$ and $a_L=a_R$ across the shock front within the symmetric phase)
\begin{align}
\frac{\tilde{v}_R(1-\tilde{v}_R^2)}{\tilde{v}_L(1-\tilde{v}_L^2)}=1
\end{align}
that simply implies $\xi_{sh}=1/\sqrt{3}$ namely the absence of shock front $v_{sh}\equiv v(\xi_{sh})=0$ or equivalently the absence of a bubble wall $\xi_w=0$. Therefore, assuming local dynamical equilibrium among plasma particles alone, the number density current of plasma fluid is only conserved across the shock front when there is no scalar wall. That is to say \textit{contra-positively}, if there indeed exists a non-vanishing scalar bubble thin wall, the local dynamical equilibrium away from the bubble wall cannot only be assumed among plasma particles but also with scalar-wall particles excited from the false and true vacua in the symmetric and broken phases, respectively. As a result, we should include the scalar-wall contribution to the total number density current in addition to the plasma-fluid particles, and their local dynamical equilibrium ensures the conservation of the total number density current away from the bubble wall including the shock front. Again, for the processes crossing the bubble wall, both the scalar-wall and plasma-fluid particles have experienced the symmetry breaking/restoration to render asymmetric cross-sections so that the local dynamical equilibrium simply cannot be established at the bubble wall, and hence the total number density current is not necessarily conserved across the bubble wall.


\section{Bubble wall velocity}\label{sec:BubbleWallVelocity}

In this section, we will determine the terminal wall velocity for a steadily expanding bubble, using the junction conditions~\eqref{eq:JClambda1},~\eqref{eq:JClambda2} and~\eqref{eq:JCnumber} or~\eqref{eq:JCGammaw}. The equivalent forms of the junction conditions~\eqref{eq:JClambda1} and~\eqref{eq:JClambda2},
\begin{align}
    w_+ \bar{v}_+ \bar{\gamma}_+^2 &= w_- \bar{v}_- \bar{\gamma}_-^2, \\
    w_+ \bar{v}_+^2 \bar{\gamma}_+^2 + p_+ &= w_- \bar{v}_-^2 \bar{\gamma}_-^2 + p_-,
\end{align}
could be rearranged to express the product and quotient the wall-frame fluid velocities as
\begin{align}
    \begin{aligned} \label{eq:vbpvbm}
        \bar{v}_+\bar{v}_- &= \frac{p_+-p_-}{\rho_+-\rho_-} = \frac{1-(1-3\alpha_+)r}{3-3(1+\alpha_+)r}, \\
        \frac{\bar{v}_+}{\bar{v}_-} &= \frac{\rho_-+p_+}{\rho_++p_-} = \frac{3+(1-3\alpha_+)r}{1+3(1+\alpha_+)r}
    \end{aligned}
\end{align}
for a bag equation of state with the abbreviations
\begin{align}
    \alpha_+=\frac{\Delta\epsilon}{a_+T_+^4}=\frac{4\Delta\epsilon}{3w_+}, \quad r=\frac{w_+}{w_-}=\frac{a_+ T_+^4}{a_-T_-^4}.
\end{align}
On the other hand, the equivalent form of the junction condition~\eqref{eq:JCnumber} reads
\begin{align} \label{eq:nRatiao_1}
    n_+ \gamma_+ (\xi_0 -v_+) = n_- \gamma_- (\xi_0 -v_-),
\end{align}
where $\xi_0$ is the interface position in the self-similar frame, and $\pm$ signs stand for the exterior and interior of the surface~\footnote{Note that, until this point, we have used $\pm$ to denote for the positions just behind and in front of the bubble wall, and $L/R$ to denote for the positions just behind and in front of the shock front. To be more informative for the use of symbols, we will uniformly use in this Section the symbols $\pm$ to stand for the positions just behind and in front of the junction interface (bubble wall or shock front) under discussion.}. When applying this condition to the shock front, $\xi_0$ is set to be $\xi_{sh}$. As we have mentioned above, the total number density current may not be conserved across the bubble wall for the possible failure of local dynamical equilibrium. Assuming this failure to be small, the ratio of the total number density currents between two sides of the wall, 
\begin{align} \label{eq:nRatio_Gamma}
    \frac{n_+ \gamma_+ (\xi_w -v_+)}{n_- \gamma_- (\xi_w -v_-)} = \Gamma_w\lessgtr1,
\end{align}
can be slightly deviated from $1$ but still could be of order $\mathcal{O}(1)$ to also maintain the hydrodynamic validity. This is where the out-of-equilibrium effects could come into play.

\subsection{Weak/Jouguet deflagration} \label{subsec:deflagration}

For the deflagration and hybrid (or weak detonation~\cite{Espinosa:2010hh}) expansion modes, the junction conditions can be applied both at the bubble wall and the shock front. It is easy to see that at the shock front, it always holds $\alpha_+=0$ from Eq.~\eqref{eq:vbpvbm} in the symmetric phase, leading to $\tilde{v}_+ \tilde{v}_- = 1/3$, where the overtilde denotes the fluid velocity $\tilde{v}=\mu(\xi_{sh},v)$ in the rest frame of the shock front. Since the fluid in front of the shock front should remain static, $v_+=0=\mu(\xi_{sh},\tilde{v}_+)$, then the shock front condition can be determined by
\begin{align}
    \xi_{sh} \mu(\xi_{sh},v_-) = 1/3, \quad \tilde{v}_+=\xi_{sh}, \quad \tilde{v}_- = \frac{1}{3\xi_{sh}}.
\end{align}
Plugging various relevant velocities into the junction condition~\eqref{eq:nRatiao_1} gives rise to
\begin{align}
     \frac{n_+}{n_-} \frac{\xi_{sh}}{\gamma\left(\mu(\xi_{sh},\frac{1}{3\xi_{sh}})\right) \cdot \left(\xi_{sh} -\mu(\xi_{sh}, \frac{1}{3\xi_{sh}})\right)} = \frac{n_+}{n_-} \frac{\xi_{sh} \sqrt{-1 + 10\xi_{sh}^2 - 9 \xi_{sh}^4}}{1-\xi_{sh}^2} = 1.
\end{align}
 
\begin{figure}
    \centering
    \includegraphics[width = 0.485\textwidth]{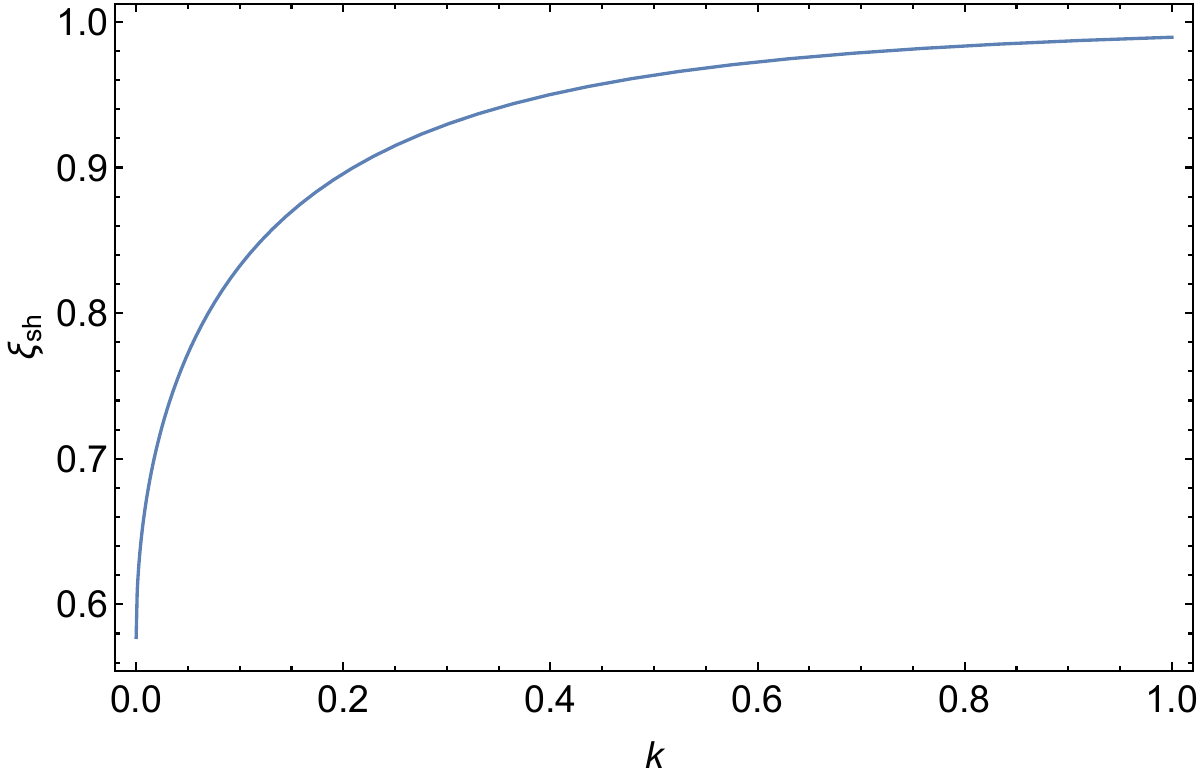}
    \includegraphics[width = 0.495\textwidth]{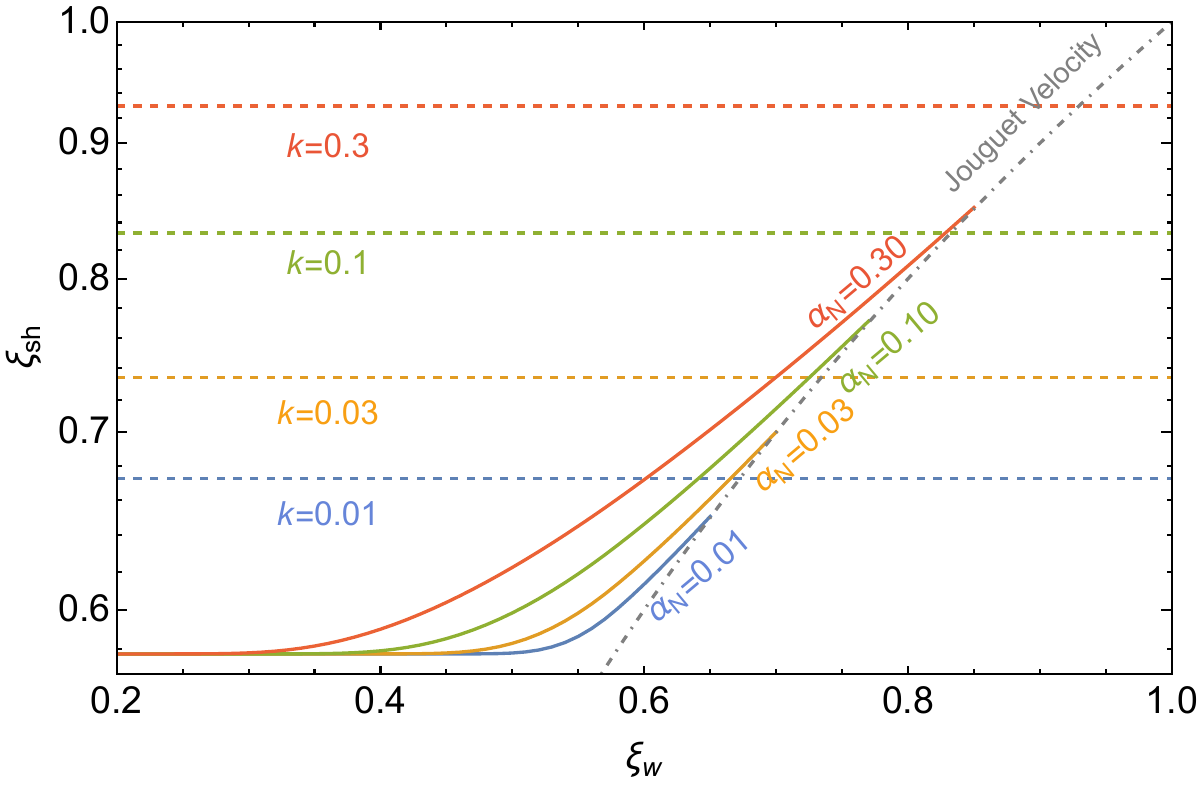}\\
    \caption{Left: $\xi_{sh}$ as a function of $k$. Right: matching the shock-front positions $\xi_{sh}(k)$ (dashed lines) with the hydrodynamic solutions $\xi_{sh}(\xi_w; \alpha_N)$ (solid curves) to locate bubble wall velocity $\xi_w$.}
    \label{fig:deflag}
\end{figure}

The total number density is the summation of the plasma-fluid contribution and scalar-wall contribution, $n = n_f+n_\phi$. An intuitive point of view is to regard the total system as a collection of interacting particles including the usual plasma particles and the excitations of $\phi$-particles on top of different vacua, though the precise definition of the ``scalar-wall number density'' is less clear than the usual ``plasma-particle number density''. Nevertheless, our results do not rely on the precise definition of the scalar-wall number density, which will be absorbed into a parameter to be determined in the future study. Recall that the particle number for the plasma-fluid with a bag EoS reads $n_f = b T^3$ with $b$ given by~\eqref{eq:b}, the number density ratio between the two sides of the shock front is
\begin{align}
    \frac{n_+}{n_-} = \cfrac{b_+ T_+^3 + n_{\phi,+}}{b_- T_-^3 + n_{\phi,-}} = \cfrac{1 + \cfrac{n_{\phi,+}}{b_+ T_+^3}}{\cfrac{b_-}{b_+} \left(\cfrac{T_-}{T_+}\right)^3 + \cfrac{n_{\phi,-}}{b_+ T_+^3}} \equiv \cfrac{1 + k_+}{ \left( w_-/w_+ \right)^{3/4} + k_-},
\end{align}
where the equation of state $w = \frac{4}{3}a T^4$ together with $a_+=a_-$ and $b_+=b_-$ have been used, again because of the whole shockwave living in the symmetric phase. The definitions of $k_\pm$ is given by $k_\pm = n_{\phi,\pm}/(b_+ T_+^3)$. In the vicinity of the shock front, the scalar field stays at the false vacuum, thus it is reasonable to assume $n_{\phi,+} = n_{\phi,-}$ and $k_+=k_-\equiv k$. Then, the enthalpy ratio can be evaluated from the junction condition~\eqref{eq:JC1} as
\begin{align}
    \frac{w_-}{w_+} = \frac{\bar{\gamma}_+^2 \bar{v}_+}{\bar{\gamma}_-^2 \bar{v}_-} = \frac{9 \xi_{sh}^2 - 1}{3(1-\xi_{sh}^2)}.
\end{align}
Finally, Eq.~\eqref{eq:nRatiao_1} can be re-written in terms of $\xi_{sh}$ and $k$,
\begin{align} \label{eq:xi_k_eq_deflag}
    \frac{n_+ \gamma_+ (\xi_{sh} -v_+) }{ n_- \gamma_- (\xi_{sh} -v_-)} = \cfrac{1+k}{\left( \cfrac{9 \xi_{sh}^2 - 1}{3(1-\xi_{sh}^2)} \right)^{3/4} + k}\cdot \frac{\xi_{sh} \sqrt{-1 + 10\xi_{sh}^2 - 9 \xi_{sh}^4}}{1-\xi_{sh}^2} = 1,
\end{align}
from which $\xi_{sh}$ can be determined for a given $k$ as shown in the left panel of Fig.~\ref{fig:deflag}.

\begin{figure}
    \centering
    \includegraphics[width = 0.53\textwidth]{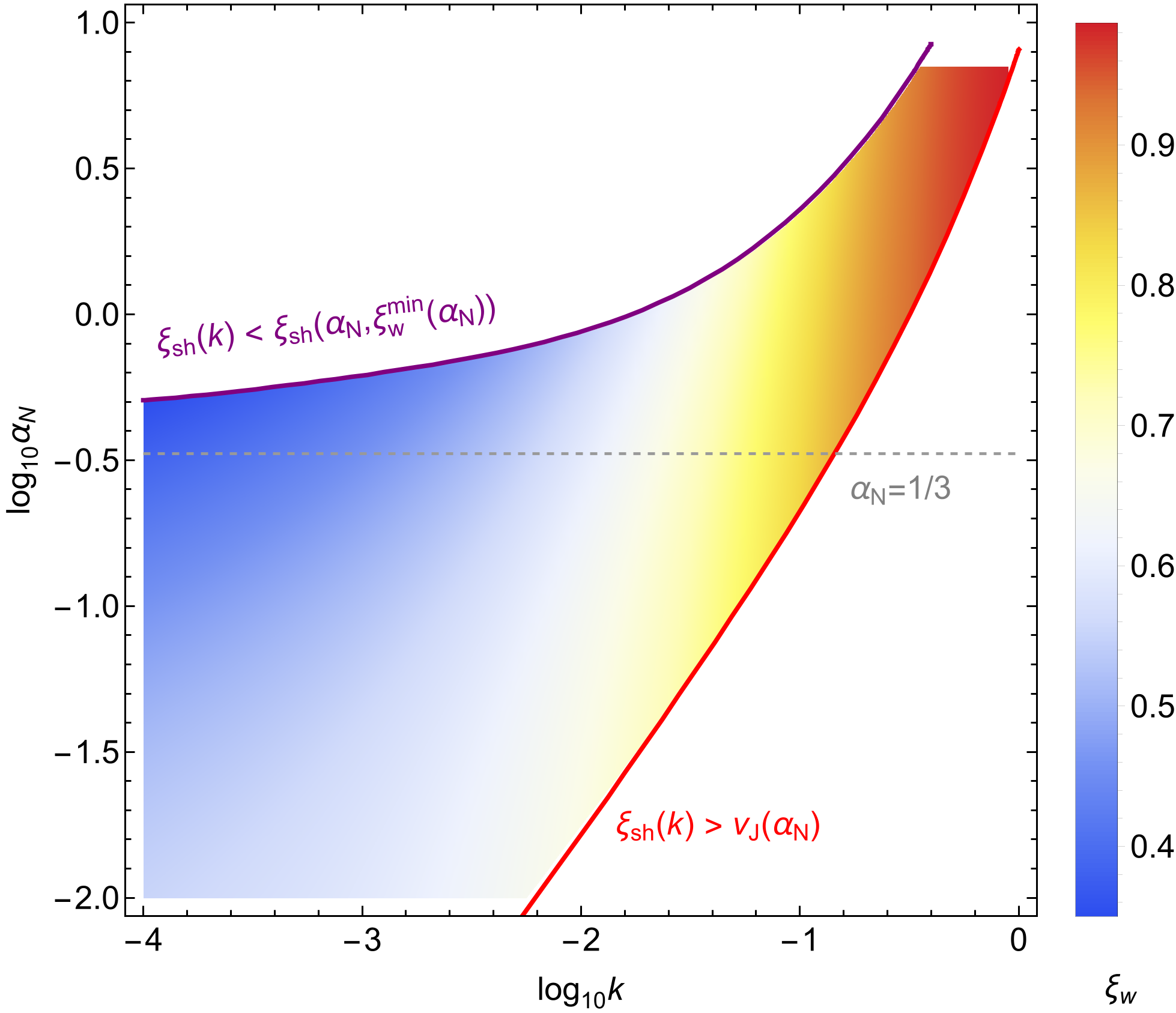}
    \includegraphics[width = 0.46\textwidth]{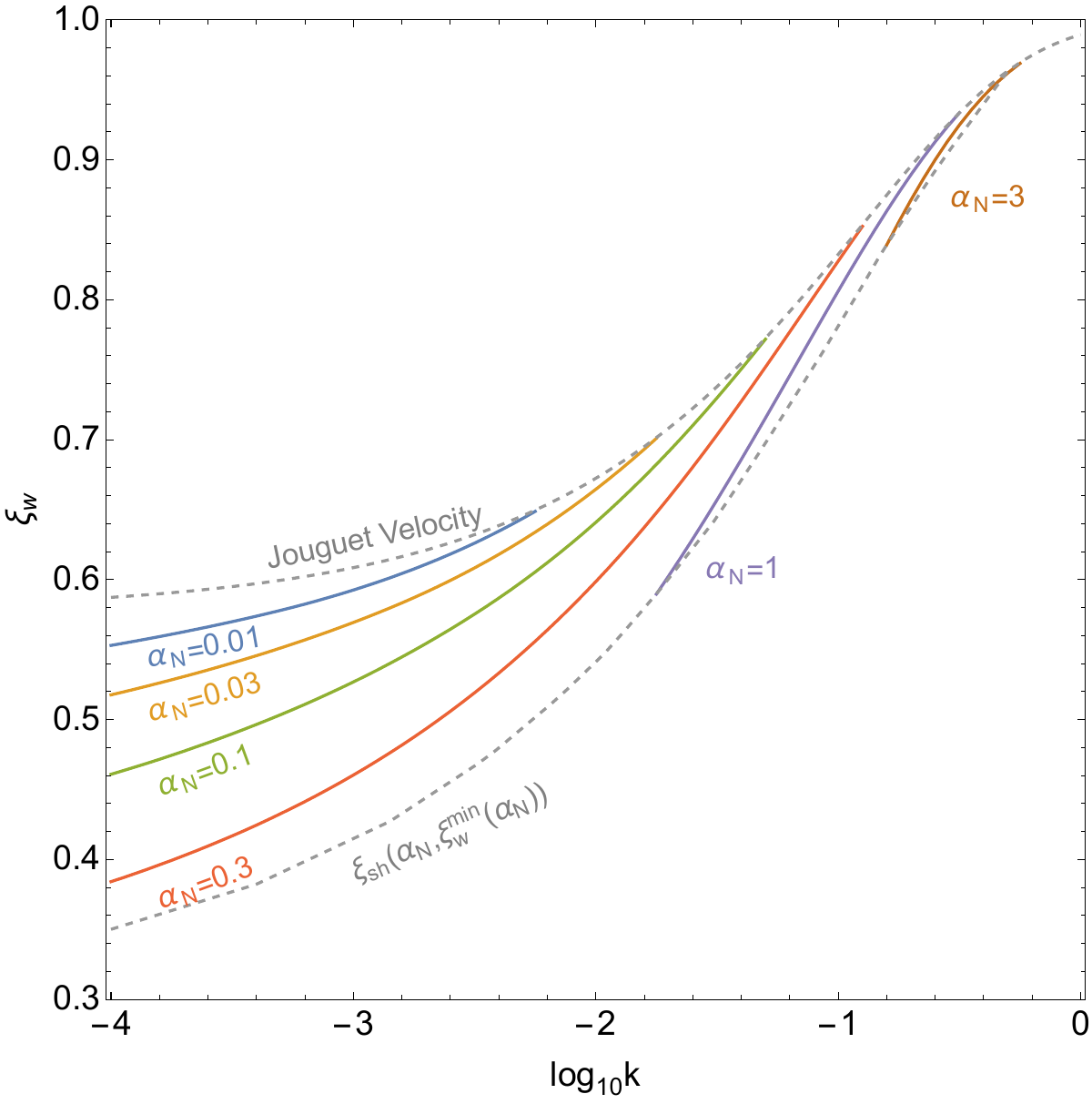}\\
    \caption{The bubble wall velocity $\xi_w$ for given $k$ and $\alpha_N$.}
    \label{fig:deflag_velocity}
\end{figure}

On the other hand, the shock front position $\xi_{sh}$ can be also determined from hydrodynamics by solving the EoM~\eqref{eq:EoMutxi} with the usual junction conditions~\cite{Espinosa:2010hh} (see also the appendix of Ref.~\cite{Wang:2022txy}), which relates $\xi_{sh}$ with $\xi_w$ for a given strength factor $\alpha_N = 4\Delta\epsilon / (3 w_N) = (w_+/w_N) \alpha_+$ at null infinity as shown in the right panel of Fig.~\ref{fig:deflag}. At the end of day, the bubble wall velocity $\xi_w(\alpha_N,k)$ can be determined by matching the hydrodynamic $\xi_{sh}(\xi_w;\alpha_N)$ solution with the root $\xi_{sh}(k)$ of our extra junction condition~\eqref{eq:xi_k_eq_deflag} as shown with the intersection point of the horizontal $\xi_{sh}(k)$-line with the $\xi_{sh}(\xi_w)$-curve for a given $\alpha_N$. It is worth noting that such a intersection point is not always possible, as there is a maximum bubble wall velocity called Jouguet velocity, 
\begin{align}
    v_J = \frac{\sqrt{\alpha_+ (2+3\alpha_+)} + 1}{\sqrt{3}(1+\alpha_+)},
\end{align}
for the Jouguet deflagration expansion with an infinitely thin shock wave, and hence if the $k$ value is too large, then the corresponding $\xi_{sh}(k)$-line is too large to intersect with the $\xi_{sh}(\xi_w)$-curve, which simply means that the bubble expansion is not in a hybrid mode, but in a detonation mode to be determined later below. Fortunately, as a ratio of number density of the scalar field to the plasma fluid particles, $k$ might be approximately the ratio for the relativistic degrees of freedom, which should be of the order $O(10^{-3}) \sim O(10^{-1})$ since $n_f=b T^3$ receives contributions from all the other particles except for $\phi$. In this case, the $\xi_{sh}(k)$-line should always possible to intersect with the $\xi_{sh}(\xi_w)$-curve for a majority range of $\alpha_N$ unless $\alpha_N$ is extremely small. This is the main difference of our approach with that from Ref.~\cite{Ai:2021kak,Ai:2023see} where the bubble wall velocity for a deflagration expansion can only be found counterintuitively within a very narrow range of $\alpha_N$ as shown in Appendix~\ref{app:subsec:spherical}. In Fig.~\ref{fig:deflag_velocity}, we present in detail the numerical results of the bubble wall velocity $\xi_w$ as a function of $\alpha_N$ and $k$. In the left panel, there are two boundaries, outside which there are no corresponding bubble wall velocities for the weak/Jouguet deflagration expansion. As we know, the shock front should move slower than the Jouguet velocity. This is the boundary $\xi_{sh}(k) = v_J(\alpha_N)$ denoted with the red curve, below which the Jouguet deflagration fails to match such a large shock front and a weak detonation or even run-away expansion is expected. Another boundary with the purple curve is found by matching $\xi_{sh}(k) = \xi_{sh}(\alpha_N, \xi_w^{\mathrm{min}}(\alpha_N))$, on the right-hand side of which $\xi_w^{\mathrm{min}}(\alpha_N)$ is the minimal possible bubble wall velocity for a given $\alpha_N$, and $\xi_{sh}(\alpha_N, \xi_w)$ denotes the hydrodynamic determination on the shock-front velocity for given $\alpha_N$ and $\xi_w$. As $\xi_w^{\mathrm{min}}(\alpha_N)$ is nonzero only if $\alpha_N > 1/3$~\cite{Espinosa:2010hh}, the purple curve approaches to the left asymptotically at $\alpha_N = 1/3$ with a vanishing $\xi_w^\mathrm{min}$ as shown in a gray dashed line. Above this boundary, the bubble wall velocity naturally cannot exist for a steady expansion. The right panel is simply an equivalent view of the left panel with interchanged $y$-axis and color remarks. We will provide in a future study the analytic fitting formulas (the difficulty lies in the irregular boundaries of the parameter space) along with a ready-to-use code for directly calculating the wall velocity for both deflagration and detonation (see below) cases.

\subsection{Weak detonation}\label{subsec:detonation}

As there is no shock front for the detonation case, all junction conditions are imposed on the bubble wall. The physically allowed detonation is achieved by a supersonic wall, where the interactions cannot propagate to the fluids in front of the wall. Thus, the fluid in front of the wall remains static, $v_+ = \mu(\xi_w, \bar{v}_+) =0$, which indicates $\bar{v}_+ = \xi_w$. Then, from the junction conditions~\eqref{eq:vbpvbm}, one can solve for
\begin{align}
    \bar{v}_- = X(\bar{v}_+) \pm \sqrt{ X(\bar{v}_+)^2 - \frac{1}{3}}, \quad X(\bar{v}_+) \equiv \frac{1+\alpha_+}{2}\bar{v}_+ + \frac{1-3\alpha_+}{6\bar{v}_+}.
\end{align}
For the detonation case, the plus-sign branch is physically allowed. Then, the fluid velocity just behind the wall can be evaluated by $v_- = \mu(\xi_w, \bar{v}_-)$. Plugging $v_+=0$ and $v_-$ into~\eqref{eq:nRatio_Gamma} results in
\begin{align}
    \frac{n_+}{n_-}\cdot \frac{\xi_w}{\gamma_-(\xi_w - v_-)} = \Gamma_w,
\end{align}
where $v_-$ is a function of $\alpha_+$ and $\xi_w$, and $\Gamma_w$ should encode for some non-equilibrium information on the number-violating interactions between the scalar wall and plasma fluid. Similar to the deflagration case, the ratio $n_+/n_-$ reads
\begin{align}
    \frac{n_+}{n_-} = \cfrac{b_+ T_+^3 + n_{\phi,+}}{b_- T_-^3 + n_{\phi,-}} = \cfrac{1 + \cfrac{n_{\phi,+}}{b_+ T_+^3}}{\cfrac{b_-}{b_+} \left(\cfrac{T_-}{T_+}\right)^3 + \cfrac{n_{\phi,-}}{b_+ T_+^3}} \equiv \cfrac{1 + k_+}{\cfrac{b_-}{b_+} \left( \left.\cfrac{w_-}{w_+}\right/ \cfrac{a_-}{a_+} \right)^{3/4} + k_-}
\end{align}
with abbreviations $k_\pm = n_{\phi,\pm}/(b_+ T_+^3)$. Here we introduce a parameter $q=(b_- a_+^{3/4}) /(b_+ a_-^{3/4})$ abbreviating the ratio in the denominator. In the Standard Model, one expects
\begin{align}
    q= \left.\frac{b_-}{b_+} \right/  \left(\frac{a_-}{a_+} \right)^{3/4} = \left.\cfrac{76.5}{95.5} \right/  \left(\cfrac{86.25}{106.75} \right)^{3/4} \simeq 0.94 \sim O(1).
\end{align}
Finally, the bubble wall velocity $\xi_w(\alpha_N;\Gamma_w,q,k_\pm)$ can be estimated from solving our extra junction condition~\eqref{eq:nRatio_Gamma} with $v_\pm$ determined from hydrodynamics and $n_+/n_-$ defined above.

\begin{figure}
    \centering
    \includegraphics[width=\textwidth]{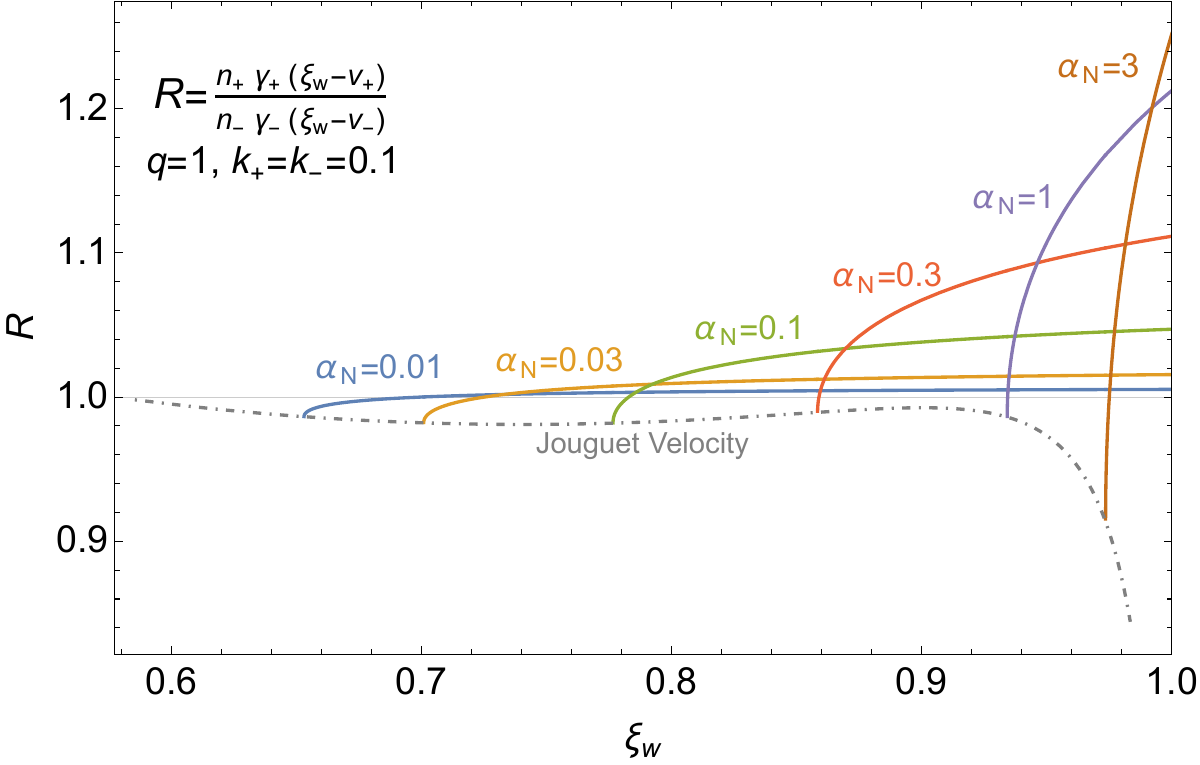}\\
    \includegraphics[width=\textwidth]{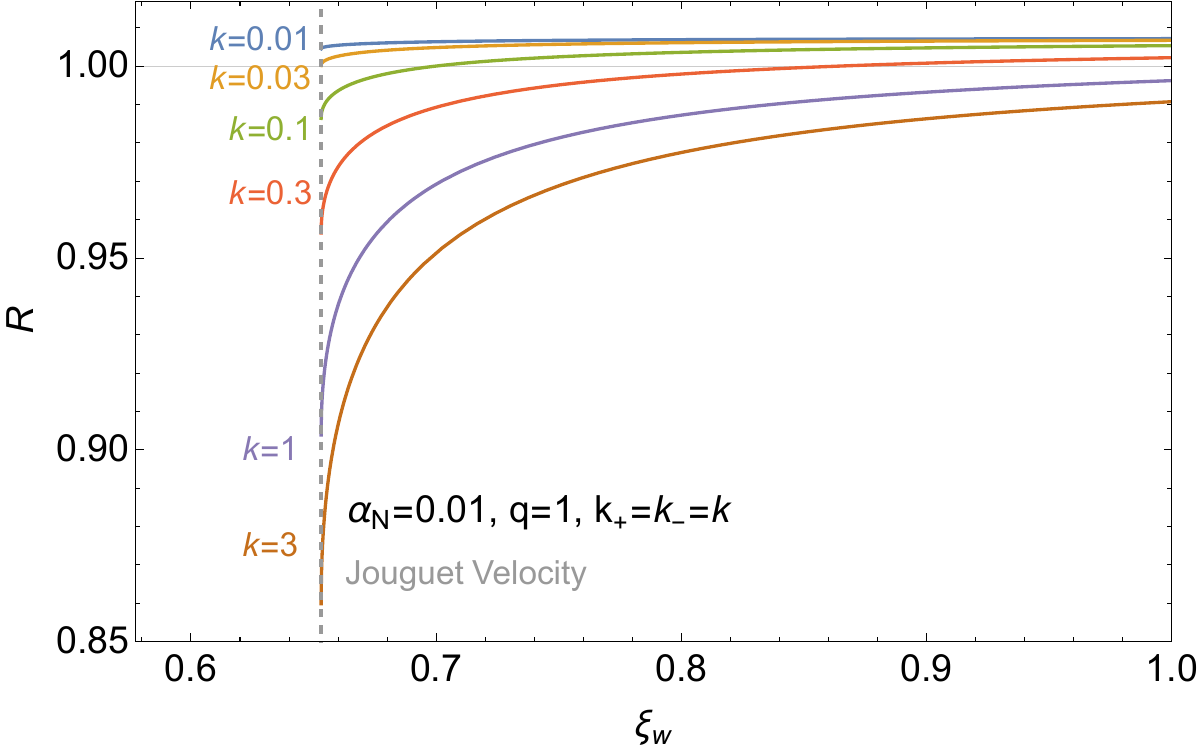}\\
    \caption{The violation degree $R$ of the number density current across the bubble wall for given $q=1$ with different $\alpha_N$ values (top) and $k_+=k_-=k$ values (bottom). The gray dashed lines denote the Jouguet velocity.}
    \label{fig:detona1}
\end{figure}

\begin{figure}
    \centering
    \includegraphics[width = 0.7\textwidth]{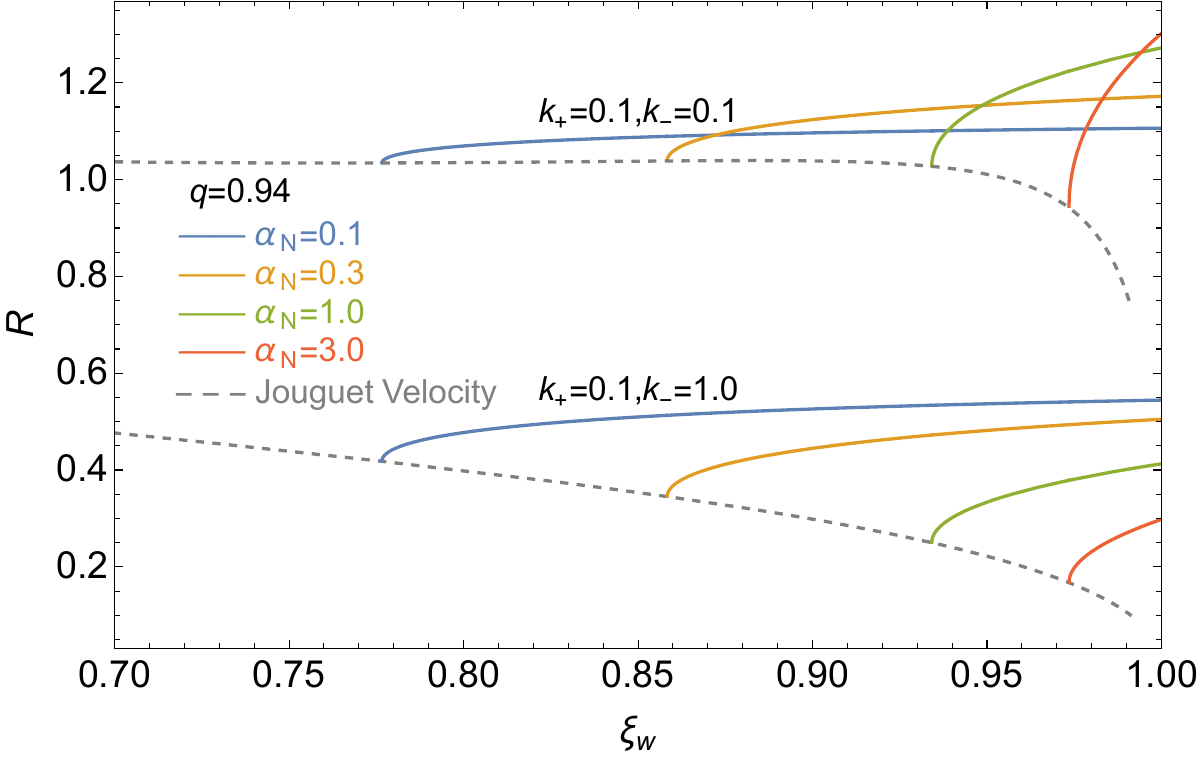}\\
    \includegraphics[width = 0.7\textwidth]{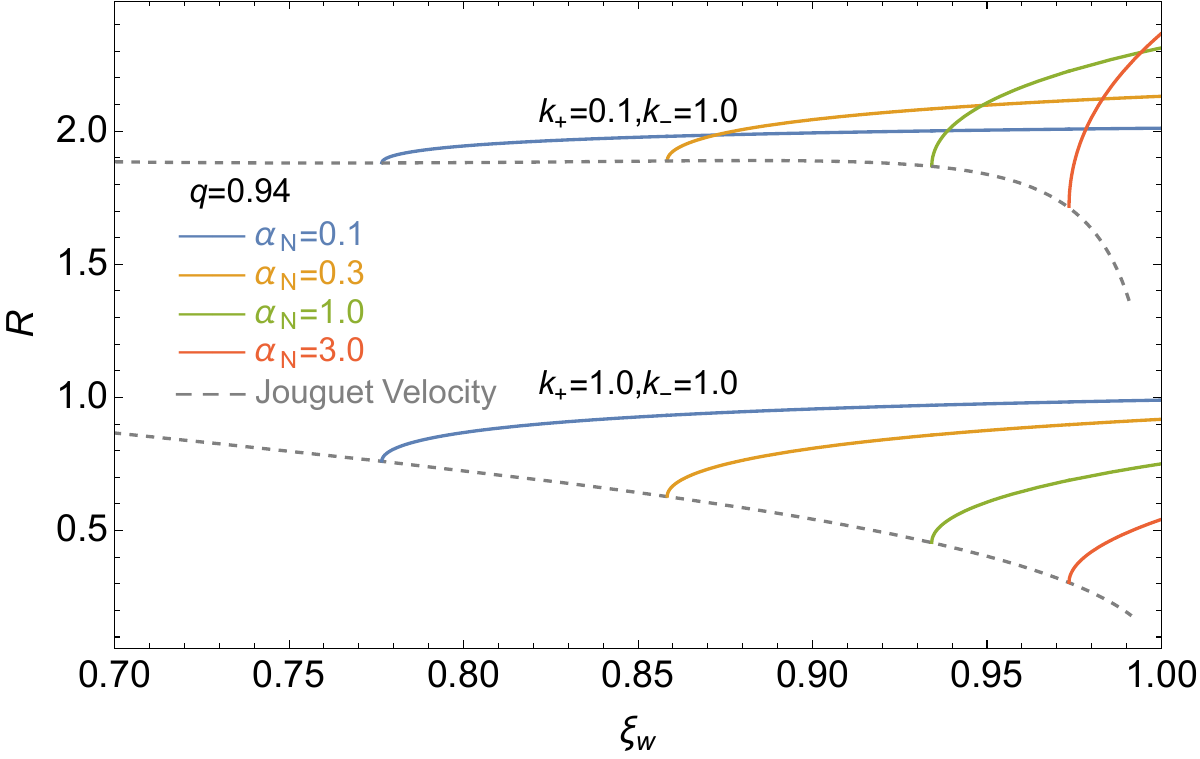}\\
    \includegraphics[width = 0.7\textwidth]{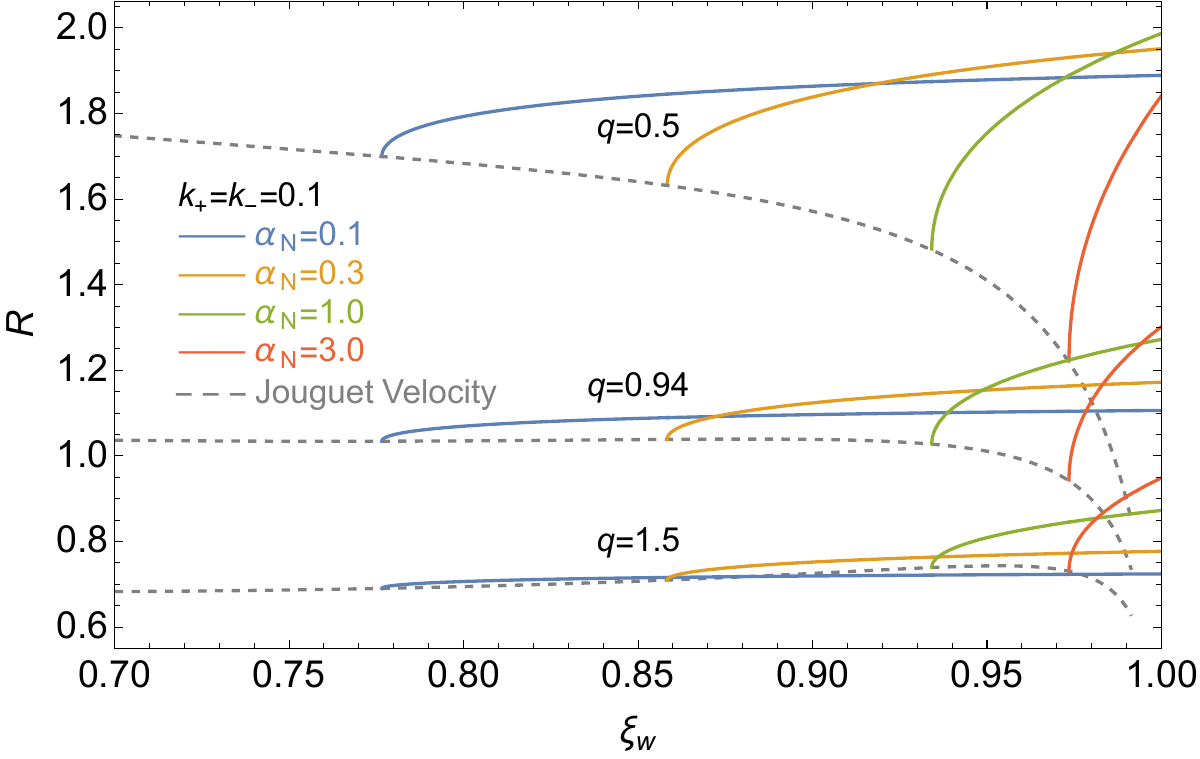}\\
    \caption{The violation degree $R$ of the number density current across the bubble wall for given $\alpha_N$ values with fixed $q=0.94$ but varying $k_+$ and $k_-$ values (top and middle) as well as fixed $k_+=k_-=0.1$ but varying $q$ values (bottom).}
    \label{fig:detona2}
\end{figure}

Now that $\xi_w(\alpha_N;\Gamma_w,q,k_\pm)$ is estimated from five parameters, we exemplify the method to determine the bubble wall velocity illustratively in Fig.~\ref{fig:detona1} and Fig.~\ref{fig:detona2} by matching the hydrodynamic evaluation on the number-density-current ratio $R\equiv[n_+\gamma_+(\xi_w-v_+)]/[n_-\gamma_-(\xi_w-v_-)]$ with a given $\Gamma_w$ value. If we assume that the scalar-wall does not participate in the particle splitting/decaying processes, we can set uniformly $k_+ = k_- \equiv k$ to some value (say, $0.1$), and then the bubble wall velocity can be found at those intersection points with $R=\Gamma_w$ for given $\alpha_N$ as shown in the top panel of Fig.~\ref{fig:detona1} with an illustrating $q=1$. It is easy to see that such an intersection point is always possible in local equilibrium with $\Gamma_w=1$ for all $\alpha_N$ values, where $\xi_w\to c_s$ for a smaller $\alpha_N$ and $\xi_w\to1$ for a larger $\alpha_N$ as one can naively expect. Including a larger non-equilibrium effect with $\Gamma_w>1$ would simply require a larger strength factor $\alpha_N$ to reach a steady-state expansion at a larger wall velocity $\xi_w$ also as one can naively expect. The bottom panel is presented for a fixed $\alpha_N$ but varying $k_\pm=k$ with the same $q=1$. There are three possible cases when matching $R=\Gamma_w$. The first case is that, a $R$-curve is always above the horizontal $\Gamma_w$-line, in which case the bubble cannot expand in a detonation mode but in a hybrid or even deflagration mode. The second case is that, a $R$-curve is always below the horizontal $\Gamma_w$-line, in which case the bubble will run away, that is, the bubble will suffer from accelerating expansion permanently and never reach a terminal velocity. The intermediate case is the one for an intersection point to locate the terminal wall velocity. On the other hand, if the scalar wall does participate in the particle splitting/decaying processes, then $k_+$ can differ from $k_-$ as shown in the first two panels of Fig.~\ref{fig:detona2}, while the last panel is presented for the effect of a varying $q$.

\section{Conclusions and discussions}\label{sec:conclusion}

The cosmological FOPTs, if ever occur in our early Universe, would impose strong constraints on the new physics beyond the standard model of particle physics. Despite that various characteristics of FOPTs can be directly evaluated from the effective potential, the terminal wall velocity of bubble expansion is in general left undetermined as a free parameter in both theoretical modelings and data analysis. Various approaches have been proposed to tackle this problem in both ultra-relativistic and non-relativistic regimes, but it is still more appealing to adopt a model-independent approach in the full regime. In this paper, we propose such a model-independent approach to calculate the bubble wall velocity by appreciating not only the conservation equations of the total energy-momentum tensor but also the conservation and violation of a total number density current from both the scalar-wall and plasma-fluid contributions. In specific, the bubble wall velocity for weak/Jouguet deflagration expansion modes can be calculated from the conservation (equilibrium) of the total number density current across the shock front when combining with hydrodynamics between the bubble wall and shock front, while the bubble wall velocity for weak detonation expansion mode can be calculated in general from the violation (non-equilibrium) of total number density current across the bubble wall when combining with hydrodynamics behind the bubble wall. In both cases, the bubble wall velocity can be determined by several model-independent parameters that could be further evaluated specifically in the future for concrete models of FOPTs. Two comments on our approach are as follows:

First, such a model-independent approach to calculate the bubble wall velocity is not new. However, the previous method by adopting the conservation of entropy density current across the bubble wall in local equilibrium fails to meet the conservation of entropy density current across the shock front as expected from the conservation of total energy-momentum tensor across the shock front for hydrodynamics with the thin-wall approximation. We trace back this inconsistency to the differential conservation equation of entropy density current, which is not a total derivative so as to be fully integrated out by the Stokes's theorem to arrive at any junction condition. 

Second, the plasma-fluid contribution to the total number density current is well-defined from the thermodynamics in equilibrium, but the scalar-wall contribution $n_\phi$ to the total number density current is undetermined. Nevertheless, we have hidden this ambiguity in the phenomenological parameters $k$ and $\Gamma_w$ so that our model-independent evaluation of terminal wall velocity does not rely on their precise definitions. However, to apply our model-independent approach for specific FOPT models, we must evaluate $n_\phi$ and $\Gamma_w$ appropriately. We will come back to this problem in future work from a field-theoretic side.

\acknowledgments
We thank the fruitful discussions (especially with Wen-Yuan Ai) during the 2024 Chengdu Symposium on Particle Physics and Cosmology from Sep. 27-30, 2024. Shao-Jiang Wang is supported by the National Key Research and Development Program of China Grant  No. 2021YFC2203004, No. 2021YFA0718304, and No. 2020YFC2201501, 
the National Natural Science Foundation of China Grants 
No. 12422502, No. 12105344, No. 12235019, and No. 12047503,
the Key Research Program of the Chinese Academy of Sciences (CAS) Grant No. XDPB15, 
the Key Research Program of Frontier Sciences of CAS, 
and the Science Research Grants from the China Manned Space Project with No. CMS-CSST-2021-B01 (supported by China Manned Space Program through its Space Application System).

\appendix

\section{The entropy density current} \label{app:EntropyFlow}

In this appendix, we will discuss in details the conservation law of the entropy density current. We will first introduce in Appendix~\ref{app:subsec:spherical} the method given in Refs.~\cite{Ai:2021kak,Ai:2023see} for deriving the bubble wall velocity with an extra junction condition from the conserved entropy density current in local equilibrium, and then we will show explicitly in Appendix~\ref{app:subsec:breakdown} the breakdown of this extra junction condition across the shock front regardless of wall geometries. In Appendix~\ref{app:subsec:conservation}, we will prove in general the conservation of the entropy density current across the shock front from the conservation of the total energy-momentum tensor under the thin-wall approximation, and hence invalidating their junction condition as the appropriate one from the conservation of the entropy density current. Finally, we reveal in Appendix~\ref{app:subsec:entropyflow} the origin of this inconsistency and forecast the possible way out.

\subsection{The wall velocity from a conserved entropy density current in local equilibrium} \label{app:subsec:spherical}

It has been argued in Refs.~\cite{Ai:2021kak,Ai:2023see} that, in addition to the usual junction condition from the conserved energy-momentum tensor, the assumption of local equilibrium around the bubble wall leads to an extra junction condition across the bubble wall from the conservation of entropy density current, and then the bubble wall velocity immediately follows from solving these junction conditions using only perfect-fluid hydrodynamics. This proposal was first tested in Ref.~\cite{Ai:2021kak} for a planar wall expansion of weak deflagration type with a bag equation of state (EoS), and then generalized in Ref.~\cite{Ai:2023see} for a spherical wall expansion of both weak and Jouguet deflagration types with a $\nu$-model EoS beyond the simple bag case. 

In specific, the central idea of their approach is that, by assuming local equilibrium around the bubble wall, the entropy density current $S^\mu\equiv su^\mu$ should be conserved across the bubble wall, $\nabla_\mu S^\mu\equiv\nabla_\mu(su^\mu)=0$, and hence $\partial_z(s(z)\bar{\gamma}(z)\bar{v}(z))=0$ along the expansion direction $z$ in the rest frame of a planar wall denoted with an overbar, leading to a new junction condition across the wall,
\begin{align}\label{eq:NewJunction}
s(z)\bar{\gamma}(z)\bar{v}(z)=\mathrm{const.}.
\end{align}
On the other hand, the conservation of the energy-momentum tensor, $\nabla_\mu T^{\mu\nu}=0$, gives rise to the usual junction condition, $w(z)\bar{\gamma}(z)^2\bar{v}(z)=\mathrm{const.}$. Using this extra input, they are able to solve the bubble wall velocity from perfect-fluid hydrodynamics for a planar wall expansion with a bag equation of state~\cite{Ai:2021kak}. Unfortunately, as we will show explicitly below in Appendixes~\ref{app:subsec:breakdown} and~\ref{app:subsec:conservation}, the very starting point~\eqref{eq:NewJunction} is incorrect. The origin of their failure will be revealed in Appendix~\ref{app:subsec:entropyflow}. Before doing so, we have to repeat the work of Ref.~\cite{Ai:2021kak} for a weak deflagration expansion of the bubble wall.

First, we recall necessary expressions from the perfect-fluid hydrodynamics of bubble expansion. The conservation of total energy-momentum tensor across the bubble wall implies the usual junction condition $w_-\bar{\gamma}_-^2\bar{v}_-=w_+\bar{\gamma}_+^2\bar{v}_+$, where $w_\pm=e_\pm+p_\pm$ and $\bar{\gamma}_\pm=(1-\bar{v}_\pm^2)^{-1/2}$ are the enthalpies and Lorentz factors of wall-frame fluid velocities just in front and back of the bubble wall, respectively. Assuming a bag EoS, $e_\pm=a_\pm T_\pm^4+V_\pm$ and $p_\pm=\frac13a_\pm T_\pm^4-V_\pm$ with $a_\pm\equiv(\pi^2/30)g_\pm$, this usual junction condition can be solved for the wall-frame fluid velocity just in front and back of the wall as
\begin{align}
\bar{v}_+(\alpha_+,r)&=\sqrt{\frac{1-(1-3\alpha_+)r}{3-3(1+\alpha_+)r}\cdot\frac{3+(1-3\alpha_+)r}{1+3(1+\alpha_+)r}},\label{eq:app:vbp}\\
\bar{v}_-(\alpha_+,r)&=\sqrt{\left.\frac{1-(1-3\alpha_+)r}{3-3(1+\alpha_+)r}\right/\frac{3+(1-3\alpha_+)r}{1+3(1+\alpha_+)r}}\label{eq:app:vbm}
\end{align}
in terms of the strength factor $\alpha_+\equiv(V_+-V_-)/(a_+T_+^4)\equiv(4\Delta V)/(3w_+)$ just in front of the wall and the enthalpy ratio $r\equiv(a_+T_+^4)/(a_-T_-^4)\equiv w_+/w_-$. 

Next, we impose the extra junction condition~\eqref{eq:NewJunction} from Ref.~\cite{Ai:2021kak} across the bubble wall as
\begin{align}\label{eq:deflagrationentropyflow}
1=\frac{s_-\bar{\gamma}_-\bar{v}_-}{s_+\bar{\gamma}_+\bar{v}_+}=\frac{s_-w_+\bar{\gamma}_+}{s_+w_-\bar{\gamma}_-}=\frac{T_+\bar{\gamma}_+}{T_-\bar{\gamma}_-}=\left(\frac{a_-w_+}{a_+w_-}\right)^\frac14\frac{\bar{\gamma}_+}{\bar{\gamma}_-}=\left(\frac{a_-}{a_+}\right)^\frac14\left(\frac{\bar{v}_-\bar{\gamma}_+^2}{\bar{v}_+\bar{\gamma}_-^2}\right)^\frac14,
\end{align}
where the usual junction condition and some thermodynamic relations,
\begin{align}
\frac{\bar{\gamma}_-\bar{v}_-}{\bar{\gamma}_+\bar{v}_+}=\frac{w_+\bar{\gamma}_+}{w_-\bar{\gamma}_-},\quad
w_\pm=T_\pm s_\pm,\quad
w_\pm=\frac43a_\pm T_\pm^4,\quad
\frac{w_+}{w_-}=\frac{\bar{\gamma}_-^2\bar{v}_-}{\bar{\gamma}_+^2\bar{v}_+},
\end{align}
are used in the second, third, fourth, and last equalities, respectively. For the deflagration expansion, the fluid velocity just behind the wall equals to the wall velocity $\xi_w$ in the rest frame of the wall, $\bar{v}_-(\alpha_+,r)=\xi_w$, leading to the ratio $r\equiv w_+/w_-$ as a function of $\alpha_+$ and $\xi_w$, thus the fluid velocity just in front of the wall $\bar{v}_+(\alpha_+, r(\alpha_+,\xi_w))\equiv\bar{v}_+(\alpha_+,\xi_w)$ is also a function of $\alpha_+$ and $\xi_w$ in the rest frame of the wall. Hence, the factor $(\bar{v}_-\bar{\gamma}_+^2)/(\bar{v}_+\bar{\gamma}_-^2)$ on the right-hand side of Eq.~\eqref{eq:deflagrationentropyflow} can be expressed as a function of $\alpha_+$ and $\xi_w$ alone. Therefore, for the usual FOPT with a decreasing number of relativistic degrees of freedom across the wall, $a_-/a_+<1$, the bubble wall velocity $\xi_w$ can be easily solved from this highly fine-tuned matching condition~\eqref{eq:deflagrationentropyflow} as a function of the strength factor $\alpha_+$. Note that, at this point, no dependence on the wall geometry has come into play.

One final step is to use the perfect-fluid hydrodynamics to express the unobservable strength factor $\alpha_+\equiv(4\Delta V)/(3w_+)$ just in front of the wall in terms of the observable strength factor $\alpha_N\equiv(4\Delta V)/(3w_N)$ at null infinity. This can be done by first relating $\alpha_+$ to the strength factor $\alpha_L\equiv(4\Delta V)/(3w_L)$ just behind the shock front with~\footnote{Here $\mu[\xi,v]=\frac{\xi-v}{1-\xi v}$ is an abbreviation for the fluid velocity in the rest frame comoving with a velocity $\xi$.}
\begin{align}\label{eq:wxi}
w_L=w_+\exp\left[4\int_{\xi_w}^{\xi_{sh}}\gamma[v(\zeta)]^2\mu[\zeta,v(\zeta)]v'(\zeta)\mathrm{d}\zeta\right]
\end{align}
via the weak deflagration solution $v(\xi)$ of different wall geometries~\footnote{Note that this is the only place where the wall geometry comes into play by solving the fluid velocity profile $v(\xi)$ from the following equation of motion~\cite{Espinosa:2010hh},
\begin{align}
D\frac{v}{\xi}=\gamma^2(1-\xi v)\left(\frac{\mu(\xi,v)^2}{c_s^2}-1\right)\frac{\mathrm{d}v}{\mathrm{d}\xi},
\end{align}
where $D=0,1,2$ correspond to planar, cylindrical, and spherical walls, respectively.}, and then further relating $\alpha_L$ to the strength factor $\alpha_R\equiv(4\Delta V)/(3w_R)\equiv\alpha_N$ just in front of the shock front via the usual junction condition $w_L\tilde{\gamma}_L^2\tilde{v}_L=w_R\tilde{\gamma}_R^2\tilde{v}_R$ across the shock front. As this determination of $\alpha_+$ in terms of $\alpha_N$ also depends on the wall velocity itself, in the end, the wall velocity $\xi_w(\alpha_N; a_-/a_+)$ of the weak deflagration expansion has to be solved from the junction condition~\eqref{eq:deflagrationentropyflow} as a function of the asymptotic strength factor $\alpha_N$ provided that $\alpha_N$ is within a narrow range $[\alpha_\mathrm{min},\alpha_\mathrm{max}]$ depending on the ratio $a_-/a_+<1$.

\begin{figure}
    \centering
    \includegraphics[width = 0.52\textwidth]{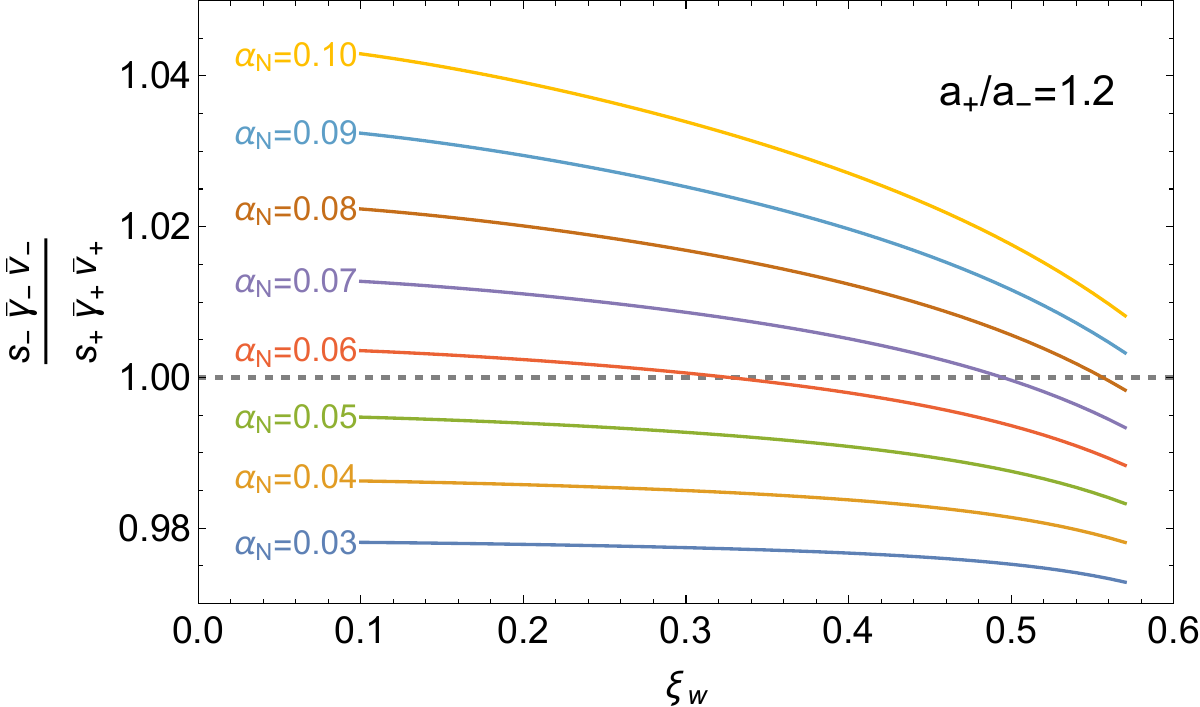}
    \includegraphics[width = 0.47\textwidth]{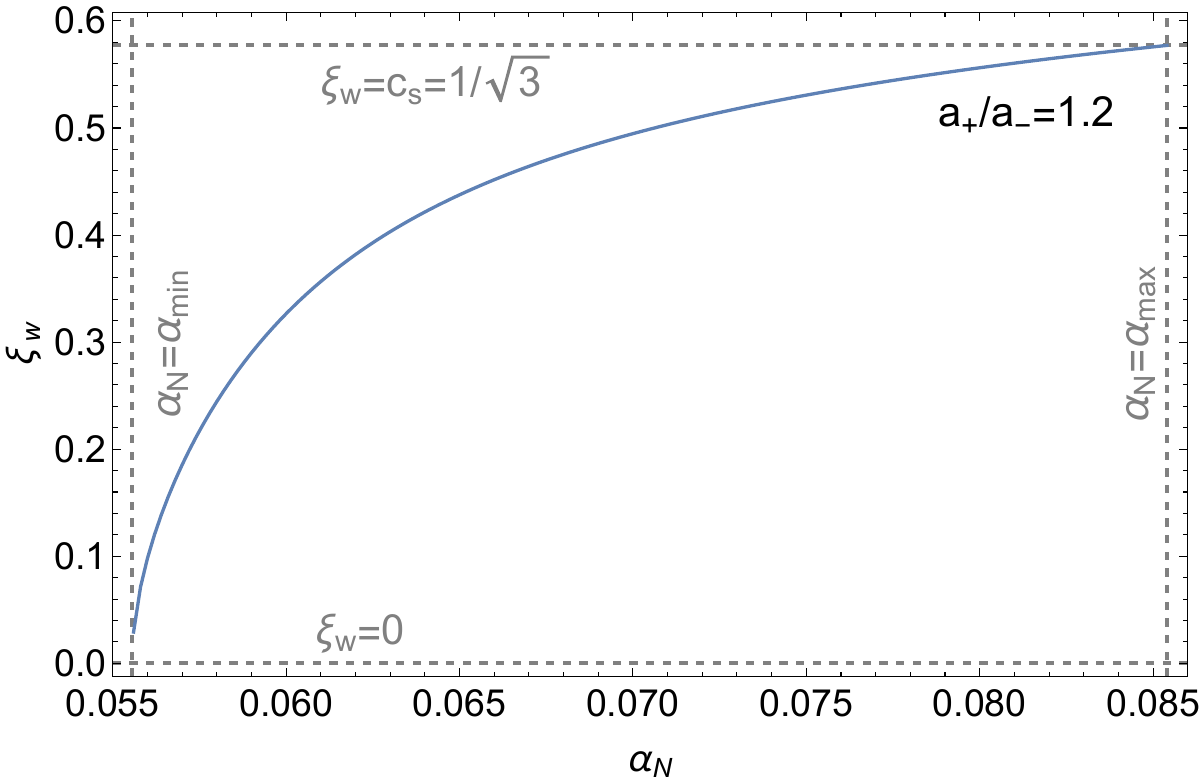}\\
    \caption{Left: the value of $(s_-\bar{\gamma}_-\bar{v}_-)/(s_+\bar{\gamma}_+\bar{v}_+)$ as a function of the spherical wall velocity $\xi_w$ for different values of $\alpha_N$ given an illustrative $a_+/a_- = 1.2$. Right: the spherical wall velocity obtained from the intersection of the curve with the horizontal dashed line in the left panel as a function of $\alpha_N$ for the same $a_+/a_- = 1.2$. For the deflagration case with $0<\xi_w<c_s = 1/\sqrt{3}$, the asymptotic strength factor is constrained between $\alpha_{\mathrm{min}}\simeq0.0556$ and $\alpha_{\mathrm{max}}\simeq0.0854$.}
    \label{fig:deflag_AI}
\end{figure}

To see why there is a bizarre narrow range $[\alpha_\mathrm{min}, \alpha_\mathrm{max}]$ for $\alpha_N$ to possibly admit a wall velocity solution from the junction condition~\eqref{eq:NewJunction}, we have directly calculated the ratio $(s_-\bar{\gamma}_-\bar{v}_-)/(s_+\bar{\gamma}_+\bar{v}_+)$ as a function of the wall velocity $\xi_w$ for different values of $\alpha_N$ given an illustrative ratio $a_+/a_-=1.2$ in the left panel of Fig.~\ref{fig:deflag_AI}. The intersection of each curve with $1$ (horizontal dashed line) solves for the bubble wall velocity as a function of $\alpha_N$ as shown in the right panel of Fig.~\ref{fig:deflag_AI} for the same $a_+/a_-=1.2$. However, such an intersection is not always possible for all values of $\alpha_N$ given $a_+/a_->1$. The minimal value of $\alpha_N$ is the one when the junction condition $s_-\bar{\gamma}_-\bar{v}_-=s_+\bar{\gamma}_+\bar{v}_+$ is found at a vanishing wall velocity $\xi_w\to0$, in which case $\alpha_N\to\alpha_+$ and $\bar{v}_+\to(1-3\alpha_+)\xi_w$, and hence the junction condition~\eqref{eq:NewJunction} can be solved for the minimal value of $\alpha_N$ as
\begin{align}
\alpha_\mathrm{min}=\frac13\left(1-\frac{a_-}{a_+}\right).
\end{align}
On the other hand, the maximum value of $\alpha_N$ is the one when the junction condition $s_-\bar{\gamma}_-\bar{v}_-=s_+\bar{\gamma}_+\bar{v}_+$ is found at a velocity near the speed of sound, $\xi_w\to c_s\equiv1/\sqrt{3}$, in which case $\bar{v}_+\to2/[\sqrt{3}(1+\alpha_+)]-\xi_J(\alpha_+)$ with the Chapman-Jouguet velocity $\xi_J(\alpha_+)=[\sqrt{\alpha_+(2+3\alpha_+)}+1]/[\sqrt{3}(1+\alpha_+)]$, and hence the junction condition ~\eqref{eq:NewJunction} can be solved for the maximum value of $\alpha_+$, then the corresponding maximum value of $\alpha_N$ can be solved from a numerical fitting formula $\alpha_+\simeq\alpha_N(0.329-0.0793\ln\alpha_N+0.0116\ln^2\alpha_N+0.00159\ln^3\alpha_N)$~\cite{Espinosa:2010hh}.

\subsection{The ``violation'' of entropy density current conservation across shockwave front}
\label{app:subsec:breakdown}

If the extra junction condition $s_-\bar{\gamma}_-\bar{v}_-=s_+\bar{\gamma}_+\bar{v}_+$ is really the correct reflection of the conservation of the entropy density current, then we can directly test it across the shock front within the symmetric phase where no entropy density current is supposed to be generated. Denoting $s_{L,R}$ and $\tilde{\gamma}_{L,R}\equiv(1-\tilde{v}_{L,R}^2)^{-1/2}$ as the entropy density and Lorentz factor of the fluid velocity $\tilde{v}_{L,R}$ in the rest frame of the shock front just behind and in front of the shock front, respectively, we can take the ratio, 
\begin{align}\label{eq:check1}
    \frac{s_L\tilde{\gamma}_L\tilde{v}_L}{s_R\tilde{\gamma}_R\tilde{v}_R}=\frac{s_L w_R\tilde{\gamma}_R}{s_Rw_L\tilde{\gamma}_L}=\frac{T_R\tilde{\gamma}_R}{T_L\tilde{\gamma}_L}=\left(\frac{w_R}{w_L}\right)^\frac14\frac{\tilde{\gamma}_R}{\tilde{\gamma}_L}=\left(\frac{\tilde{v}_L\tilde{\gamma}_L^2}{\tilde{v}_R\tilde{\gamma}_R^2}\right)^\frac14\frac{\tilde{\gamma}_R}{\tilde{\gamma}_L}=\left(\frac{\tilde{v}_L\tilde{\gamma}_R^2}{\tilde{v}_R\tilde{\gamma}_L^2}\right)^\frac14,
\end{align}
where the usual junction condition and some thermodynamic relations,
\begin{align}\label{eq:check2}
    \frac{\tilde{\gamma}_L\tilde{v}_L}{\tilde{\gamma}_R\tilde{v}_R}=\frac{w_R\tilde{\gamma}_R}{w_L\tilde{\gamma}_L},\quad
    w_{L/R}=T_{L/R}s_{L/R},\quad
    w_{L/R}=\frac43a_+T_{L/R}^4,\quad
    \frac{w_R}{w_L}=\frac{\tilde{v}_L\tilde{\gamma}_L^2}{\tilde{v}_R\tilde{\gamma}_R^2},
\end{align}
are used in the first, second, third, and fourth equalities, respectively. Since the fluid velocity just in front of the shock front equals to the shock front velocity, $\tilde{v}_R=\xi_{sh}$ in the rest frame of the shock front, and the fluid velocity just behind the shock front reads $\tilde{v}_L=1/(3\tilde{v}_R)=1/(3\xi_{sh})$ in the rest frame of the shock front, then it is easy to see that
\begin{align}\label{eq:check3}
    \frac{s_L\tilde{\gamma}_L\tilde{v}_L}{s_R\tilde{\gamma}_R\tilde{v}_R}=\left(\frac{9\xi_{sh}^2-1}{27\xi_{sh}^4(1-\xi_{sh}^2)}\right)^\frac14>1,
\end{align}
which is always larger than $1$ since $(9\xi_{sh}^2-1)-27\xi_{sh}^4(1-\xi_{sh}^2)=(3\xi_{sh}^2-1)^3>0$ for $\xi_{sh}>1/\sqrt{3}$. Unless $\xi_{sh}=1/\sqrt{3}$, namely $\xi_w=0$ (no bubble wall), the proposed junction condition~\eqref{eq:NewJunction} for the conserved entropy density current is obviously violated across the shock front. This violation was also observed in the footnote 7 of Ref.~\cite{Ai:2021kak} but was argued to be likely cured by going beyond the planar wall approximation. However, as we can easily see, all the derivations above, Eqs.~\eqref{eq:check1}, \eqref{eq:check2}, and~\eqref{eq:check3}, actually do not depend on the global geometric shape of the wall. Therefore, the proposed junction condition~\eqref{eq:NewJunction} for the conserved entropy density current is generally not correct across the shock front regardless of the wall geometries. On the other hand, we can actually prove the conservation of entropy density current across the shock front from the conservation of energy-momentum tensor under the thin-wall approximation.

\subsection{The conservation of the entropy density current from the conserved stress tensor}
\label{app:subsec:conservation}

Intuitively, the entropy density current itself should be conserved anyway across the shock front as naively expected from the simple physical picture that there is no net entropy generation across the shock front, namely $a_L=a_R=a_+$, since they are all in the symmetric phase so as to maintain local dynamical equilibrium as required in the early Universe. Indeed, we can actually prove the conservation for the entropy density current across the shock front directly from the conservation of the total energy-momentum tensor by noting that,
\begin{align}
    \begin{aligned}
        0&=u_\nu\nabla_\mu T^{\mu\nu}=u_\nu\nabla_\mu(wu^\mu u^\nu+p\eta^{\mu\nu})\\
        &=(u_\nu\nabla_\mu u^\nu)(wu^\mu)+(u_\nu u^\nu)\nabla_\mu(wu^\mu)+u^\mu\nabla_\mu p\\
        &=-\nabla_\mu(wu^\mu)+u^\mu\left(\nabla_\mu T\frac{\partial p}{\partial T}+\nabla_\mu\phi\frac{\partial p}{\partial\phi}\right)\\
        &=-T\nabla_\mu(s u^\mu)-(su^\mu)\nabla_\mu T+u^\mu(\nabla_\mu T)s+u^\mu\nabla_\mu\phi\frac{\partial p}{\partial\phi}\\
        &=-T\nabla_\mu(s u^\mu)+u^\mu\nabla_\mu\phi\frac{\partial p}{\partial\phi},
    \end{aligned}  
\end{align}
where only the conserved perfect-fluid ansatz $T^{\mu\nu}=wu^\mu u^\nu+p\eta^{\mu\nu}$, the four-velocity properties $u_\nu\nabla_\mu u^\nu=0$ and $u_\nu u^\nu=-1$, the chain rule for derivative of total pressure $p(\phi,T)$, and the definition for the entropy density $s=\partial p/\partial T$ are used in the first, third, and forth lines, respectively. Hence the conservation equation for the entropy density current now reads
\begin{align}\label{eq:entropyflowconservation}
T\nabla_\mu(s u^\mu)=u^\mu\nabla_\mu\phi\frac{\partial p}{\partial\phi}.
\end{align}
It is easy to see that, with the thin-wall approximation, $\nabla_\mu\phi$ is everywhere zero but a Dirac-delta function at the bubble wall, thus, it always holds $\nabla_\mu\phi=0$ at the shock front, and $\partial p/\partial\phi$ is simply a normal function of $\phi$ and $T$ at the shock front due to $p=-V_\mathrm{eff}(\phi,T)$ in local equilibrium. Hence, we generally prove the conservation of the entropy density current $T\nabla_\mu(su^\mu)=0$ across the shock front in local equilibrium regardless of thin-wall geometries as a result of the conservation of the perfect-fluid energy-momentum tensor under the thin-wall approximation. It is worth noting that, this relation between the conservation of entropy density current and the thin-wall approximation across the shock front is also obvious from previous phenomenological parameterization Eq.(5.28) of Ref.~\cite{Hindmarsh:2020hop},
\begin{align}
\partial_\mu S^\mu=\frac{\tilde{\eta}}{T}(u^\mu\partial_\mu\phi)^2.
\end{align}

On the other hand, we have explicitly checked in the last subsection with the thin-wall perfect-fluid bag-like hydrodynamics that the proposed junction condition $s_L\tilde{\gamma}_L\tilde{v}_L=s_R\tilde{\gamma}_R\tilde{v}_R$ is always violated across the shock front regardless of wall geometries. Unlike the number density current case, we cannot solve this contradiction by including the scalar-wall contribution to the entropy density current since the entropy density of a thin wall is simply zero, $s=\partial p_\phi/\partial T=0$ as $p_\phi=-\frac12(\nabla\phi)^2-V_0$ is independent of temperature $T$ for a bag EoS~\footnote{This also suggests that, for a EoS beyond the bag model, we might be able to resolve this contradiction by including the scalar-wall contribution to the entropy density current, or put in another way, since there is no clear separation between the scalar-wall and plasma fluid contributions, one naturally use the total entropy density current to begin with.}. Therefore, as long as we work with hydrodynamics under the thin-wall approximation for a bag EoS, we will always encounter this contradiction, and one can only conclude that the proposed junction condition $s_L\tilde{\gamma}_L\tilde{v}_L=s_R\tilde{\gamma}_R\tilde{v}_R$ is not an appropriate reflection for a conserved entropy density current across the shock front, and hence one should also not apply it as $s_-\bar{\gamma}_-\bar{v}_-=s_+\bar{\gamma}_+\bar{v}_+$ across the bubble wall in local equilibrium.

\subsection{The difficulty in deriving junction conditions from entropy current conservation}
\label{app:subsec:entropyflow}

As we can see from Sec.~\ref{subsec:JunctionConditions},  the usual junction conditions~\eqref{eq:junctionwall1}, \eqref{eq:junctionwall2} and~\eqref{eq:junctionshock1}, \eqref{eq:junctionshock2} are not directly obtained from the differential form $\nabla_\mu T^{\mu\nu}=0$, but are actually derived from the integral form~\eqref{eq:TIntegralStokes} across a discontinuous interface. Only because the integrand $\nabla_\mu(T^{\mu\nu}\lambda_\nu)$ is a total derivative so that we can fully integrate it as~\eqref{eq:TIntegralStokes} with the Stokes's theorem. However, the conservation equation~\eqref{eq:entropyflowconservation} of the entropy density current across the shock front, $T\nabla_\mu(su^\mu)=0$, is not a total derivative of anything and hence cannot be fully integrated by the Stokes's theorem into the proposed junction condition $s_L\tilde{\gamma}_L\tilde{v}_L=s_R\tilde{\gamma}_R\tilde{v}_R$, nor should we apply it to the bubble wall as $s_-\bar{\gamma}_-\bar{v}_-=s_+\bar{\gamma}_+\bar{v}_+$. Note here that the temperature $T$ on the left hand side cannot be simply removed by division. This is because that the temperature profile $T$ is a discontinuous function across the bubble wall/shock front, which, after multiplied by a Dirac-delta-like function $\nabla_\mu(s u^\mu)$, renders a highly non-trivial integral~\footnote{To see why such an integral on the left-hand side is highly non-trivial, let's look at a simplified example,
\begin{align}
I=\int_a^b \mathrm{d}x \, f(x) \delta(x-c),\quad a<c<b,
\end{align}
which is simply $f(c)$ if $f(x)$ is a continuous function within the integration domain. However, if $f(x)$ is also discontinuous at $x=c$, then we should find another discontinuous function $g(x)$ at $x=c$ so that $g'(x)=\delta(x-c)$ so that 
\begin{align}
I=\int_{g(a)}^{g(b)}\mathrm{d}g(x)\,f(g(x)).
\end{align}
Now if we can provide a junction condition at $x=c$ relating $f$ and $g$ by a continuous function $f=h(g)$, then this non-trivial integral can be consistently calculated as
\begin{align}
I=\lim_{\delta\to0}\int_{g(c-\delta)}^{g(c+\delta)}\mathrm{d}g\, h(g).
\end{align}
This mathematical trick was first introduced in our previous study~\cite{Wang:2022txy} with numerical checks and further tested in our recent study~\cite{Wang:2023kux} with analytical validations.} that is not equivalent to the integral over $\nabla_\mu(s u^\mu)$ without $T$,
\begin{align}
\int_\mathcal{V}T\nabla_\mu(s u^\mu)\mathrm{d}^4x=0\nLeftrightarrow \int_\mathcal{V}\nabla_\mu(s u^\mu)\mathrm{d}^4x=0.
\end{align}
when the integration region $\mathcal{V}$ contains such a discontinuous interface. The integral on the right-hand side, even if it were to be zero, would simply give rise to the proposed junction condition $s_-\bar{\gamma}_-\bar{v}_-=s_+\bar{\gamma}_+\bar{v}_+$ or $s_L\tilde{\gamma}_L\tilde{v}_L=s_R\tilde{\gamma}_R\tilde{v}_R$, which, however, has nothing to do with the true conservation equation of the entropy density current as shown on the left-hand side with an integral form. Unfortunately, this highly non-trivial integral cannot be fully integrated by the Stokes's theorem to arrive at any form of junction conditions. 

One way to remedy this is to add some term to $T\nabla_\mu(su^\mu)$ to produce a total derivative, for example, $T\nabla_\mu(su^\mu)+su^\mu\nabla_\mu T=\nabla_\mu(Tsu^\mu)=\nabla_\mu(wu^\mu)$, which is the total enthalpy current, but it is not conserved across the bubble wall or shock front even in local equilibrium as one can explicitly check with hydrodynamics. Here we provide and prove another but non-trivial example that is conserved in local equilibrium,
\begin{align}
\nabla_\mu(w_\phi u^\mu)=T\nabla_\mu(su^\mu)+u^\mu\nabla_\mu T\frac{\partial\delta p_\phi}{\partial T}.
\end{align}
With the most general decomposition in non-equilibrium, $p=p_\phi+p_f=(-V_0+\delta p_\phi)+(-\Delta V_T+\delta p_f)=-(V_0+\Delta V_T)+(\delta p_\phi+\delta p_f)=-V_\mathrm{eff}+\delta p$, our starting point is the plasma EoM~\eqref{eq:EOMplasma}, which, after projected along the fluid velocity,
\begin{align}
u_\nu\nabla_\mu(w_fu^\mu u^\nu+p_f\eta^{\mu\nu})+u_\nu\nabla^\nu\phi\frac{\partial\Delta V_T}{\partial\phi}=u_\nu\nabla^\nu\phi\frac{\partial p_{\delta f}}{\partial\phi},
\end{align}
becomes
\begin{align}
(u_\nu\nabla_\mu u^\nu)(w_fu^\mu)+(u_\nu u^\nu)\nabla_\mu(w_f u^\mu)+u^\mu\nabla_\mu p_f+u^\mu\nabla_\mu\phi\frac{\partial\Delta V_T}{\partial\phi}=u^\mu\nabla_\mu\phi\frac{\partial p_{\delta f}}{\partial\phi}.
\end{align}
The first term is vanished as $u_\nu\nabla_\mu u^\nu=0$, the second term after normalization $u_\nu u^\nu=-1$ will be split into two terms $-\nabla_\mu(w_f u^\mu)=\nabla_\mu(w_\phi u^\mu)-\nabla_\mu(w u^\mu)$ by $w=w_\phi+w_f$. The third term can also be split into two terms $u^\mu\nabla_\mu p_f=u^\mu\nabla_\mu(-\Delta V_T+\delta p_f)$ with further splitting for its first contribution $-u^\mu\nabla_\mu\Delta V_T=-u^\mu\nabla_\mu T\partial_T\Delta V_T-u^\mu\nabla_\mu\phi\partial_\phi\Delta V_T$, where the second splitting term can be canceled out by the fourth term of left-hand side of above equation. Then, the projected plasma EoM looks like
\begin{align}
\nabla_\mu(w_\phi u^\mu)-\nabla_\mu(w u^\mu)-u^\mu\nabla_\mu T\frac{\partial\Delta V_T}{\partial T}+u^\mu\nabla_\mu\delta p_f=u^\mu\nabla_\mu\phi\frac{\partial p_{\delta f}}{\partial\phi},
\end{align}
where the second term can be split into two terms $-\nabla_\mu(wu^\mu)=-T\nabla_\mu(su^\mu)-su^\mu\nabla_\mu T$, and the $\partial_T\Delta V_T$ factor in the third term becomes $\partial_T\Delta V_T=\partial_TV_\mathrm{eff}=\partial_T(-p+\delta p)=-s+\partial_T\delta p$. Then, the $-su^\mu\nabla_\mu T$ contribution from the second term will be canceled out by the $-u^\mu(\nabla_\mu T)(-s)$ contribution from the third term. The fourth term on the left-hand side can also split into two terms $u^\mu\nabla_\mu\delta p_f=u^\mu\nabla_\mu T\partial_T\delta p_f+u^\mu\nabla_\mu\phi\partial_\phi\delta p_f$. Now the projected plasma EoM arrives at
\begin{align}
\nabla_\mu(w_\phi u^\mu)-T\nabla_\mu(su^\mu)-u^\mu\nabla_\mu T\frac{\partial\delta p}{\partial T}+u^\mu\nabla_\mu T\frac{\partial\delta p_f}{\partial T}+u^\mu\nabla_\mu\phi\frac{\partial\delta p_f}{\partial\phi}=u^\mu\nabla_\mu\phi\frac{\partial p_{\delta f}}{\partial\phi},
\end{align}
where the third term and fourth term can be merged into $-u^\mu\nabla_\mu T\partial_T\delta p+u^\mu\nabla_\mu T\partial_T\delta p_f=-u^\mu\nabla_\mu T\partial_T\delta p_\phi$. The fifth term on the left-hand side can be canceled out by the term on the right-hand side by recalling that we have proved $\partial_\phi\delta p_f=\partial_\phi p_{\delta f}$ in the footnote 2. Therefore, the final form of the projected plasma EoM reads
\begin{align}
\nabla_\mu(w_\phi u^\mu)=T\nabla_\mu(su^\mu)+u^\mu\nabla_\mu T\frac{\partial\delta p_\phi}{\partial T},
\end{align}
where both the first and second terms on the right-hand side would be vanished in local equilibrium, leading to a conserved scalar enthalpy current across the bubble wall as well as shock front, if any. However, the junction condition given by the weak-form integral of the left-hand side, $w_{\phi-}\bar{\gamma}_-\bar{v}_-=w_{\phi+}\bar{\gamma}_+\bar{v}_+$, is nothing but an identity since both $w_{\phi\pm}\equiv w_\phi(\xi\to\xi_w+0^\pm)=0$ are zero under the thin-wall approximation. Therefore, this non-trivial example does not give anything useful from the junction condition of the scalar enthalpy current conservation, but it might work when the bubble wall has a finite width. Further investigations beyond the thin-wall approximation will be reserved for future works.

In the end, the only remaining way out is to seek for a totally different integration over a total derivative but analog to the entropy density current conservation in local equilibrium, for example, the conservation equation of the number density current and its possible violation across the bubble wall, which is the main motivation and starting point of this paper.

\bibliographystyle{JHEP}
\bibliography{ref}

\providecommand{\href}[2]{#2}\begingroup\raggedright\begin{thebibliography}{100}

\bibitem{Mazumdar:2018dfl}
A.~Mazumdar and G.~White, \emph{{Review of cosmic phase transitions: their
  significance and experimental signatures}},
  \href{http://dx.doi.org/10.1088/1361-6633/ab1f55}{\emph{Rept. Prog. Phys.}
  {\bf 82} (2019) 076901}, [\href{https://arxiv.org/abs/1811.01948}{{\tt
  1811.01948}}].

\bibitem{Hindmarsh:2020hop}
M.~B. Hindmarsh, M.~L\"uben, J.~Lumma and M.~Pauly, \emph{{Phase transitions in
  the early universe}},
  \href{http://dx.doi.org/10.21468/SciPostPhysLectNotes.24}{\emph{SciPost Phys.
  Lect. Notes} {\bf 24} (2021) 1},
  [\href{https://arxiv.org/abs/2008.09136}{{\tt 2008.09136}}].

\bibitem{Caldwell:2022qsj}
R.~Caldwell et~al., \emph{{Detection of early-universe gravitational-wave
  signatures and fundamental physics}},
  \href{http://dx.doi.org/10.1007/s10714-022-03027-x}{\emph{Gen. Rel. Grav.}
  {\bf 54} (2022) 156}, [\href{https://arxiv.org/abs/2203.07972}{{\tt
  2203.07972}}].

\bibitem{Athron:2023xlk}
P.~Athron, C.~Bal\'azs, A.~Fowlie, L.~Morris and L.~Wu, \emph{{Cosmological
  phase transitions: From perturbative particle physics to gravitational
  waves}}, \href{http://dx.doi.org/10.1016/j.ppnp.2023.104094}{\emph{Prog.
  Part. Nucl. Phys.} {\bf 135} (2024) 104094},
  [\href{https://arxiv.org/abs/2305.02357}{{\tt 2305.02357}}].

\bibitem{Cai:2017cbj}
R.-G. Cai, Z.~Cao, Z.-K. Guo, S.-J. Wang and T.~Yang, \emph{{The
  Gravitational-Wave Physics}},
  \href{http://dx.doi.org/10.1093/nsr/nwx029}{\emph{Natl. Sci. Rev.} {\bf 4}
  (2017) 687--706}, [\href{https://arxiv.org/abs/1703.00187}{{\tt
  1703.00187}}].

\bibitem{Bian:2021ini}
L.~Bian et~al., \emph{{The Gravitational-wave physics II: Progress}},
  \href{http://dx.doi.org/10.1007/s11433-021-1781-x}{\emph{Sci. China Phys.
  Mech. Astron.} {\bf 64} (2021) 120401},
  [\href{https://arxiv.org/abs/2106.10235}{{\tt 2106.10235}}].

\bibitem{Caprini:2015zlo}
C.~Caprini et~al., \emph{{Science with the space-based interferometer eLISA.
  II: Gravitational waves from cosmological phase transitions}},
  \href{http://dx.doi.org/10.1088/1475-7516/2016/04/001}{\emph{JCAP} {\bf 1604}
  (2016) 001}, [\href{https://arxiv.org/abs/1512.06239}{{\tt 1512.06239}}].

\bibitem{Caprini:2019egz}
C.~Caprini et~al., \emph{{Detecting gravitational waves from cosmological phase
  transitions with LISA: an update}},
  \href{http://dx.doi.org/10.1088/1475-7516/2020/03/024}{\emph{JCAP} {\bf 2003}
  (2020) 024}, [\href{https://arxiv.org/abs/1910.13125}{{\tt 1910.13125}}].

\bibitem{Liu:2022lvz}
J.~Liu, L.~Bian, R.-G. Cai, Z.-K. Guo and S.-J. Wang, \emph{{Constraining
  First-Order Phase Transitions with Curvature Perturbations}},
  \href{http://dx.doi.org/10.1103/PhysRevLett.130.051001}{\emph{Phys. Rev.
  Lett.} {\bf 130} (2023) 051001},
  [\href{https://arxiv.org/abs/2208.14086}{{\tt 2208.14086}}].

\bibitem{Elor:2023xbz}
G.~Elor, R.~Jinno, S.~Kumar, R.~McGehee and Y.~Tsai, \emph{{Finite Bubble
  Statistics Constrain Late Cosmological Phase Transitions}},
  \href{https://arxiv.org/abs/2311.16222}{{\tt 2311.16222}}.

\bibitem{Lewicki:2024ghw}
M.~Lewicki, P.~Toczek and V.~Vaskonen, \emph{{Black holes and gravitational
  waves from slow phase transitions}},
  \href{https://arxiv.org/abs/2402.04158}{{\tt 2402.04158}}.

\bibitem{Cai:2024nln}
R.-G. Cai, Y.-S. Hao and S.-J. Wang, \emph{{Primordial black holes and
  curvature perturbations from false vacuum islands}},
  \href{http://dx.doi.org/10.1007/s11433-024-2416-3}{\emph{Sci. China Phys.
  Mech. Astron.} {\bf 67} (2024) 290411},
  [\href{https://arxiv.org/abs/2404.06506}{{\tt 2404.06506}}].

\bibitem{Hawking:1982ga}
S.~W. Hawking, I.~G. Moss and J.~M. Stewart, \emph{{Bubble Collisions in the
  Very Early Universe}},
  \href{http://dx.doi.org/10.1103/PhysRevD.26.2681}{\emph{Phys. Rev.} {\bf D26}
  (1982) 2681}.

\bibitem{Moss:1994pi}
I.~G. Moss, \emph{{Black hole formation from colliding bubbles}},
  \href{https://arxiv.org/abs/gr-qc/9405045}{{\tt gr-qc/9405045}}.

\bibitem{Moss:1994iq}
I.~G. Moss, \emph{{Singularity formation from colliding bubbles}},
  \href{http://dx.doi.org/10.1103/PhysRevD.50.676}{\emph{Phys. Rev.} {\bf D50}
  (1994) 676--681}.

\bibitem{Jung:2021mku}
T.~H. Jung and T.~Okui, \emph{{Primordial black holes from bubble collisions
  during a first-order phase transition}},
  \href{https://arxiv.org/abs/2110.04271}{{\tt 2110.04271}}.

\bibitem{Baker:2021nyl}
M.~J. Baker, M.~Breitbach, J.~Kopp and L.~Mittnacht, \emph{{Primordial Black
  Holes from First-Order Cosmological Phase Transitions}},
  \href{https://arxiv.org/abs/2105.07481}{{\tt 2105.07481}}.

\bibitem{Baker:2021sno}
M.~J. Baker, M.~Breitbach, J.~Kopp and L.~Mittnacht, \emph{{Detailed
  Calculation of Primordial Black Hole Formation During First-Order
  Cosmological Phase Transitions}},
  \href{https://arxiv.org/abs/2110.00005}{{\tt 2110.00005}}.

\bibitem{Cline:2022xhx}
J.~M. Cline, B.~Laurent, S.~Raby and J.-S. Roux, \emph{{PeV-scale leptogenesis,
  gravitational waves, and black holes from a SUSY-breaking phase transition}},
  \href{http://dx.doi.org/10.1103/PhysRevD.107.095011}{\emph{Phys. Rev. D} {\bf
  107} (2023) 095011}, [\href{https://arxiv.org/abs/2211.00422}{{\tt
  2211.00422}}].

\bibitem{Kawana:2021tde}
K.~Kawana and K.-P. Xie, \emph{{Primordial black holes from a cosmic phase
  transition: The collapse of Fermi-balls}},
  \href{http://dx.doi.org/10.1016/j.physletb.2021.136791}{\emph{Phys. Lett. B}
  {\bf 824} (2022) 136791}, [\href{https://arxiv.org/abs/2106.00111}{{\tt
  2106.00111}}].

\bibitem{Lu:2022paj}
P.~Lu, K.~Kawana and K.-P. Xie, \emph{{Old phase remnants in first-order phase
  transitions}},
  \href{http://dx.doi.org/10.1103/PhysRevD.105.123503}{\emph{Phys. Rev. D} {\bf
  105} (2022) 123503}, [\href{https://arxiv.org/abs/2202.03439}{{\tt
  2202.03439}}].

\bibitem{Kawana:2022lba}
K.~Kawana, P.~Lu and K.-P. Xie, \emph{{First-order phase transition and fate of
  false vacuum remnants}},
  \href{http://dx.doi.org/10.1088/1475-7516/2022/10/030}{\emph{JCAP} {\bf 10}
  (2022) 030}, [\href{https://arxiv.org/abs/2206.09923}{{\tt 2206.09923}}].

\bibitem{Huang:2022him}
P.~Huang and K.-P. Xie, \emph{{Primordial black holes from an electroweak phase
  transition}},  \href{https://arxiv.org/abs/2201.07243}{{\tt 2201.07243}}.

\bibitem{Kodama:1982sf}
H.~Kodama, M.~Sasaki and K.~Sato, \emph{{Abundance of Primordial Holes Produced
  by Cosmological First Order Phase Transition}},
  \href{http://dx.doi.org/10.1143/PTP.68.1979}{\emph{Prog. Theor. Phys.} {\bf
  68} (1982) 1979}.

\bibitem{Liu:2021svg}
J.~Liu, L.~Bian, R.-G. Cai, Z.-K. Guo and S.-J. Wang, \emph{{Primordial black
  hole production during first-order phase transitions}},
  \href{http://dx.doi.org/10.1103/PhysRevD.105.L021303}{\emph{Phys. Rev. D}
  {\bf 105} (2022) L021303}, [\href{https://arxiv.org/abs/2106.05637}{{\tt
  2106.05637}}].

\bibitem{Hashino:2021qoq}
K.~Hashino, S.~Kanemura and T.~Takahashi, \emph{{Primordial black holes as a
  probe of strongly first-order electroweak phase transition}},
  \href{https://arxiv.org/abs/2111.13099}{{\tt 2111.13099}}.

\bibitem{He:2022amv}
S.~He, L.~Li, Z.~Li and S.-J. Wang, \emph{{Gravitational waves and primordial
  black hole productions from gluodynamics by holography}},
  \href{http://dx.doi.org/10.1007/s11433-023-2293-2}{\emph{Sci. China Phys.
  Mech. Astron.} {\bf 67} (2024) 240411},
  [\href{https://arxiv.org/abs/2210.14094}{{\tt 2210.14094}}].

\bibitem{Kawana:2022olo}
K.~Kawana, T.~Kim and P.~Lu, \emph{{PBH formation from overdensities in delayed
  vacuum transitions}},
  \href{http://dx.doi.org/10.1103/PhysRevD.108.103531}{\emph{Phys. Rev. D} {\bf
  108} (2023) 103531}, [\href{https://arxiv.org/abs/2212.14037}{{\tt
  2212.14037}}].

\bibitem{Gouttenoire:2023naa}
Y.~Gouttenoire and T.~Volansky, \emph{{Primordial Black Holes from Supercooled
  Phase Transitions}},  \href{https://arxiv.org/abs/2305.04942}{{\tt
  2305.04942}}.

\bibitem{Lewicki:2023ioy}
M.~Lewicki, P.~Toczek and V.~Vaskonen, \emph{{Primordial black holes from
  strong first-order phase transitions}},
  \href{http://dx.doi.org/10.1007/JHEP09(2023)092}{\emph{JHEP} {\bf 09} (2023)
  092}, [\href{https://arxiv.org/abs/2305.04924}{{\tt 2305.04924}}].

\bibitem{Kanemura:2024pae}
S.~Kanemura, M.~Tanaka and K.-P. Xie, \emph{{Primordial black holes from slow
  phase transitions: a model-building perspective}},
  \href{http://dx.doi.org/10.1007/JHEP06(2024)036}{\emph{JHEP} {\bf 06} (2024)
  036}, [\href{https://arxiv.org/abs/2404.00646}{{\tt 2404.00646}}].

\bibitem{Flores:2024lng}
M.~M. Flores, A.~Kusenko and M.~Sasaki, \emph{{Revisiting formation of
  primordial black holes in a supercooled first-order phase transition}},
  \href{http://dx.doi.org/10.1103/PhysRevD.110.015005}{\emph{Phys. Rev. D} {\bf
  110} (2024) 015005}, [\href{https://arxiv.org/abs/2402.13341}{{\tt
  2402.13341}}].

\bibitem{Cohen:1990it}
A.~G. Cohen, D.~B. Kaplan and A.~E. Nelson, \emph{{Baryogenesis at the weak
  phase transition}},
  \href{http://dx.doi.org/10.1016/0550-3213(91)90395-E}{\emph{Nucl. Phys.} {\bf
  B349} (1991) 727--742}.

\bibitem{Cohen:1990py}
A.~G. Cohen, D.~B. Kaplan and A.~E. Nelson, \emph{{WEAK SCALE BARYOGENESIS}},
  \href{http://dx.doi.org/10.1016/0370-2693(90)90690-8}{\emph{Phys. Lett.} {\bf
  B245} (1990) 561--564}.

\bibitem{Cohen:1993nk}
A.~G. Cohen, D.~B. Kaplan and A.~E. Nelson, \emph{{Progress in electroweak
  baryogenesis}},
  \href{http://dx.doi.org/10.1146/annurev.ns.43.120193.000331}{\emph{Ann. Rev.
  Nucl. Part. Sci.} {\bf 43} (1993) 27--70},
  [\href{https://arxiv.org/abs/hep-ph/9302210}{{\tt hep-ph/9302210}}].

\bibitem{Cohen:1994ss}
A.~G. Cohen, D.~B. Kaplan and A.~E. Nelson, \emph{{Diffusion enhances
  spontaneous electroweak baryogenesis}},
  \href{http://dx.doi.org/10.1016/0370-2693(94)00935-X}{\emph{Phys. Lett.} {\bf
  B336} (1994) 41--47}, [\href{https://arxiv.org/abs/hep-ph/9406345}{{\tt
  hep-ph/9406345}}].

\bibitem{Cohen:2012zza}
T.~Cohen, D.~E. Morrissey and A.~Pierce, \emph{{Electroweak Baryogenesis and
  Higgs Signatures}},
  \href{http://dx.doi.org/10.1103/PhysRevD.86.013009}{\emph{Phys. Rev.} {\bf
  D86} (2012) 013009}, [\href{https://arxiv.org/abs/1203.2924}{{\tt
  1203.2924}}].

\bibitem{Hogan:1983zz}
C.~J. Hogan, \emph{{Magnetohydrodynamic Effects of a First-Order Cosmological
  Phase Transition}},
  \href{http://dx.doi.org/10.1103/PhysRevLett.51.1488}{\emph{Phys. Rev. Lett.}
  {\bf 51} (1983) 1488--1491}.

\bibitem{Quashnock:1988vs}
J.~M. Quashnock, A.~Loeb and D.~N. Spergel, \emph{{Magnetic Field Generation
  During the Cosmological QCD Phase Transition}},
  \href{http://dx.doi.org/10.1086/185528}{\emph{Astrophys. J. Lett.} {\bf 344}
  (1989) L49--L51}.

\bibitem{Vachaspati:1991nm}
T.~Vachaspati, \emph{{Magnetic fields from cosmological phase transitions}},
  \href{http://dx.doi.org/10.1016/0370-2693(91)90051-Q}{\emph{Phys. Lett. B}
  {\bf 265} (1991) 258--261}.

\bibitem{Cheng:1994yr}
B.-l. Cheng and A.~V. Olinto, \emph{{Primordial magnetic fields generated in
  the quark - hadron transition}},
  \href{http://dx.doi.org/10.1103/PhysRevD.50.2421}{\emph{Phys. Rev.} {\bf D50}
  (1994) 2421--2424}.

\bibitem{Di:2020kbw}
Y.~Di, J.~Wang, R.~Zhou, L.~Bian, R.-G. Cai and J.~Liu, \emph{{Magnetic Field
  and Gravitational Waves from the First-Order Phase Transition}},
  \href{http://dx.doi.org/10.1103/PhysRevLett.126.251102}{\emph{Phys. Rev.
  Lett.} {\bf 126} (2021) 251102},
  [\href{https://arxiv.org/abs/2012.15625}{{\tt 2012.15625}}].

\bibitem{Yang:2021uid}
J.~Yang and L.~Bian, \emph{{Magnetic field generation from bubble collisions
  during first-order phase transition}},
  \href{http://dx.doi.org/10.1103/PhysRevD.106.023510}{\emph{Phys. Rev. D} {\bf
  106} (2022) 023510}, [\href{https://arxiv.org/abs/2102.01398}{{\tt
  2102.01398}}].

\bibitem{Romero:2021kby}
A.~Romero, K.~Martinovic, T.~A. Callister, H.-K. Guo, M.~Mart\'\i{}nez,
  M.~Sakellariadou et~al., \emph{{Implications for First-Order Cosmological
  Phase Transitions from the Third LIGO-Virgo Observing Run}},
  \href{http://dx.doi.org/10.1103/PhysRevLett.126.151301}{\emph{Phys. Rev.
  Lett.} {\bf 126} (2021) 151301},
  [\href{https://arxiv.org/abs/2102.01714}{{\tt 2102.01714}}].

\bibitem{Huang:2021rrk}
F.~Huang, V.~Sanz, J.~Shu and X.~Xue, \emph{{LIGO as a probe of dark sectors}},
  \href{http://dx.doi.org/10.1103/PhysRevD.104.095001}{\emph{Phys. Rev. D} {\bf
  104} (2021) 095001}, [\href{https://arxiv.org/abs/2102.03155}{{\tt
  2102.03155}}].

\bibitem{Jiang:2022mzt}
Y.~Jiang and Q.-G. Huang, \emph{{Constraining the gravitational-wave spectrum
  from cosmological first-order phase transitions using data from LIGO-Virgo
  first three observing runs}},
  \href{http://dx.doi.org/10.1088/1475-7516/2023/06/053}{\emph{JCAP} {\bf 06}
  (2023) 053}, [\href{https://arxiv.org/abs/2203.11781}{{\tt 2203.11781}}].

\bibitem{Badger:2022nwo}
C.~Badger et~al., \emph{{Probing early Universe supercooled phase transitions
  with gravitational wave data}},
  \href{http://dx.doi.org/10.1103/PhysRevD.107.023511}{\emph{Phys. Rev. D} {\bf
  107} (2023) 023511}, [\href{https://arxiv.org/abs/2209.14707}{{\tt
  2209.14707}}].

\bibitem{Yu:2022xdw}
W.-W. Yu and S.-J. Wang, \emph{{Searching for double-peak and doubly broken
  gravitational-wave spectra from Advanced LIGO-Virgo\textquoteright{}s first
  three observing runs}},
  \href{http://dx.doi.org/10.1103/PhysRevD.108.063526}{\emph{Phys. Rev. D} {\bf
  108} (2023) 063526}, [\href{https://arxiv.org/abs/2211.13111}{{\tt
  2211.13111}}].

\bibitem{Leitao:2015fmj}
L.~Leitao and A.~Megevand, \emph{{Gravitational waves from a very strong
  electroweak phase transition}},
  \href{http://dx.doi.org/10.1088/1475-7516/2016/05/037}{\emph{JCAP} {\bf 1605}
  (2016) 037}, [\href{https://arxiv.org/abs/1512.08962}{{\tt 1512.08962}}].

\bibitem{Megevand:2016lpr}
A.~Megevand and S.~Ramirez, \emph{{Bubble nucleation and growth in very strong
  cosmological phase transitions}},
  \href{http://dx.doi.org/10.1016/j.nuclphysb.2017.03.009}{\emph{Nucl. Phys. B}
  {\bf 919} (2017) 74--109}, [\href{https://arxiv.org/abs/1611.05853}{{\tt
  1611.05853}}].

\bibitem{Megevand:2017vtb}
A.~M\'egevand and S.~Ram\'\i{}rez, \emph{{Bubble nucleation and growth in slow
  cosmological phase transitions}},
  \href{http://dx.doi.org/10.1016/j.nuclphysb.2018.01.012}{\emph{Nucl. Phys. B}
  {\bf 928} (2018) 38--71}, [\href{https://arxiv.org/abs/1710.06279}{{\tt
  1710.06279}}].

\bibitem{Jinno:2017ixd}
R.~Jinno, S.~Lee, H.~Seong and M.~Takimoto, \emph{{Gravitational waves from
  first-order phase transitions: Towards model separation by bubble nucleation
  rate}}, \href{http://dx.doi.org/10.1088/1475-7516/2017/11/050}{\emph{JCAP}
  {\bf 1711} (2017) 050}, [\href{https://arxiv.org/abs/1708.01253}{{\tt
  1708.01253}}].

\bibitem{Cutting:2018tjt}
D.~Cutting, M.~Hindmarsh and D.~J. Weir, \emph{{Gravitational waves from vacuum
  first-order phase transitions: from the envelope to the lattice}},
  \href{http://dx.doi.org/10.1103/PhysRevD.97.123513}{\emph{Phys. Rev.} {\bf
  D97} (2018) 123513}, [\href{https://arxiv.org/abs/1802.05712}{{\tt
  1802.05712}}].

\bibitem{Kobakhidze:2017mru}
A.~Kobakhidze, C.~Lagger, A.~Manning and J.~Yue, \emph{{Gravitational waves
  from a supercooled electroweak phase transition and their detection with
  pulsar timing arrays}},
  \href{http://dx.doi.org/10.1140/epjc/s10052-017-5132-y}{\emph{Eur. Phys. J.}
  {\bf C77} (2017) 570}, [\href{https://arxiv.org/abs/1703.06552}{{\tt
  1703.06552}}].

\bibitem{Cai:2017tmh}
R.-G. Cai, M.~Sasaki and S.-J. Wang, \emph{{The gravitational waves from the
  first-order phase transition with a dimension-six operator}},
  \href{http://dx.doi.org/10.1088/1475-7516/2017/08/004}{\emph{JCAP} {\bf 1708}
  (2017) 004}, [\href{https://arxiv.org/abs/1707.03001}{{\tt 1707.03001}}].

\bibitem{Witten:1984rs}
E.~Witten, \emph{{Cosmic Separation of Phases}},
  \href{http://dx.doi.org/10.1103/PhysRevD.30.272}{\emph{Phys. Rev.} {\bf D30}
  (1984) 272--285}.

\bibitem{Hogan:1986qda}
C.~J. Hogan, \emph{{Gravitational radiation from cosmological phase
  transitions}}, \href{http://dx.doi.org/10.1093/mnras/218.4.629}{\emph{Mon.
  Not. Roy. Astron. Soc.} {\bf 218} (1986) 629--636}.

\bibitem{Kosowsky:1991ua}
A.~Kosowsky, M.~S. Turner and R.~Watkins, \emph{{Gravitational radiation from
  colliding vacuum bubbles}},
  \href{http://dx.doi.org/10.1103/PhysRevD.45.4514}{\emph{Phys. Rev.} {\bf D45}
  (1992) 4514--4535}.

\bibitem{Kosowsky:1992rz}
A.~Kosowsky, M.~S. Turner and R.~Watkins, \emph{{Gravitational waves from first
  order cosmological phase transitions}},
  \href{http://dx.doi.org/10.1103/PhysRevLett.69.2026}{\emph{Phys. Rev. Lett.}
  {\bf 69} (1992) 2026--2029}.

\bibitem{Kosowsky:1992vn}
A.~Kosowsky and M.~S. Turner, \emph{{Gravitational radiation from colliding
  vacuum bubbles: envelope approximation to many bubble collisions}},
  \href{http://dx.doi.org/10.1103/PhysRevD.47.4372}{\emph{Phys. Rev.} {\bf D47}
  (1993) 4372--4391}, [\href{https://arxiv.org/abs/astro-ph/9211004}{{\tt
  astro-ph/9211004}}].

\bibitem{Kamionkowski:1993fg}
M.~Kamionkowski, A.~Kosowsky and M.~S. Turner, \emph{{Gravitational radiation
  from first order phase transitions}},
  \href{http://dx.doi.org/10.1103/PhysRevD.49.2837}{\emph{Phys. Rev.} {\bf D49}
  (1994) 2837--2851}, [\href{https://arxiv.org/abs/astro-ph/9310044}{{\tt
  astro-ph/9310044}}].

\bibitem{Huber:2008hg}
S.~J. Huber and T.~Konstandin, \emph{{Gravitational Wave Production by
  Collisions: More Bubbles}},
  \href{http://dx.doi.org/10.1088/1475-7516/2008/09/022}{\emph{JCAP} {\bf 0809}
  (2008) 022}, [\href{https://arxiv.org/abs/0806.1828}{{\tt 0806.1828}}].

\bibitem{Caprini:2007xq}
C.~Caprini, R.~Durrer and G.~Servant, \emph{{Gravitational wave generation from
  bubble collisions in first-order phase transitions: An analytic approach}},
  \href{http://dx.doi.org/10.1103/PhysRevD.77.124015}{\emph{Phys. Rev.} {\bf
  D77} (2008) 124015}, [\href{https://arxiv.org/abs/0711.2593}{{\tt
  0711.2593}}].

\bibitem{Caprini:2009fx}
C.~Caprini, R.~Durrer, T.~Konstandin and G.~Servant, \emph{{General Properties
  of the Gravitational Wave Spectrum from Phase Transitions}},
  \href{http://dx.doi.org/10.1103/PhysRevD.79.083519}{\emph{Phys. Rev.} {\bf
  D79} (2009) 083519}, [\href{https://arxiv.org/abs/0901.1661}{{\tt
  0901.1661}}].

\bibitem{Hindmarsh:2013xza}
M.~Hindmarsh, S.~J. Huber, K.~Rummukainen and D.~J. Weir, \emph{{Gravitational
  waves from the sound of a first order phase transition}},
  \href{http://dx.doi.org/10.1103/PhysRevLett.112.041301}{\emph{Phys. Rev.
  Lett.} {\bf 112} (2014) 041301}, [\href{https://arxiv.org/abs/1304.2433}{{\tt
  1304.2433}}].

\bibitem{Hindmarsh:2015qta}
M.~Hindmarsh, S.~J. Huber, K.~Rummukainen and D.~J. Weir, \emph{{Numerical
  simulations of acoustically generated gravitational waves at a first order
  phase transition}},
  \href{http://dx.doi.org/10.1103/PhysRevD.92.123009}{\emph{Phys. Rev.} {\bf
  D92} (2015) 123009}, [\href{https://arxiv.org/abs/1504.03291}{{\tt
  1504.03291}}].

\bibitem{Hindmarsh:2017gnf}
M.~Hindmarsh, S.~J. Huber, K.~Rummukainen and D.~J. Weir, \emph{{Shape of the
  acoustic gravitational wave power spectrum from a first order phase
  transition}}, \href{http://dx.doi.org/10.1103/PhysRevD.96.103520,
  10.1103/PhysRevD.101.089902}{\emph{Phys. Rev.} {\bf D96} (2017) 103520},
  [\href{https://arxiv.org/abs/1704.05871}{{\tt 1704.05871}}].

\bibitem{Kosowsky:2001xp}
A.~Kosowsky, A.~Mack and T.~Kahniashvili, \emph{{Gravitational radiation from
  cosmological turbulence}},
  \href{http://dx.doi.org/10.1103/PhysRevD.66.024030}{\emph{Phys. Rev.} {\bf
  D66} (2002) 024030}, [\href{https://arxiv.org/abs/astro-ph/0111483}{{\tt
  astro-ph/0111483}}].

\bibitem{Dolgov:2002ra}
A.~D. Dolgov, D.~Grasso and A.~Nicolis, \emph{{Relic backgrounds of
  gravitational waves from cosmic turbulence}},
  \href{http://dx.doi.org/10.1103/PhysRevD.66.103505}{\emph{Phys. Rev.} {\bf
  D66} (2002) 103505}, [\href{https://arxiv.org/abs/astro-ph/0206461}{{\tt
  astro-ph/0206461}}].

\bibitem{Nicolis:2003tg}
A.~Nicolis, \emph{{Relic gravitational waves from colliding bubbles and cosmic
  turbulence}},
  \href{http://dx.doi.org/10.1088/0264-9381/21/4/L05}{\emph{Class. Quant.
  Grav.} {\bf 21} (2004) L27}, [\href{https://arxiv.org/abs/gr-qc/0303084}{{\tt
  gr-qc/0303084}}].

\bibitem{Caprini:2006jb}
C.~Caprini and R.~Durrer, \emph{{Gravitational waves from stochastic
  relativistic sources: Primordial turbulence and magnetic fields}},
  \href{http://dx.doi.org/10.1103/PhysRevD.74.063521}{\emph{Phys. Rev.} {\bf
  D74} (2006) 063521}, [\href{https://arxiv.org/abs/astro-ph/0603476}{{\tt
  astro-ph/0603476}}].

\bibitem{Gogoberidze:2007an}
G.~Gogoberidze, T.~Kahniashvili and A.~Kosowsky, \emph{{The Spectrum of
  Gravitational Radiation from Primordial Turbulence}},
  \href{http://dx.doi.org/10.1103/PhysRevD.76.083002}{\emph{Phys. Rev.} {\bf
  D76} (2007) 083002}, [\href{https://arxiv.org/abs/0705.1733}{{\tt
  0705.1733}}].

\bibitem{Caprini:2009yp}
C.~Caprini, R.~Durrer and G.~Servant, \emph{{The stochastic gravitational wave
  background from turbulence and magnetic fields generated by a first-order
  phase transition}},
  \href{http://dx.doi.org/10.1088/1475-7516/2009/12/024}{\emph{JCAP} {\bf 0912}
  (2009) 024}, [\href{https://arxiv.org/abs/0909.0622}{{\tt 0909.0622}}].

\bibitem{Niksa:2018ofa}
P.~Niksa, M.~Schlederer and G.~Sigl, \emph{{Gravitational Waves produced by
  Compressible MHD Turbulence from Cosmological Phase Transitions}},
  \href{http://dx.doi.org/10.1088/1361-6382/aac89c}{\emph{Class. Quant. Grav.}
  {\bf 35} (2018) 144001}, [\href{https://arxiv.org/abs/1803.02271}{{\tt
  1803.02271}}].

\bibitem{Pol:2019yex}
A.~Roper~Pol, S.~Mandal, A.~Brandenburg, T.~Kahniashvili and A.~Kosowsky,
  \emph{{Numerical simulations of gravitational waves from early-universe
  turbulence}},
  \href{http://dx.doi.org/10.1103/PhysRevD.102.083512}{\emph{Phys. Rev.} {\bf
  D102} (2020) 083512}, [\href{https://arxiv.org/abs/1903.08585}{{\tt
  1903.08585}}].

\bibitem{Brandenburg:2021tmp}
A.~Brandenburg, E.~Clarke, Y.~He and T.~Kahniashvili, \emph{{Can we observe the
  QCD phase transition-generated gravitational waves through pulsar timing
  arrays?}}, \href{http://dx.doi.org/10.1103/PhysRevD.104.043513}{\emph{Phys.
  Rev. D} {\bf 104} (2021) 043513},
  [\href{https://arxiv.org/abs/2102.12428}{{\tt 2102.12428}}].

\bibitem{Brandenburg:2021bvg}
A.~Brandenburg, G.~Gogoberidze, T.~Kahniashvili, S.~Mandal, A.~Roper~Pol and
  N.~Shenoy, \emph{{The scalar, vector, and tensor modes in gravitational wave
  turbulence simulations}},
  \href{http://dx.doi.org/10.1088/1361-6382/ac011c}{\emph{Class. Quant. Grav.}
  {\bf 38} (2021) 145002}, [\href{https://arxiv.org/abs/2103.01140}{{\tt
  2103.01140}}].

\bibitem{Jinno:2016vai}
R.~Jinno and M.~Takimoto, \emph{{Gravitational waves from bubble collisions: An
  analytic derivation}},
  \href{http://dx.doi.org/10.1103/PhysRevD.95.024009}{\emph{Phys. Rev.} {\bf
  D95} (2017) 024009}, [\href{https://arxiv.org/abs/1605.01403}{{\tt
  1605.01403}}].

\bibitem{Jinno:2017fby}
R.~Jinno and M.~Takimoto, \emph{{Gravitational waves from bubble dynamics:
  Beyond the Envelope}},
  \href{http://dx.doi.org/10.1088/1475-7516/2019/01/060}{\emph{JCAP} {\bf 1901}
  (2019) 060}, [\href{https://arxiv.org/abs/1707.03111}{{\tt 1707.03111}}].

\bibitem{Konstandin:2017sat}
T.~Konstandin, \emph{{Gravitational radiation from a bulk flow model}},
  \href{http://dx.doi.org/10.1088/1475-7516/2018/03/047}{\emph{JCAP} {\bf 1803}
  (2018) 047}, [\href{https://arxiv.org/abs/1712.06869}{{\tt 1712.06869}}].

\bibitem{Cutting:2020nla}
D.~Cutting, E.~G. Escartin, M.~Hindmarsh and D.~J. Weir, \emph{{Gravitational
  waves from vacuum first order phase transitions II: from thin to thick
  walls}}, \href{http://dx.doi.org/10.1103/PhysRevD.103.023531}{\emph{Phys.
  Rev. D} {\bf 103} (2021) 023531},
  [\href{https://arxiv.org/abs/2005.13537}{{\tt 2005.13537}}].

\bibitem{Ellis:2019oqb}
J.~Ellis, M.~Lewicki, J.~M. No and V.~Vaskonen, \emph{{Gravitational wave
  energy budget in strongly supercooled phase transitions}},
  \href{http://dx.doi.org/10.1088/1475-7516/2019/06/024}{\emph{JCAP} {\bf 1906}
  (2019) 024}, [\href{https://arxiv.org/abs/1903.09642}{{\tt 1903.09642}}].

\bibitem{Ellis:2020nnr}
J.~Ellis, M.~Lewicki and V.~Vaskonen, \emph{{Updated predictions for
  gravitational waves produced in a strongly supercooled phase transition}},
  \href{http://dx.doi.org/10.1088/1475-7516/2020/11/020}{\emph{JCAP} {\bf 2011}
  (2020) 020}, [\href{https://arxiv.org/abs/2007.15586}{{\tt 2007.15586}}].

\bibitem{Cai:2020djd}
R.-G. Cai and S.-J. Wang, \emph{{Effective picture of bubble expansion}},
  \href{http://dx.doi.org/10.1088/1475-7516/2021/03/096}{\emph{JCAP} {\bf 2021}
  (2021) 096}, [\href{https://arxiv.org/abs/2011.11451}{{\tt 2011.11451}}].

\bibitem{Lewicki:2022pdb}
M.~Lewicki and V.~Vaskonen, \emph{{Gravitational waves from bubble collisions
  and fluid motion in strongly supercooled phase transitions}},
  \href{http://dx.doi.org/10.1140/epjc/s10052-023-11241-3}{\emph{Eur. Phys. J.
  C} {\bf 83} (2023) 109}, [\href{https://arxiv.org/abs/2208.11697}{{\tt
  2208.11697}}].

\bibitem{Espinosa:2010hh}
J.~R. Espinosa, T.~Konstandin, J.~M. No and G.~Servant, \emph{{Energy Budget of
  Cosmological First-order Phase Transitions}},
  \href{http://dx.doi.org/10.1088/1475-7516/2010/06/028}{\emph{JCAP} {\bf 1006}
  (2010) 028}, [\href{https://arxiv.org/abs/1004.4187}{{\tt 1004.4187}}].

\bibitem{Cai:2018teh}
R.-G. Cai and S.-J. Wang, \emph{{Energy budget of cosmological first-order
  phase transition in FLRW background}},
  \href{http://dx.doi.org/10.1007/s11433-018-9216-7}{\emph{Sci. China Phys.
  Mech. Astron.} {\bf 61} (2018) 080411},
  [\href{https://arxiv.org/abs/1803.03002}{{\tt 1803.03002}}].

\bibitem{Giese:2020rtr}
F.~Giese, T.~Konstandin and J.~van~de Vis, \emph{{Model-independent energy
  budget of cosmological first-order phase transitions—A sound argument to go
  beyond the bag model}},
  \href{http://dx.doi.org/10.1088/1475-7516/2020/07/057}{\emph{JCAP} {\bf 2007}
  (2020) 057}, [\href{https://arxiv.org/abs/2004.06995}{{\tt 2004.06995}}].

\bibitem{Giese:2020znk}
F.~Giese, T.~Konstandin, K.~Schmitz and J.~van~de Vis, \emph{{Model-independent
  energy budget for LISA}},
  \href{http://dx.doi.org/10.1088/1475-7516/2021/01/072}{\emph{JCAP} {\bf 01}
  (2021) 072}, [\href{https://arxiv.org/abs/2010.09744}{{\tt 2010.09744}}].

\bibitem{Wang:2020nzm}
X.~Wang, F.~P. Huang and X.~Zhang, \emph{{Energy budget and the gravitational
  wave spectra beyond the bag model}},
  \href{http://dx.doi.org/10.1103/PhysRevD.103.103520}{\emph{Phys. Rev. D} {\bf
  103} (2021) 103520}, [\href{https://arxiv.org/abs/2010.13770}{{\tt
  2010.13770}}].

\bibitem{Tenkanen:2022tly}
T.~V.~I. Tenkanen and J.~van~de Vis, \emph{{Speed of sound in cosmological
  phase transitions and effect on gravitational waves}},
  \href{http://dx.doi.org/10.1007/JHEP08(2022)302}{\emph{JHEP} {\bf 08} (2022)
  302}, [\href{https://arxiv.org/abs/2206.01130}{{\tt 2206.01130}}].

\bibitem{Wang:2022lyd}
S.-J. Wang and Z.-Y. Yuwen, \emph{{The energy budget of cosmological
  first-order phase transitions beyond the bag equation of state}},
  \href{http://dx.doi.org/10.1088/1475-7516/2022/10/047}{\emph{JCAP} {\bf 10}
  (2022) 047}, [\href{https://arxiv.org/abs/2206.01148}{{\tt 2206.01148}}].

\bibitem{Wang:2023jto}
X.~Wang, C.~Tian and F.~P. Huang, \emph{{Model-dependent analysis method for
  energy budget of the cosmological first-order phase transition}},
  \href{http://dx.doi.org/10.1088/1475-7516/2023/07/006}{\emph{JCAP} {\bf 07}
  (2023) 006}, [\href{https://arxiv.org/abs/2301.12328}{{\tt 2301.12328}}].

\bibitem{Hindmarsh:2016lnk}
M.~Hindmarsh, \emph{{Sound shell model for acoustic gravitational wave
  production at a first-order phase transition in the early Universe}},
  \href{http://dx.doi.org/10.1103/PhysRevLett.120.071301}{\emph{Phys. Rev.
  Lett.} {\bf 120} (2018) 071301},
  [\href{https://arxiv.org/abs/1608.04735}{{\tt 1608.04735}}].

\bibitem{Hindmarsh:2019phv}
M.~Hindmarsh and M.~Hijazi, \emph{{Gravitational waves from first order
  cosmological phase transitions in the Sound Shell Model}},
  \href{http://dx.doi.org/10.1088/1475-7516/2019/12/062}{\emph{JCAP} {\bf 1912}
  (2019) 062}, [\href{https://arxiv.org/abs/1909.10040}{{\tt 1909.10040}}].

\bibitem{Guo:2020grp}
H.-K. Guo, K.~Sinha, D.~Vagie and G.~White, \emph{{Phase Transitions in an
  Expanding Universe: Stochastic Gravitational Waves in Standard and
  Non-Standard Histories}},
  \href{http://dx.doi.org/10.1088/1475-7516/2021/01/001}{\emph{JCAP} {\bf 01}
  (2021) 001}, [\href{https://arxiv.org/abs/2007.08537}{{\tt 2007.08537}}].

\bibitem{Cai:2023guc}
R.-G. Cai, S.-J. Wang and Z.-Y. Yuwen, \emph{{Hydrodynamic sound shell model}},
  \href{http://dx.doi.org/10.1103/PhysRevD.108.L021502}{\emph{Phys. Rev. D}
  {\bf 108} (2023) L021502}, [\href{https://arxiv.org/abs/2305.00074}{{\tt
  2305.00074}}].

\bibitem{Cai:2019cdl}
R.-G. Cai, S.~Pi and M.~Sasaki, \emph{{Universal infrared scaling of
  gravitational wave background spectra}},
  \href{http://dx.doi.org/10.1103/PhysRevD.102.083528}{\emph{Phys. Rev. D} {\bf
  102} (2020) 083528}, [\href{https://arxiv.org/abs/1909.13728}{{\tt
  1909.13728}}].

\bibitem{Sharma:2023mao}
R.~Sharma, J.~Dahl, A.~Brandenburg and M.~Hindmarsh, \emph{{Shallow relic
  gravitational wave spectrum with acoustic peak}},
  \href{http://dx.doi.org/10.1088/1475-7516/2023/12/042}{\emph{JCAP} {\bf 12}
  (2023) 042}, [\href{https://arxiv.org/abs/2308.12916}{{\tt 2308.12916}}].

\bibitem{RoperPol:2023dzg}
A.~Roper~Pol, S.~Procacci and C.~Caprini, \emph{{Characterization of the
  gravitational wave spectrum from sound waves within the sound shell model}},
  \href{http://dx.doi.org/10.1103/PhysRevD.109.063531}{\emph{Phys. Rev. D} {\bf
  109} (2024) 063531}, [\href{https://arxiv.org/abs/2308.12943}{{\tt
  2308.12943}}].

\bibitem{Turok:1992jp}
N.~Turok, \emph{{Electroweak bubbles: Nucleation and growth}},
  \href{http://dx.doi.org/10.1103/PhysRevLett.68.1803}{\emph{Phys. Rev. Lett.}
  {\bf 68} (1992) 1803--1806}.

\bibitem{Moore:2000wx}
G.~D. Moore, \emph{{Electroweak bubble wall friction: Analytic results}},
  \href{http://dx.doi.org/10.1088/1126-6708/2000/03/006}{\emph{JHEP} {\bf 03}
  (2000) 006}, [\href{https://arxiv.org/abs/hep-ph/0001274}{{\tt
  hep-ph/0001274}}].

\bibitem{Bodeker:2009qy}
D.~Bodeker and G.~D. Moore, \emph{{Can electroweak bubble walls run away?}},
  \href{http://dx.doi.org/10.1088/1475-7516/2009/05/009}{\emph{JCAP} {\bf 0905}
  (2009) 009}, [\href{https://arxiv.org/abs/0903.4099}{{\tt 0903.4099}}].

\bibitem{Bodeker:2017cim}
D.~Bodeker and G.~D. Moore, \emph{{Electroweak Bubble Wall Speed Limit}},
  \href{http://dx.doi.org/10.1088/1475-7516/2017/05/025}{\emph{JCAP} {\bf 1705}
  (2017) 025}, [\href{https://arxiv.org/abs/1703.08215}{{\tt 1703.08215}}].

\bibitem{Hoche:2020ysm}
S.~H\"oche, J.~Kozaczuk, A.~J. Long, J.~Turner and Y.~Wang, \emph{{Towards an
  all-orders calculation of the electroweak bubble wall velocity}},
  \href{http://dx.doi.org/10.1088/1475-7516/2021/03/009}{\emph{JCAP} {\bf 03}
  (2021) 009}, [\href{https://arxiv.org/abs/2007.10343}{{\tt 2007.10343}}].

\bibitem{Gouttenoire:2021kjv}
Y.~Gouttenoire, R.~Jinno and F.~Sala, \emph{{Friction pressure on relativistic
  bubble walls}}, \href{http://dx.doi.org/10.1007/JHEP05(2022)004}{\emph{JHEP}
  {\bf 05} (2022) 004}, [\href{https://arxiv.org/abs/2112.07686}{{\tt
  2112.07686}}].

\bibitem{Ai:2023suz}
W.-Y. Ai, \emph{{Logarithmically divergent friction on ultrarelativistic bubble
  walls}}, \href{http://dx.doi.org/10.1088/1475-7516/2023/10/052}{\emph{JCAP}
  {\bf 10} (2023) 052}, [\href{https://arxiv.org/abs/2308.10679}{{\tt
  2308.10679}}].

\bibitem{Long:2024sqg}
A.~J. Long and J.~Turner, \emph{{Thermal pressure on ultrarelativistic bubbles
  from a semiclassical formalism}},
  \href{https://arxiv.org/abs/2407.18196}{{\tt 2407.18196}}.

\bibitem{GarciaGarcia:2022yqb}
I.~Garcia~Garcia, G.~Koszegi and R.~Petrossian-Byrne, \emph{{Reflections on
  bubble walls}}, \href{http://dx.doi.org/10.1007/JHEP09(2023)013}{\emph{JHEP}
  {\bf 09} (2023) 013}, [\href{https://arxiv.org/abs/2212.10572}{{\tt
  2212.10572}}].

\bibitem{Azatov:2023xem}
A.~Azatov, G.~Barni, R.~Petrossian-Byrne and M.~Vanvlasselaer,
  \emph{{Quantisation across bubble walls and friction}},
  \href{http://dx.doi.org/10.1007/JHEP05(2024)294}{\emph{JHEP} {\bf 05} (2024)
  294}, [\href{https://arxiv.org/abs/2310.06972}{{\tt 2310.06972}}].

\bibitem{BarrosoMancha:2020fay}
M.~Barroso~Mancha, T.~Prokopec and B.~Swiezewska, \emph{{Field-theoretic
  derivation of bubble-wall force}},
  \href{http://dx.doi.org/10.1007/JHEP01(2021)070}{\emph{JHEP} {\bf 01} (2021)
  070}, [\href{https://arxiv.org/abs/2005.10875}{{\tt 2005.10875}}].

\bibitem{Azatov:2020ufh}
A.~Azatov and M.~Vanvlasselaer, \emph{{Bubble wall velocity: heavy physics
  effects}}, \href{http://dx.doi.org/10.1088/1475-7516/2021/01/058}{\emph{JCAP}
  {\bf 01} (2021) 058}, [\href{https://arxiv.org/abs/2010.02590}{{\tt
  2010.02590}}].

\bibitem{Balaji:2020yrx}
S.~Balaji, M.~Spannowsky and C.~Tamarit, \emph{{Cosmological bubble friction in
  local equilibrium}},
  \href{http://dx.doi.org/10.1088/1475-7516/2021/03/051}{\emph{JCAP} {\bf 03}
  (2021) 051}, [\href{https://arxiv.org/abs/2010.08013}{{\tt 2010.08013}}].

\bibitem{Joyce:1994zt}
M.~Joyce, T.~Prokopec and N.~Turok, \emph{{Nonlocal electroweak baryogenesis.
  Part 2: The Classical regime}},
  \href{http://dx.doi.org/10.1103/PhysRevD.53.2958}{\emph{Phys. Rev. D} {\bf
  53} (1996) 2958--2980}, [\href{https://arxiv.org/abs/hep-ph/9410282}{{\tt
  hep-ph/9410282}}].

\bibitem{Joyce:1994zn}
M.~Joyce, T.~Prokopec and N.~Turok, \emph{{Nonlocal electroweak baryogenesis.
  Part 1: Thin wall regime}},
  \href{http://dx.doi.org/10.1103/PhysRevD.53.2930}{\emph{Phys. Rev. D} {\bf
  53} (1996) 2930--2957}, [\href{https://arxiv.org/abs/hep-ph/9410281}{{\tt
  hep-ph/9410281}}].

\bibitem{Moore:1995ua}
G.~D. Moore and T.~Prokopec, \emph{{Bubble wall velocity in a first order
  electroweak phase transition}},
  \href{http://dx.doi.org/10.1103/PhysRevLett.75.777}{\emph{Phys. Rev. Lett.}
  {\bf 75} (1995) 777--780}, [\href{https://arxiv.org/abs/hep-ph/9503296}{{\tt
  hep-ph/9503296}}].

\bibitem{Moore:1995si}
G.~D. Moore and T.~Prokopec, \emph{{How fast can the wall move? A Study of the
  electroweak phase transition dynamics}},
  \href{http://dx.doi.org/10.1103/PhysRevD.52.7182}{\emph{Phys. Rev.} {\bf D52}
  (1995) 7182--7204}, [\href{https://arxiv.org/abs/hep-ph/9506475}{{\tt
  hep-ph/9506475}}].

\bibitem{Huber:2013kj}
S.~J. Huber and M.~Sopena, \emph{{An efficient approach to electroweak bubble
  velocities}},  \href{https://arxiv.org/abs/1302.1044}{{\tt 1302.1044}}.

\bibitem{Konstandin:2014zta}
T.~Konstandin, G.~Nardini and I.~Rues, \emph{{From Boltzmann equations to
  steady wall velocities}},
  \href{http://dx.doi.org/10.1088/1475-7516/2014/09/028}{\emph{JCAP} {\bf 1409}
  (2014) 028}, [\href{https://arxiv.org/abs/1407.3132}{{\tt 1407.3132}}].

\bibitem{Cline:2020jre}
J.~M. Cline and K.~Kainulainen, \emph{{Electroweak baryogenesis at high bubble
  wall velocities}},
  \href{http://dx.doi.org/10.1103/PhysRevD.101.063525}{\emph{Phys. Rev. D} {\bf
  101} (2020) 063525}, [\href{https://arxiv.org/abs/2001.00568}{{\tt
  2001.00568}}].

\bibitem{Laurent:2020gpg}
B.~Laurent and J.~M. Cline, \emph{{Fluid equations for fast-moving electroweak
  bubble walls}},
  \href{http://dx.doi.org/10.1103/PhysRevD.102.063516}{\emph{Phys. Rev.} {\bf
  D102} (2020) 063516}, [\href{https://arxiv.org/abs/2007.10935}{{\tt
  2007.10935}}].

\bibitem{Dorsch:2021ubz}
G.~C. Dorsch, S.~J. Huber and T.~Konstandin, \emph{{On the wall velocity
  dependence of electroweak baryogenesis}},
  \href{http://dx.doi.org/10.1088/1475-7516/2021/08/020}{\emph{JCAP} {\bf 08}
  (2021) 020}, [\href{https://arxiv.org/abs/2106.06547}{{\tt 2106.06547}}].

\bibitem{Dorsch:2021nje}
G.~C. Dorsch, S.~J. Huber and T.~Konstandin, \emph{{A sonic boom in bubble wall
  friction}},
  \href{http://dx.doi.org/10.1088/1475-7516/2022/04/010}{\emph{JCAP} {\bf 04}
  (2022) 010}, [\href{https://arxiv.org/abs/2112.12548}{{\tt 2112.12548}}].

\bibitem{DeCurtis:2022hlx}
S.~De~Curtis, L.~D. Rose, A.~Guiggiani, A.~G. Muyor and G.~Panico,
  \emph{{Bubble wall dynamics at the electroweak phase transition}},
  \href{http://dx.doi.org/10.1007/JHEP03(2022)163}{\emph{JHEP} {\bf 03} (2022)
  163}, [\href{https://arxiv.org/abs/2201.08220}{{\tt 2201.08220}}].

\bibitem{Laurent:2022jrs}
B.~Laurent and J.~M. Cline, \emph{{First principles determination of bubble
  wall velocity}},
  \href{http://dx.doi.org/10.1103/PhysRevD.106.023501}{\emph{Phys. Rev. D} {\bf
  106} (2022) 023501}, [\href{https://arxiv.org/abs/2204.13120}{{\tt
  2204.13120}}].

\bibitem{DeCurtis:2023hil}
S.~De~Curtis, L.~Delle~Rose, A.~Guiggiani, A.~Gil~Muyor and G.~Panico,
  \emph{{Collision integrals for cosmological phase transitions}},
  \href{http://dx.doi.org/10.1007/JHEP05(2023)194}{\emph{JHEP} {\bf 05} (2023)
  194}, [\href{https://arxiv.org/abs/2303.05846}{{\tt 2303.05846}}].

\bibitem{Dorsch:2023tss}
G.~C. Dorsch and D.~A. Pinto, \emph{{Bubble wall velocities with an extended
  fluid Ansatz}},
  \href{http://dx.doi.org/10.1088/1475-7516/2024/04/027}{\emph{JCAP} {\bf 04}
  (2024) 027}, [\href{https://arxiv.org/abs/2312.02354}{{\tt 2312.02354}}].

\bibitem{DeCurtis:2024hvh}
S.~De~Curtis, L.~Delle~Rose, A.~Guiggiani, A.~Gil~Muyor and G.~Panico,
  \emph{{Non-linearities in cosmological bubble wall dynamics}},
  \href{http://dx.doi.org/10.1007/JHEP05(2024)009}{\emph{JHEP} {\bf 05} (2024)
  009}, [\href{https://arxiv.org/abs/2401.13522}{{\tt 2401.13522}}].

\bibitem{John:2000zq}
P.~John and M.~G. Schmidt, \emph{{Do stops slow down electroweak bubble
  walls?}}, \href{http://dx.doi.org/10.1016/S0550-3213(00)00768-9,
  10.1016/S0550-3213(02)01014-3}{\emph{Nucl. Phys.} {\bf B598} (2001)
  291--305}, [\href{https://arxiv.org/abs/hep-ph/0002050}{{\tt
  hep-ph/0002050}}].

\bibitem{Cline:2000nw}
J.~M. Cline, M.~Joyce and K.~Kainulainen, \emph{{Supersymmetric electroweak
  baryogenesis}},
  \href{http://dx.doi.org/10.1088/1126-6708/2000/07/018}{\emph{JHEP} {\bf 07}
  (2000) 018}, [\href{https://arxiv.org/abs/hep-ph/0006119}{{\tt
  hep-ph/0006119}}].

\bibitem{Carena:2000id}
M.~Carena, J.~M. Moreno, M.~Quiros, M.~Seco and C.~E.~M. Wagner,
  \emph{{Supersymmetric CP violating currents and electroweak baryogenesis}},
  \href{http://dx.doi.org/10.1016/S0550-3213(01)00032-3}{\emph{Nucl. Phys.}
  {\bf B599} (2001) 158--184},
  [\href{https://arxiv.org/abs/hep-ph/0011055}{{\tt hep-ph/0011055}}].

\bibitem{Huber:2001xf}
S.~J. Huber, P.~John and M.~G. Schmidt, \emph{{Bubble walls, CP violation and
  electroweak baryogenesis in the MSSM}},
  \href{http://dx.doi.org/10.1007/PL00022989}{\emph{Eur. Phys. J. C} {\bf 20}
  (2001) 695--711}, [\href{https://arxiv.org/abs/hep-ph/0101249}{{\tt
  hep-ph/0101249}}].

\bibitem{Carena:2002ss}
M.~Carena, M.~Quiros, M.~Seco and C.~E.~M. Wagner, \emph{{Improved Results in
  Supersymmetric Electroweak Baryogenesis}},
  \href{http://dx.doi.org/10.1016/S0550-3213(02)01065-9}{\emph{Nucl. Phys.}
  {\bf B650} (2003) 24--42}, [\href{https://arxiv.org/abs/hep-ph/0208043}{{\tt
  hep-ph/0208043}}].

\bibitem{Konstandin:2005cd}
T.~Konstandin, T.~Prokopec, M.~G. Schmidt and M.~Seco, \emph{{MSSM electroweak
  baryogenesis and flavor mixing in transport equations}},
  \href{http://dx.doi.org/10.1016/j.nuclphysb.2005.11.028}{\emph{Nucl. Phys.}
  {\bf B738} (2006) 1--22}, [\href{https://arxiv.org/abs/hep-ph/0505103}{{\tt
  hep-ph/0505103}}].

\bibitem{Cirigliano:2006dg}
V.~Cirigliano, S.~Profumo and M.~J. Ramsey-Musolf, \emph{{Baryogenesis,
  Electric Dipole Moments and Dark Matter in the MSSM}},
  \href{http://dx.doi.org/10.1088/1126-6708/2006/07/002}{\emph{JHEP} {\bf 07}
  (2006) 002}, [\href{https://arxiv.org/abs/hep-ph/0603246}{{\tt
  hep-ph/0603246}}].

\bibitem{Kozaczuk:2015owa}
J.~Kozaczuk, \emph{{Bubble Expansion and the Viability of Singlet-Driven
  Electroweak Baryogenesis}},
  \href{http://dx.doi.org/10.1007/JHEP10(2015)135}{\emph{JHEP} {\bf 10} (2015)
  135}, [\href{https://arxiv.org/abs/1506.04741}{{\tt 1506.04741}}].

\bibitem{Dorsch:2018pat}
G.~C. Dorsch, S.~J. Huber and T.~Konstandin, \emph{{Bubble wall velocities in
  the Standard Model and beyond}},
  \href{http://dx.doi.org/10.1088/1475-7516/2018/12/034}{\emph{JCAP} {\bf 12}
  (2018) 034}, [\href{https://arxiv.org/abs/1809.04907}{{\tt 1809.04907}}].

\bibitem{Friedlander:2020tnq}
A.~Friedlander, I.~Banta, J.~M. Cline and D.~Tucker-Smith, \emph{{Wall speed
  and shape in singlet-assisted strong electroweak phase transitions}},
  \href{https://arxiv.org/abs/2009.14295}{{\tt 2009.14295}}.

\bibitem{Wang:2020zlf}
X.~Wang, F.~P. Huang and X.~Zhang, \emph{{Bubble wall velocity beyond
  leading-log approximation in electroweak phase transition}},
  \href{https://arxiv.org/abs/2011.12903}{{\tt 2011.12903}}.

\bibitem{Lewicki:2021pgr}
M.~Lewicki, M.~Merchand and M.~Zych, \emph{{Electroweak bubble wall expansion:
  gravitational waves and baryogenesis in Standard Model-like thermal plasma}},
  \href{http://dx.doi.org/10.1007/JHEP02(2022)017}{\emph{JHEP} {\bf 02} (2022)
  017}, [\href{https://arxiv.org/abs/2111.02393}{{\tt 2111.02393}}].

\bibitem{Cline:2021iff}
J.~M. Cline, A.~Friedlander, D.-M. He, K.~Kainulainen, B.~Laurent and
  D.~Tucker-Smith, \emph{{Baryogenesis and gravity waves from a UV-completed
  electroweak phase transition}},
  \href{http://dx.doi.org/10.1103/PhysRevD.103.123529}{\emph{Phys. Rev. D} {\bf
  103} (2021) 123529}, [\href{https://arxiv.org/abs/2102.12490}{{\tt
  2102.12490}}].

\bibitem{Jiang:2022btc}
S.~Jiang, F.~P. Huang and X.~Wang, \emph{{Bubble wall velocity during
  electroweak phase transition in the inert doublet model}},
  \href{http://dx.doi.org/10.1103/PhysRevD.107.095005}{\emph{Phys. Rev. D} {\bf
  107} (2023) 095005}, [\href{https://arxiv.org/abs/2211.13142}{{\tt
  2211.13142}}].

\bibitem{Li:2023xto}
L.~Li, S.-J. Wang and Z.-Y. Yuwen, \emph{{Bubble expansion at strong
  coupling}}, \href{http://dx.doi.org/10.1103/PhysRevD.108.096033}{\emph{Phys.
  Rev. D} {\bf 108} (2023) 096033},
  [\href{https://arxiv.org/abs/2302.10042}{{\tt 2302.10042}}].

\bibitem{Wang:2023lam}
J.-C. Wang, Z.-Y. Yuwen, Y.-S. Hao and S.-J. Wang, \emph{{General bubble
  expansion at strong coupling}},
  \href{http://dx.doi.org/10.1103/PhysRevD.109.096012}{\emph{Phys. Rev. D} {\bf
  109} (2024) 096012}, [\href{https://arxiv.org/abs/2311.07347}{{\tt
  2311.07347}}].

\bibitem{Kang:2024xqk}
Z.~Kang and J.~Zhu, \emph{{Confinement Bubble Wall Velocity via Quasiparticle
  Determination}},  \href{https://arxiv.org/abs/2401.03849}{{\tt 2401.03849}}.

\bibitem{Bea:2021zsu}
Y.~Bea, J.~Casalderrey-Solana, T.~Giannakopoulos, D.~Mateos,
  M.~Sanchez-Garitaonandia and M.~Zilh\~ao, \emph{{Bubble wall velocity from
  holography}},
  \href{http://dx.doi.org/10.1103/PhysRevD.104.L121903}{\emph{Phys. Rev. D}
  {\bf 104} (2021) L121903}, [\href{https://arxiv.org/abs/2104.05708}{{\tt
  2104.05708}}].

\bibitem{Bigazzi:2021ucw}
F.~Bigazzi, A.~Caddeo, T.~Canneti and A.~L. Cotrone, \emph{{Bubble wall
  velocity at strong coupling}},
  \href{http://dx.doi.org/10.1007/JHEP08(2021)090}{\emph{JHEP} {\bf 08} (2021)
  090}, [\href{https://arxiv.org/abs/2104.12817}{{\tt 2104.12817}}].

\bibitem{Janik:2022wsx}
R.~A. Janik, M.~Jarvinen, H.~Soltanpanahi and J.~Sonnenschein, \emph{{Perfect
  Fluid Hydrodynamic Picture of Domain Wall Velocities at Strong Coupling}},
  \href{http://dx.doi.org/10.1103/PhysRevLett.129.081601}{\emph{Phys. Rev.
  Lett.} {\bf 129} (2022) 081601},
  [\href{https://arxiv.org/abs/2205.06274}{{\tt 2205.06274}}].

\bibitem{Bea:2022mfb}
Y.~Bea, J.~Casalderrey-Solana, T.~Giannakopoulos, A.~Jansen, D.~Mateos,
  M.~Sanchez-Garitaonandia et~al., \emph{{Holographic bubbles with Jecco:
  expanding, collapsing and critical}},
  \href{http://dx.doi.org/10.1007/JHEP09(2022)008}{\emph{JHEP} {\bf 09} (2022)
  008}, [\href{https://arxiv.org/abs/2202.10503}{{\tt 2202.10503}}].

\bibitem{Bigazzi:2020avc}
F.~Bigazzi, A.~Caddeo, A.~L. Cotrone and A.~Paredes, \emph{{Dark Holograms and
  Gravitational Waves}},
  \href{http://dx.doi.org/10.1007/JHEP04(2021)094}{\emph{JHEP} {\bf 04} (2021)
  094}, [\href{https://arxiv.org/abs/2011.08757}{{\tt 2011.08757}}].

\bibitem{Ares:2020lbt}
F.~R. Ares, M.~Hindmarsh, C.~Hoyos and N.~Jokela, \emph{{Gravitational waves
  from a holographic phase transition}},
  \href{http://dx.doi.org/10.1007/JHEP04(2021)100}{\emph{JHEP} {\bf 21} (2020)
  100}, [\href{https://arxiv.org/abs/2011.12878}{{\tt 2011.12878}}].

\bibitem{Bigazzi:2020phm}
F.~Bigazzi, A.~Caddeo, A.~L. Cotrone and A.~Paredes, \emph{{Fate of false vacua
  in holographic first-order phase transitions}},
  \href{http://dx.doi.org/10.1007/JHEP12(2020)200}{\emph{JHEP} {\bf 12} (2020)
  200}, [\href{https://arxiv.org/abs/2008.02579}{{\tt 2008.02579}}].

\bibitem{Zhu:2021vkj}
Z.-R. Zhu, J.~Chen and D.~Hou, \emph{{Gravitational waves from holographic QCD
  phase transition with gluon condensate}},
  \href{http://dx.doi.org/10.1140/epja/s10050-022-00754-2}{\emph{Eur. Phys. J.
  A} {\bf 58} (2022) 104}, [\href{https://arxiv.org/abs/2109.09933}{{\tt
  2109.09933}}].

\bibitem{Ares:2021ntv}
F.~R. Ares, O.~Henriksson, M.~Hindmarsh, C.~Hoyos and N.~Jokela,
  \emph{{Effective actions and bubble nucleation from holography}},
  \href{http://dx.doi.org/10.1103/PhysRevD.105.066020}{\emph{Phys. Rev. D} {\bf
  105} (2022) 066020}, [\href{https://arxiv.org/abs/2109.13784}{{\tt
  2109.13784}}].

\bibitem{Ares:2021nap}
F.~R. Ares, O.~Henriksson, M.~Hindmarsh, C.~Hoyos and N.~Jokela,
  \emph{{Gravitational Waves at Strong Coupling from an Effective Action}},
  \href{http://dx.doi.org/10.1103/PhysRevLett.128.131101}{\emph{Phys. Rev.
  Lett.} {\bf 128} (2022) 131101},
  [\href{https://arxiv.org/abs/2110.14442}{{\tt 2110.14442}}].

\bibitem{Morgante:2022zvc}
E.~Morgante, N.~Ramberg and P.~Schwaller, \emph{{Gravitational waves from dark
  SU(3) Yang-Mills theory}},
  \href{http://dx.doi.org/10.1103/PhysRevD.107.036010}{\emph{Phys. Rev. D} {\bf
  107} (2023) 036010}, [\href{https://arxiv.org/abs/2210.11821}{{\tt
  2210.11821}}].

\bibitem{Cai:2022omk}
R.-G. Cai, S.~He, L.~Li and Y.-X. Wang, \emph{{Probing QCD critical point and
  induced gravitational wave by black hole physics}},
  \href{http://dx.doi.org/10.1103/PhysRevD.106.L121902}{\emph{Phys. Rev. D}
  {\bf 106} (2022) L121902}, [\href{https://arxiv.org/abs/2201.02004}{{\tt
  2201.02004}}].

\bibitem{Zhao:2022uxc}
Y.-Q. Zhao, S.~He, D.~Hou, L.~Li and Z.~Li, \emph{{Phase diagram of holographic
  thermal dense QCD matter with rotation}},
  \href{http://dx.doi.org/10.1007/JHEP04(2023)115}{\emph{JHEP} {\bf 04} (2023)
  115}, [\href{https://arxiv.org/abs/2212.14662}{{\tt 2212.14662}}].

\bibitem{Li:2023mpv}
Z.~Li, J.~Liang, S.~He and L.~Li, \emph{{Holographic study of higher-order
  baryon number susceptibilities at finite temperature and density}},
  \href{http://dx.doi.org/10.1103/PhysRevD.108.046008}{\emph{Phys. Rev. D} {\bf
  108} (2023) 046008}, [\href{https://arxiv.org/abs/2305.13874}{{\tt
  2305.13874}}].

\bibitem{He:2023ado}
S.~He, L.~Li, S.~Wang and S.-J. Wang, \emph{{Constraints on holographic QCD
  phase transitions from PTA observations}},
  \href{https://arxiv.org/abs/2308.07257}{{\tt 2308.07257}}.

\bibitem{Wang:2022txy}
S.-J. Wang and Z.-Y. Yuwen, \emph{{Hydrodynamic backreaction force of
  cosmological bubble expansion}},
  \href{http://dx.doi.org/10.1103/PhysRevD.107.023501}{\emph{Phys. Rev. D} {\bf
  107} (2023) 023501}, [\href{https://arxiv.org/abs/2205.02492}{{\tt
  2205.02492}}].

\bibitem{Wang:2023kux}
J.-C. Wang, Z.-Y. Yuwen, Y.-S. Hao and S.-J. Wang, \emph{{General backreaction
  force of cosmological bubble expansion}},
  \href{http://dx.doi.org/10.1103/PhysRevD.110.016031}{\emph{Phys. Rev. D} {\bf
  110} (2024) 016031}, [\href{https://arxiv.org/abs/2310.07691}{{\tt
  2310.07691}}].

\bibitem{Enqvist:1991xw}
K.~Enqvist, J.~Ignatius, K.~Kajantie and K.~Rummukainen, \emph{{Nucleation and
  bubble growth in a first order cosmological electroweak phase transition}},
  \href{http://dx.doi.org/10.1103/PhysRevD.45.3415}{\emph{Phys. Rev.} {\bf D45}
  (1992) 3415--3428}.

\bibitem{Dine:1992wr}
M.~Dine, R.~G. Leigh, P.~Y. Huet, A.~D. Linde and D.~A. Linde, \emph{{Towards
  the theory of the electroweak phase transition}},
  \href{http://dx.doi.org/10.1103/PhysRevD.46.550}{\emph{Phys. Rev.} {\bf D46}
  (1992) 550--571}, [\href{https://arxiv.org/abs/hep-ph/9203203}{{\tt
  hep-ph/9203203}}].

\bibitem{Liu:1992tn}
B.-H. Liu, L.~D. McLerran and N.~Turok, \emph{{Bubble nucleation and growth at
  a baryon number producing electroweak phase transition}},
  \href{http://dx.doi.org/10.1103/PhysRevD.46.2668}{\emph{Phys. Rev.} {\bf D46}
  (1992) 2668--2688}.

\bibitem{Laine:1993ey}
M.~Laine, \emph{{Bubble growth as a detonation}},
  \href{http://dx.doi.org/10.1103/PhysRevD.49.3847}{\emph{Phys. Rev.} {\bf D49}
  (1994) 3847--3853}, [\href{https://arxiv.org/abs/hep-ph/9309242}{{\tt
  hep-ph/9309242}}].

\bibitem{Ignatius:1993qn}
J.~Ignatius, K.~Kajantie, H.~Kurki-Suonio and M.~Laine, \emph{{The growth of
  bubbles in cosmological phase transitions}},
  \href{http://dx.doi.org/10.1103/PhysRevD.49.3854}{\emph{Phys. Rev.} {\bf D49}
  (1994) 3854--3868}, [\href{https://arxiv.org/abs/astro-ph/9309059}{{\tt
  astro-ph/9309059}}].

\bibitem{Carrington:1993ng}
M.~E. Carrington and J.~I. Kapusta, \emph{{Dynamics of the electroweak phase
  transition}}, \href{http://dx.doi.org/10.1103/PhysRevD.47.5304}{\emph{Phys.
  Rev.} {\bf D47} (1993) 5304--5315}.

\bibitem{Heckler:1994uu}
A.~F. Heckler, \emph{{The Effects of electroweak phase transition dynamics on
  baryogenesis and primordial nucleosynthesis}},
  \href{http://dx.doi.org/10.1103/PhysRevD.51.405}{\emph{Phys. Rev.} {\bf D51}
  (1995) 405--428}, [\href{https://arxiv.org/abs/astro-ph/9407064}{{\tt
  astro-ph/9407064}}].

\bibitem{Kurki-Suonio:1995rrv}
H.~Kurki-Suonio and M.~Laine, \emph{{Supersonic deflagrations in cosmological
  phase transitions}},
  \href{http://dx.doi.org/10.1103/PhysRevD.51.5431}{\emph{Phys. Rev. D} {\bf
  51} (1995) 5431--5437}, [\href{https://arxiv.org/abs/hep-ph/9501216}{{\tt
  hep-ph/9501216}}].

\bibitem{KurkiSuonio:1996rk}
H.~Kurki-Suonio and M.~Laine, \emph{{Real time history of the cosmological
  electroweak phase transition}},
  \href{http://dx.doi.org/10.1103/PhysRevLett.77.3951}{\emph{Phys. Rev. Lett.}
  {\bf 77} (1996) 3951--3954},
  [\href{https://arxiv.org/abs/hep-ph/9607382}{{\tt hep-ph/9607382}}].

\bibitem{Lewicki:2022nba}
M.~Lewicki, V.~Vaskonen and H.~Veerm\"ae, \emph{{Bubble dynamics in fluids with
  N-body simulations}},
  \href{http://dx.doi.org/10.1103/PhysRevD.106.103501}{\emph{Phys. Rev. D} {\bf
  106} (2022) 103501}, [\href{https://arxiv.org/abs/2205.05667}{{\tt
  2205.05667}}].

\bibitem{Krajewski:2023clt}
T.~Krajewski, M.~Lewicki and M.~Zych, \emph{{Hydrodynamical constraints on the
  bubble wall velocity}},
  \href{http://dx.doi.org/10.1103/PhysRevD.108.103523}{\emph{Phys. Rev. D} {\bf
  108} (2023) 103523}, [\href{https://arxiv.org/abs/2303.18216}{{\tt
  2303.18216}}].

\bibitem{Krajewski:2024gma}
T.~Krajewski, M.~Lewicki and M.~Zych, \emph{{Bubble-wall velocity in local
  thermal equilibrium: hydrodynamical simulations vs analytical treatment}},
  \href{http://dx.doi.org/10.1007/JHEP05(2024)011}{\emph{JHEP} {\bf 05} (2024)
  011}, [\href{https://arxiv.org/abs/2402.15408}{{\tt 2402.15408}}].

\bibitem{Megevand:2009ut}
A.~Megevand and A.~D. Sanchez, \emph{{Detonations and deflagrations in
  cosmological phase transitions}},
  \href{http://dx.doi.org/10.1016/j.nuclphysb.2009.05.007}{\emph{Nucl. Phys.}
  {\bf B820} (2009) 47--74}, [\href{https://arxiv.org/abs/0904.1753}{{\tt
  0904.1753}}].

\bibitem{Megevand:2009gh}
A.~Megevand and A.~D. Sanchez, \emph{{Velocity of electroweak bubble walls}},
  \href{http://dx.doi.org/10.1016/j.nuclphysb.2009.09.019}{\emph{Nucl. Phys.}
  {\bf B825} (2010) 151--176}, [\href{https://arxiv.org/abs/0908.3663}{{\tt
  0908.3663}}].

\bibitem{Leitao:2010yw}
L.~Leitao and A.~Megevand, \emph{{Spherical and non-spherical bubbles in
  cosmological phase transitions}},
  \href{http://dx.doi.org/10.1016/j.nuclphysb.2010.11.012}{\emph{Nucl. Phys. B}
  {\bf 844} (2011) 450--470}, [\href{https://arxiv.org/abs/1010.2134}{{\tt
  1010.2134}}].

\bibitem{Huber:2011aa}
S.~J. Huber and M.~Sopena, \emph{{The bubble wall velocity in the minimal
  supersymmetric light stop scenario}},
  \href{http://dx.doi.org/10.1103/PhysRevD.85.103507}{\emph{Phys. Rev.} {\bf
  D85} (2012) 103507}, [\href{https://arxiv.org/abs/1112.1888}{{\tt
  1112.1888}}].

\bibitem{Megevand:2012rt}
A.~Megevand and A.~D. Sanchez, \emph{{Analytic approach to the motion of
  cosmological phase transition fronts}},
  \href{http://dx.doi.org/10.1016/j.nuclphysb.2012.08.001}{\emph{Nucl. Phys. B}
  {\bf 865} (2012) 217--237}, [\href{https://arxiv.org/abs/1206.2339}{{\tt
  1206.2339}}].

\bibitem{Megevand:2013hwa}
A.~Mégevand, \emph{{Friction forces on phase transition fronts}},
  \href{http://dx.doi.org/10.1088/1475-7516/2013/07/045}{\emph{JCAP} {\bf 1307}
  (2013) 045}, [\href{https://arxiv.org/abs/1303.4233}{{\tt 1303.4233}}].

\bibitem{Megevand:2013yua}
A.~Megevand and F.~A. Membiela, \emph{{Stability of cosmological deflagration
  fronts}}, \href{http://dx.doi.org/10.1103/PhysRevD.89.103507}{\emph{Phys.
  Rev.} {\bf D89} (2014) 103507}, [\href{https://arxiv.org/abs/1311.2453}{{\tt
  1311.2453}}].

\bibitem{Megevand:2014yua}
A.~Megevand and F.~A. Membiela, \emph{{Stability of cosmological detonation
  fronts}}, \href{http://dx.doi.org/10.1103/PhysRevD.89.103503}{\emph{Phys.
  Rev.} {\bf D89} (2014) 103503}, [\href{https://arxiv.org/abs/1402.5791}{{\tt
  1402.5791}}].

\bibitem{Leitao:2014pda}
L.~Leitao and A.~Megevand, \emph{{Hydrodynamics of phase transition fronts and
  the speed of sound in the plasma}},
  \href{http://dx.doi.org/10.1016/j.nuclphysb.2014.12.008}{\emph{Nucl. Phys.}
  {\bf B891} (2015) 159--199}, [\href{https://arxiv.org/abs/1410.3875}{{\tt
  1410.3875}}].

\bibitem{Megevand:2014dua}
A.~Megevand, F.~A. Membiela and A.~D. Sanchez, \emph{{Lower bound on the
  electroweak wall velocity from hydrodynamic instability}},
  \href{http://dx.doi.org/10.1088/1475-7516/2015/03/051}{\emph{JCAP} {\bf 1503}
  (2015) 051}, [\href{https://arxiv.org/abs/1412.8064}{{\tt 1412.8064}}].

\bibitem{Leitao:2015ola}
L.~Leitao and A.~Megevand, \emph{{Hydrodynamics of ultra-relativistic bubble
  walls}}, \href{http://dx.doi.org/10.1016/j.nuclphysb.2016.02.009}{\emph{Nucl.
  Phys.} {\bf B905} (2016) 45--72},
  [\href{https://arxiv.org/abs/1510.07747}{{\tt 1510.07747}}].

\bibitem{Ai:2021kak}
W.-Y. Ai, B.~Garbrecht and C.~Tamarit, \emph{{Bubble wall velocities in local
  equilibrium}},
  \href{http://dx.doi.org/10.1088/1475-7516/2022/03/015}{\emph{JCAP} {\bf 03}
  (2022) 015}, [\href{https://arxiv.org/abs/2109.13710}{{\tt 2109.13710}}].

\bibitem{Ai:2023see}
W.-Y. Ai, B.~Laurent and J.~van~de Vis, \emph{{Model-independent bubble wall
  velocities in local thermal equilibrium}},
  \href{http://dx.doi.org/10.1088/1475-7516/2023/07/002}{\emph{JCAP} {\bf 07}
  (2023) 002}, [\href{https://arxiv.org/abs/2303.10171}{{\tt 2303.10171}}].

\bibitem{Ai:2024shx}
W.-Y. Ai, X.~Nagels and M.~Vanvlasselaer, \emph{{Criterion for ultra-fast
  bubble walls: the impact of hydrodynamic obstruction}},
  \href{http://dx.doi.org/10.1088/1475-7516/2024/03/037}{\emph{JCAP} {\bf 03}
  (2024) 037}, [\href{https://arxiv.org/abs/2401.05911}{{\tt 2401.05911}}].

\bibitem{Jackiw:1974cv}
R.~Jackiw, \emph{{Functional evaluation of the effective potential}},
  \href{http://dx.doi.org/10.1103/PhysRevD.9.1686}{\emph{Phys. Rev.} {\bf D9}
  (1974) 1686}.

\bibitem{Dolan:1973qd}
L.~Dolan and R.~Jackiw, \emph{{Symmetry Behavior at Finite Temperature}},
  \href{http://dx.doi.org/10.1103/PhysRevD.9.3320}{\emph{Phys. Rev.} {\bf D9}
  (1974) 3320--3341}.

\bibitem{Taub:1948zz}
A.~H. Taub, \emph{{Relativistic Rankine-Hugoniot Equations}},
  \href{http://dx.doi.org/10.1103/PhysRev.74.328}{\emph{Phys. Rev.} {\bf 74}
  (1948) 328--334}.

\bibitem{Thorne:1973}
K.~S. {Thorne}, \emph{{Relativistic Shocks: the Taub Adiabat}},
  \href{http://dx.doi.org/10.1086/151927}{\emph{Astrophys.J.} {\bf 179} (Feb.,
  1973) 897--908}.

\bibitem{Mallick:2022ubp}
R.~Mallick and A.~Verma, \emph{{Study of general relativistic shocks and their
  propagation in neutron stars}},
  \href{http://dx.doi.org/10.1016/j.jheap.2022.07.005}{\emph{JHEAp} {\bf 36}
  (2022) 36--47}.

\end{thebibliography}\endgroup

\end{document}